\documentclass{agujournal2019}
\usepackage{url} % this package should fix any errors with uRLs in refs.
\usepackage{lineno}
\usepackage{bm}
\usepackage{layouts, microtype}
\usepackage{amsmath, amssymb, amsfonts}

\usepackage{soul} 
\usepackage{scalerel, stackengine}
\setstackEOL{\#}
\stackMath
\def\hatgap{2pt}
\def\subdown{-2pt}
\newcommand\reallywidehat[2][]{ \renewcommand\stackalignment{l} \stackon[\hatgap]{#2}{ \stretchto{
    \scalerel*[\widthof{$#2$}]{\kern-.6pt\bigwedge\kern-.6pt}
    {\rule[-\textheight/2]{1ex}{\textheight}}}
    {0.5ex}_{\smash{ \belowbaseline[\subdown]{\scriptstyle#1} }}
}}

\newcommand{\mystrut}[1]{\vrule width0pt height0pt depth#1\relax}

\renewcommand{\b}[1]    {\boldsymbol{#1}}

\renewcommand{\r}[1]    {\mathrm{#1}}

\renewcommand{\d}       {\partial}
\newcommand{\bu}        {\boldsymbol u}

\newcommand{\bUS}       {\boldsymbol U^\mathrm{S}}
\newcommand{\bxh}       {\hspace{0.1em}\boldsymbol{\hat x}} 
 
\newcommand{\bzh}       {\hspace{0.1em}\boldsymbol{\hat z}} 

\newcommand{\ee}        {\r{e}}

\newcommand{\dd}        {\r{d}} 
\newcommand{\di}        {{\, \dd}}

\newcommand{\half}      {\tfrac{1}{2}}

\newcommand{\beq}       {\begin{linenomath*}\begin{equation}}
\newcommand{\eeq}       {\end{equation}\end{linenomath*}}
\newcommand{\beqs}      {\begin{gather}}
\newcommand{\eeqs}      {\end{gather}}

\newcommand{\defn}      {\ensuremath{\stackrel{\r{def}}{=}}}

\newcommand{\bnabla}    {\b{\nabla}}

\newcommand{\lap}       {\nabla^2}
\newcommand{\bcdot}     {\b{\cdot}}

\newcommand{\Ri}        {Ri}
\newcommand{\Ei}        {\mathrm{Ei}}
\newcommand{\La}        {La}
\renewcommand{\Pr}      {Pr}
\newcommand{\Sc}        {Sc}
\newcommand{\K}         {\kappa}
\newcommand{\s}{\mathbb{S}}

\newcommand{\lo}{0}
\newcommand{\hi}{\infty}

\newcommand{\com}       {\, ,}
\newcommand{\per}       {\, .}

\newcommand{\bUL}       {\b{U}^\mathrm{L}}
\newcommand{\US}        {U^\mathrm{S}}

\newcommand{\UL}        {U^\mathrm{L}}
\newcommand{\VL}        {V^\mathrm{L}}
\newcommand{\WL}        {W^\mathrm{L}}

\newcommand{\wL}        {w^\mathrm{L}}

\newcommand{\C}[2]      {\mathbb{C}^{#1}_{#2}}
\newcommand{\y}         {\mathcal{Y}}
\newcommand{\cov}       {\mathcal{K}}
\renewcommand{\t}       {\mathcal{T}}
\renewcommand{\c}       {\mathbb{C}}
\newcommand{\CC}       {\mathbf{C}}
\newcommand{\GC}       {\mathbf{G}}
\newcommand{\G}        {\mathcal{G}}
\newcommand{\M}        {\mathcal{M}}

\newcommand{\E}{\mathcal{E}}

\newcommand{\B}       {\mathcal{B}}

\newcommand{\erf}{{\rm erf}}

\usepackage{diagbox}
\usepackage{xfrac}

\draftfalse

\journalname{Journal of Advances in Modeling Earth Systems (JAMES)}

\hypersetup{breaklinks, colorlinks=true, linkcolor=blue, urlcolor=blue, citecolor=RoyalPurple}

\begin{document}
\justify

%\title{Formulation and calibration of a turbulence parameterization with dynamic convective adjusment and prognostic turbulent kinetic energy}

\title{Formulation and calibration of CATKE, a one-equation parameterization for microscale ocean mixing}

\authors{
    Gregory~LeClaire~Wagner\affil{1},
    Adeline~Hillier\affil{1},
    Navid~C.~Constantinou\affil{2, 3},
    Simone~Silvestri\affil{1},
    Andre~Souza\affil{1},
    Keaton~J.~Burns\affil{1},
    Chris~Hill\affil{1},
    Jean-Michel~Campin\affil{1},
    John~Marshall\affil{1}, and
    Raffaele~Ferrari\affil{1}
}

\affiliation{1}{Massachusetts Institute of Technology, Cambridge, MA, USA}
\affiliation{2}{University of Melbourne, Parkville, VIC, Australia}
\affiliation{3}{ARC Center of Excellence for the Weather of the 21st Century, Parkville, VIC, Australia}

\correspondingauthor{Gregory L. Wagner}{\href{mailto:wagner.greg@gmail.com}{wagner.greg@gmail.com}}

%% Keypoints, final entry on title page.

%  List up to three key points (at least one is required)
%  Key Points summarize the main points and conclusions of the article
%  Each must be 140 characters or fewer with no special characters or punctuation and must be complete sentences

% Example:
\begin{keypoints}
\item We describe a new parameterization called CATKE with a convective adjustment (CA) component and prognostic turbulent kinetic energy (TKE).
\item We use Ensemble Kalman Inversion to calibrate CATKE's free parameters against 21 idealized large eddy simulations (LES).
\item We validate CATKE by interpreting its free parameters and comparing to additional idealized and realistic LES.
\end{keypoints}

\begin{abstract}
We describe CATKE, a parameterization for fluxes associated with small-scale or ``microscale'' ocean turbulent mixing on scales between 1 and 100 meters.
CATKE uses a downgradient formulation that depends on a prognostic turbulent kinetic energy (TKE) variable and a diagnostic mixing length scale that includes a dynamic convective adjustment (CA) component. 
With its dynamic convective mixing length, CATKE predicts not just the depth spanned by convective plumes but also the characteristic convective mixing timescale, an important aspect of turbulent convection not captured by simpler static convective adjustment schemes.
As a result, CATKE can describe the competition between convection and other processes such as shear-driven mixing and baroclinic restratification.
To calibrate CATKE, we use Ensemble Kalman Inversion to minimize the error between 21~large eddy simulations (LES) and predictions of the LES data by CATKE-parameterized single column simulations at three different vertical resolutions.
We find that CATKE makes accurate predictions of both idealized and realistic LES compared to microscale turbulence parameterizations commonly used in climate models.\vspace{-1em}
\end{abstract}

\section*{Plain Language Summary}
Turbulence is everywhere in the Earth's ocean, from ephemeral swirls no bigger than a fingertip to gigantic eddies larger than Iceland. Ocean models used in climate studies simulate currents by dividing the ocean into grid cells between 10 and 100 kilometers wide. As a result, ocean models do a decent job simulating eddies that are significantly larger than a single grid cell. But models do far worse at incorporating the effects of eddies that are person- to building-sized --- because these ``microscale' eddies are smaller than a grid cell and therefore must be represented more approximately. This is a problem because these small yet mighty eddies mix heat and carbon deep into the ocean, and thus help keep the atmosphere from getting too hot, and too rich in CO$_2$. In this paper, we propose a new model component called ``CATKE' (pronounced \textit{k$\breve{a}$t-kee}) that does a decent job at approximately incorporating the effect of such relatively small ocean eddies in global ocean models. CATKE stands for ``Convective Adjustment and Turbulent Kinetic Energy". Basically, CATKE keeps track of the \textit{energy} of small-scale turbulence --- a measure of how vigorous it is, and thus how much it mixes the ocean --- to predict ocean mixing rates.
%% ------------------------------------------------------------------------ %%
%
%  TEXT
%
%% ------------------------------------------------------------------------ %%

\section{Introduction}
\label{introduction}

Vertical mixing by ``microscale'' ocean turbulence, with scales between 1 and 100 meters, is an important process affecting, for example, ocean uptake of atmospheric heat and carbon~\cite{price1986diurnal, large1994oceanic, omand2015eddy}, the structure of the ocean interior~\cite{luyten1983ventilated, williams1991role}, and ocean circulation on decadal to millennial time-scales~\cite{WunschFerrari04,meletetal2025}.
In large-scale ocean models ---~from regional models covering tens of kilometers to global ocean models~--- microscale turbulent vertical fluxes are approximately modeled by parameterizations.
Imperfect predictions by turbulence parameterizations contribute to biases in tropical sea surface temperature \cite{li2014tropical}, Southern Ocean boundary layer depth \cite{sallee2013assessment, duvivier2018argo}, and water mass transformation rates~\cite{groeskamp2019water}.
These errors degrade the accuracy of climate projections that depend on accurate air-sea fluxes \cite<sensitive to sea surface temperature,>{large1994oceanic} and the effective heat capacity of the upper ocean \cite<which scales with the boundary layer depth,>{gregory2000vertical, held2010probing}.

This paper documents the development, calibration, and preliminary validation of a new parameterization for vertical mixing by ocean microscale turbulence.
Our goal is to use the new parameterization in a GPU-based climate model that is automatically calibrated to observations, reports quantified uncertainties, and has an ocean component with a high, $O(10 \, \mathrm{km})$ or finer resolution that fully resolves ocean mesoscale turbulence.
The dynamical core of the GPU-based ocean component is described by \citeA{silvestri2024mesoscaleGPU}.
In service of this ultimate goal, the work documented in this paper prioritizes not just accurate predictions, but also efficiency on GPUs in high-resolution configurations.
We also invest in automated calibration that constrains all of the parameterization's free parameters to 21~large eddy simulations (LESs) simultaneously, accounting for the peculiarities of our specific numerical implementation of the parameterization in a single column model.
The 21~LES we use to calibrate and the additional 14~LES we use to validate the parameterization are described in section~\ref{sec:les-data}.
%The experience of this paper shows that automated calibration is essential for informing the development of ``goldilocks'' parameterization that are simple and accurate, having neither too many --- nor too few! --- free parameters.

Our new parameterization, which we call ``CATKE'', uses a downgradient formulation that estimates eddy diffusivities in terms of a prognostic turbulent kinetic energy (TKE) variable and a diagnostic mixing length with a novel dynamic convective adjustment (CA) component.
CATKE is a ``one-equation'' model (because it includes an additional equation for TKE) that bears resemblance to a family of battle-tested parameterizations long used in European climate models~\cite{gaspar1990simple, blanke1993variability, kuhlbrodt2018low, madec2017nemo, gutjahr2021comparison, jungclaus2022icon}.
One-equation downgradient parameterizations are appropriate for high-resolution ocean modeling and amenable to GPU performance optimization due to their spatially-local formulation.
In contrast, the main benefit of ``$K$-profile'' schemes used in many global ocean models --- accommodating hours-long time steps \cite{reichl2018simplified} --- is not realized in high-resolution simulations that require short time-steps anyways to resolve advection by mesoscale turbulence.
Moreover, $K$-profile schemes achieve this time-step flexibility by solving nonlinear algebraic equations to determine boundary layer depth diagnostically~\cite{large1994oceanic, reichl2018simplified, reichl2019parameterization}, which may require significant optimization to achieve good performance on GPU-like systems \cite<as experienced by>{zhang2020optimizing}.
As for two-equation or ``$k$--$\epsilon$''-type models~\cite{mellor1982development, kantha1994improved, canuto2001ocean, umlauf2003generic, harcourt2015improved}, CATKE is less expensive merely by having one fewer prognostic variable.
%We can't help but point out that one-equation models are also used for LES \cite{deardorff1980stratocumulus, sullivan1994subgrid} --- suggesting that in principle a single one-equation model could be designed for ocean models of any resolution, including those spanning regimes between nonhydrostatic LES and mesoscale ocean turbulence.
The primary downside of any downgradient parameterization is unavoidable biases when instantaneously non-local, non-downgradient fluxes dominate, such as during free convection.

We therefore devote special attention to free convection during CATKE's formulation, which is described in section~\ref{sec:CATKE-formulation}, to minimize this downgradient bias and assess its importance.
Section~\ref{sec:convective-mixing-length} describes CATKE's diagnostic convective length scale and primary novelty, which uses dimensional analysis \cite{deardorff1970convective} to predict the convective boundary layer depth in terms of the \textit{local} TKE in order to estimate a dynamically evolving convective diffusivity.
This improves on the constant ``convective adjustment'' diffusivity typically used with one-equation parameterizations in ocean climate models \cite<typically $0.1 \, \rm{m^2 \, s^{-1}}$;>{madec2017nemo, gutjahr2021comparison, jungclaus2022icon}, which cannot describe how the convective mixing rate \textit{varies} with both boundary layer depth and the intensity of the destabilizing surface buoyancy flux over the wide range of conditions observed in Earth's ocean.
As a result, CATKE might be able to represent scenarios where mixing competes with other dynamics such as submesoscale restratification.
We also implement different mixing lengths for momentum, tracer, TKE, and the TKE dissipation rate in shear-driven turbulence that all vary as a function of the local gradient Richardson number.
This contrasts with typical approaches that estimate the TKE diffusivity as a constant multiple of the eddy viscosity \cite{blanke1993variability, madec2017nemo, umlauf2003generic}, or which allow only the tracer mixing length to vary with Richardson number \cite{blanke1993variability, madec2017nemo}.
%turbulent Prandtl number to vary rather than modeling simultaneous variation of the momentum and tracer mixing lengths.

CATKE's formulation could not be realized without an effective method for constraining CATKE's free parameters against observational or LES data.
Section~\ref{sec:calibration} describes how we calibrate CATKE's free parameters by minimizing the error between 21 variously-forced LES and the predictions of the LES data made by forward CATKE-parameterized single column simulations.
Because this calibration method is posed in terms of forward simulations, rather than an \textit{a priori} analysis of parameters or isolated subcomponents of the parameterization, it is sometimes called ``\textit{a posteriori}'' calibration \cite{duraisamy2021perspectives, frezatposteriori2022}.
Because \textit{a posteriori} calibration computes errors based on simulated time-series, it can incorporate numerical errors that accumulate during time stepping and can leverage even indirect observational data if it can be computed from model output.
For example, we leverage \textit{a posteriori} calibration to specifically minimize CATKE's dependence on vertical resolution.
We solve the calibration problem using Ensemble Kalman Inversion \cite<EKI; see>{iglesias2013ensemble}, which does not require gradients of the error with respect to free parameters.

We validate CATKE by a variety of methods in section~\ref{sec:validation}.
We first diagnose quantities with known physical interpretations such as CATKE's steady-state Richardson number and ``similarity layer constant'' (analogous to the von K\'arm\'an constant) in terms of CATKE's calibrated free parameters, and assess their consistency with observations or other measurements.
%Because these quantities are not directly constrained by calibration, this evaluation provides LES-based evidence that CATKE's formulation invokes reasonable physical hypotheses.
%For example, we find that calibration to LES produces a steady-state Richardson number $\Ri^\dagger \approx 0.23$ --- comfortingly close to the critical value for laminar instability, 1/4, and exactly equal to the value used by \citeA{blanke1993variability}.
Second, we compare CATKE's predictions versus idealized LES, both including those used in calibration and additional LES that are more strongly and more weakly forced than the calibration cases.
In this way we test whether CATKE can reproduce the training data as well as CATKE's capacity for extrapolation.
Third, we compare CATKE predictions to LES of a long 34 day deep cycle turbulence case, which is forced by realistic winds, heat fluxes, salinity fluxes, solar insolation, and lateral flux divergences derived from a regional ocean model.
This case illustrates CATKE's ability to extrapolate to cases with time-dependent forcing.
Fourth, we evaluate the sensitivity of CATKE's predictions to vertical resolution and time-step size.
After finding that CATKE can be sensitive to time steps longer than 1 minute if the forcing is very strong and the vertical resolution is 1 meter or finer, we describe a split-explicit substepping scheme for turbulent kinetic energy that nearly eliminates time step sensitivity while preserving the ability to step forward momentum and tracers with a relatively long time step.

We also compare CATKE to the $K$-profile parameterization \cite<KPP;>{large1994oceanic} and the second-moment closure of Langmuir turbulence \cite<Langmuir Turbulence Second Moment Closure, or ``SMC-LT'';>{harcourt2015improved}, which are implemented in the General Ocean Turbulence Model \cite<GOTM; see>{umlauf2005second, li2019comparing}.
CATKE outperforms both of these in almost all cases --- though the results must be taken with a grain of salt, because both KPP and SMC-LT have been calibrated to different data.
Despite this caveat, the comparison contributes context to CATKE's small but finite biases versus constant forcing LES.

We conclude in section~\ref{sec:discussion} with comments about future efforts to calibrate CATKE against more comprehensive data sets and future model development efforts to capture physics not considered in this work, such as the effect of surface wave fields that vary independently from winds and the modulation of turbulence by lateral density fronts.
The most important piece of future work is the construction of a global calibration context to further calibrate CATKE's free parameters against satellite and in-situ ocean observations.

\section{Large eddy simulations of turbulent mixing beneath surface waves}
\label{sec:les-data}

We begin by concretely defining the parameterization problem that drives the cyclical process of formulating, calibrating, and validating CATKE.
In this paper, the problem is posed by comparing high-fidelity and three-dimensional large eddy simulations (LES) of turbulent mixing with one-dimensional parameterized models for the horizontally-averaged dynamics of the LES.
Our LES integrate the rotating, wave-averaged Boussinesq equations simplified for a steady surface wave field \cite{craik1976rational, huang1979surface, suzuki2016understanding},
\begin{linenomath*}
\begin{gather}
\label{momentum}
\d_t \bUL + \left ( \bUL \bcdot \bnabla \right ) \bUL
    + \left ( f \bzh - \bnabla \times \bUS \right ) \times \bUL + \bnabla P = B \bzh + \d_t \bUS + \b{F}_u \com \\
\label{continuity}
\bnabla \bcdot \bUL = 0 \com \\
\label{tracers}
\d_t C + \left ( \bUL \bcdot \bnabla \right ) C = - \bnabla \bcdot \b{J}_c + F_c \com
\end{gather}
\end{linenomath*}
where $\bUL = (\UL, \VL, \WL)$ is the Lagrangian-mean velocity, $\bUS$ is the Stokes drift associated with surface waves (which are always steady and oriented in the $\bxh$-direction in this paper), $P$ is Eulerian-mean pressure, $B$ is Eulerian-mean buoyancy, $f$ is the Coriolis parameter, $\b{F}_u$ is a momentum forcing term representing surface wind stress, $C$ is any tracer such as temperature or salinity, and $F_c$ is forcing term for $C$ representing boundary conditions, solar insolation, and other other imposed body forcing.
The Lagrangian-mean velocity $\bUL$ is defined as the sum of the Eulerian-mean velocity and Stokes drift, and setting $\bUS=0$ reduces equation~\eqref{momentum} to the ordinary Navier--Stokes equations.
Note that we have neglected molecular diffusion from~\eqref{momentum} and~\eqref{tracers}, as well as diffusion by a hypothetical LES closure, to simplify the ensuing discussion.
In this work we use buoyancy $B$ itself as a tracer, which is tantamount to using a linear equation of state with a single constituent.

We conduct 35 LES of~\eqref{momentum}--\eqref{tracers} forced by constant, horizontally-uniform fluxes of momentum and buoyancy in a $512 \, \rm{m} \times 512 \, \rm{m} \times 256 \, \rm{m}$ horizontally-periodic domain with $O(1 \, \rm{m})$ resolution using Oceananigans~\cite{OceananigansJOSS}.
All 35 LES are initialized with the same piecewise-constant density stratification given in equation~\ref{initial-buoyancy}, which has a weakly-stratified near-surface layer, a more strongly stratified middle layer, and a weakly-stratified lower layer.
The surface momentum flux or ``wind stress'' $\tau_x$ is defined via $\b{F}_u$ in~\eqref{momentum} as
\beq
\b{F}_u = - \d_z \left [ \tau_x \, \delta(z) \right ] \bxh \com
\eeq
where $\delta(z)$ is a delta function concentrate at $z=0$, such that negative stress $\tau_x < 0$ forces a current in the $+x$-direction.
Two types of buoyancy fluxes are used: a destabilizing surface flux $J_b > 0$ representing cooling or heat loss, which. isdefined via $F_b$ in equation~\eqref{tracers} via
\beq
F_b = - \d_z \left [ J_b \, \delta(z) \right ] \per
\eeq
We also include 5 LES forced by both wind stress and stabilizing buoyancy forcing that represents heating by solar insolation.
In these ``sunny'' cases, the flux divergence of buoyancy $F_b$ is given by
\beq \label{solar-radiation}
F_b = - \d_z I \com
\qquad \text{where}
\qquad
I(z) = J_b \left [ \epsilon_1 \ee^{z/\lambda_1} + \left ( 1 - \epsilon_1 \right ) \ee^{z / \lambda_2} \right ] \per
\eeq
In~\eqref{solar-radiation}, $I(z)$ is the buoyancy flux profile associated with penetrating solar insolation, $J_b < 0$ is the surface solar insolation, $\epsilon_1$ is the fraction of penetrating radiation absorbed over the vertical scale $\lambda_1$, and $(1 - \epsilon_1)$ is the remaining fraction absorbed over $\lambda_2$.
All simulations use $\epsilon_1 = 0.6$, $\lambda_1 = 1$ m, and $\lambda_2 = 16$ m \cite<see for example the solar insolation used by>{whitt2022simulation}.

The forcing strength for each case is rationalized by categorizing the LES into 6-, 12-, 24-, 48-, and 72-hour ``suites'' according to their duration.
Because all the LES are initialized identically and run until the boundary layer is roughly half the depth of the domain, duration indicates forcing strength: the 6-hour-suite are the most strongly forced and the 72-hour suite simulations are the most weakly forced.
The intermediately-forced 12-, 24-, and 48-hour suites are used for calibration.
The 35 LES are divided into 5 ``suites'' with 7 cases each, according to their duration and the intensity of the surface fluxes: the 6-hour suite exhibits extreme forcing, while the 72-hour suite exhibits relatively weak forcing.
Each suite consists of 7~physical scenarios that represent different forcing regimes:
\begin{itemize}
\item ``free convection'', which has pure destabilizing buoyancy forcing and no winds,
\item ``weak wind strong cooling'',
\item ``medium wind medium cooling'',
\item ``strong wind weak cooling'',
\item ``strong wind'', with no buoyancy forcing,
\item ``strong wind no rotation'' with no buoyancy forcing and $f = 0$.
\item ``strong wind and sunny'' with penetrative heating, wind forcing, and $f = 0$.
\end{itemize}
The ``strong wind no rotation'' and ``strong wind and sunny'' are non-rotating with $f=0$, and the rest are rotating with Coriolis parameter $f = 10^{-4} \, \mathrm{s^{-1}}$.
The range of buoyancy fluxes roughly corresponds to cooling between 156--2000 $\mathrm{W \, m^{-2}}$ or heating by penetrating solar insolation between 104--1250 $\mathrm{W \, m^{-2}}$, and the momentum fluxes correspond to 10-meter atmospheric winds of approximately 9--25 $\mathrm{m \, s^{-1}}$ and oriented in the $\bxh$-direction.
The fluxes associated with each case are summarized in tables~\ref{table:les-summary} and~\ref{table:les-validate-summary}.

\begin{figure}[ht]
    \includegraphics[width = 1\textwidth]{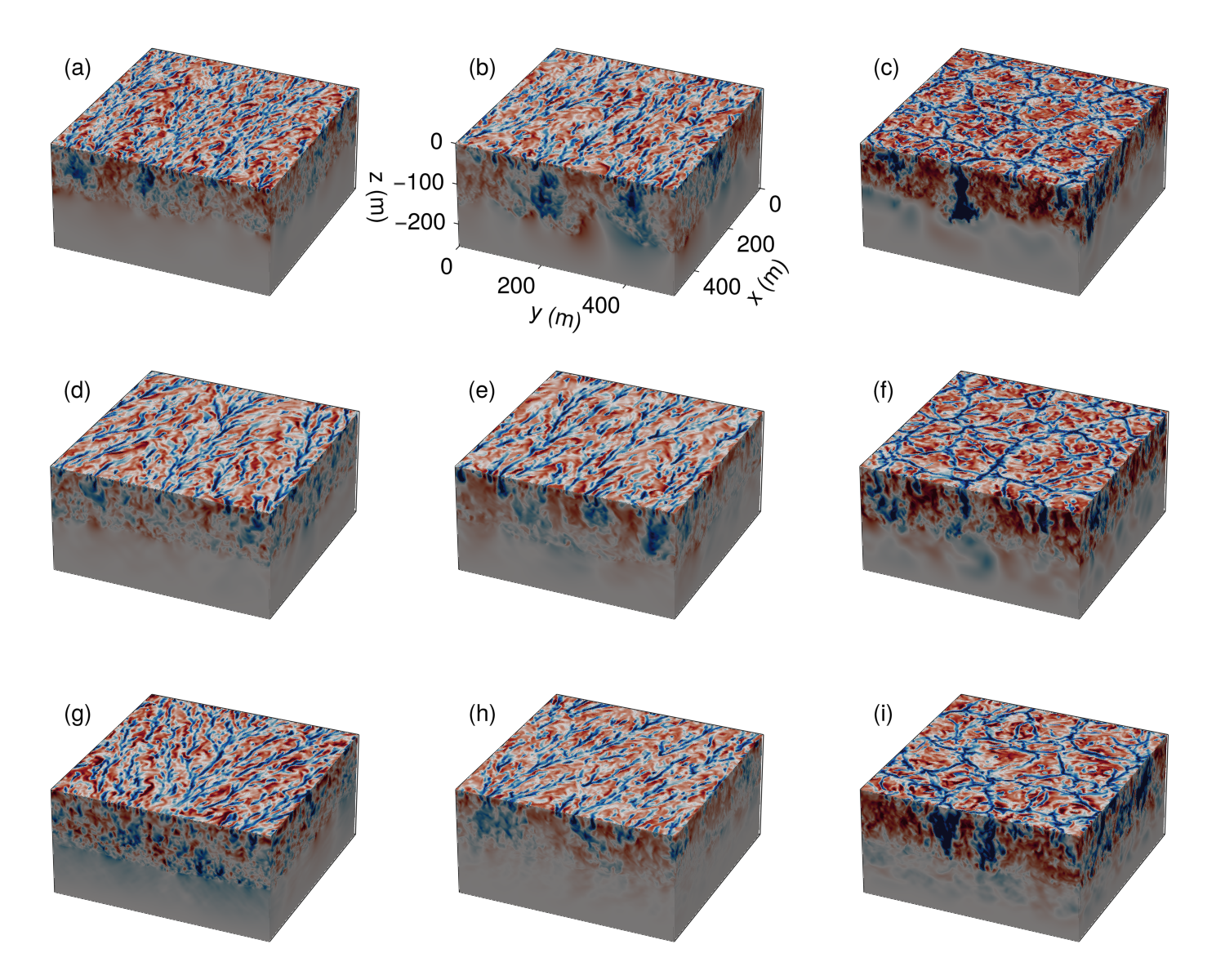}
    \caption{
    Visualization of vertical velocity in 9 of 35 large eddy simulations (LES) of the ocean surface boundary layer used in this paper, forced variously by winds, surface waves, and heat fluxes.
    All LES, which are summarized in tables~\ref{table:les-summary} and~\ref{table:les-validate-summary} and described in more detail in~\ref{les-description}, are initialized with the same density stratification.
    (a)--(c) show strongly-forced LES after just 6 hours of simulation, (d)--(f) show LES driven by medium-strength forcing after 24 hours, and (g)--(i) show weakly forced LES after 72 hours.
    (a), (d), and (g) show a purely wind and wave driven case, (b), (e), (h) are forced by a mixture of winds, waves, and cooling, and (c), (f), and (i) are ``free convection'' cases forced only by cooling with no winds and waves.
    All simulations are rotating with Coriolis parameter $f = 10^{-4} \, \mathrm{s^{-1}}$.
    The colorscale for each panel saturates at $\tfrac{1}{2} \max|w|$, which for each panel is (a) 0.26, (b) 0.29, (c) 0.086, (d) 0.20, (e) 0.23, (f) 0.070, (g) 0.056, (h) 0.14, and (i) 0.041 $\mathrm{m \, s^{-1}}$.
    }
    \label{fig:les-summary}
\end{figure}

In any LES with wind forcing, we also include the effect of wind-driven surface waves through an estimate of $\d_z \bUS = \d_z \US \bxh$ in~\eqref{momentum} for equilibrium waves \cite{lenain2020contribution}.
The equilibrium wave model depends on the peak wavenumber of the surface wave field, which is chosen so that the Langmuir number $\La$ is
\beq \label{langmuir-number}
\La \defn \sqrt{\frac{u_\star}{\US(z=0)}} \approx 0.3 \com
\eeq
close to the peak of its global distribution \cite{belcher2012global}.
In~\eqref{langmuir-number}, $u_\star$ is the friction velocity computed from the surface wind stress (here $u_\star = \sqrt{|\tau_x|}$, where $\boldsymbol{\tau} = \tau_x \bxh$ is the wind stress).
All LES are initialized from rest with $\bUL = 0$.
The LES also include a forced passive tracer, providing additional information about the time scales of mixing in the interior of the boundary layer.
The initial density stratification, numerical methods, Stokes drift model, effects of including Stokes drift, and the sensitivity of the LES to resolution are described in~\ref{les-description}.
Out of the 35 LES cases, 21 are used for calibration, while another 14 are reserved for validation.
Figure~\ref{fig:les-summary} visualizes vertical velocity in 9 of the 35 cases.

\begin{table}  
\centering
\begin{tabular}{l l c c c c c}
\bf{Suite} & \bf{Case} & $J_b \, \mathrm{(m^2 \, s^{-3})}$ & $|\tau_x| \, \mathrm{(m^2 \, s^{-2})}$ & $Q$ $\mathrm{\left ( \tfrac{W}{m^2} \right )}$ & $u_{10} \, \mathrm{\left ( \tfrac{m}{s} \right )}$ \\[1ex] 
\hline \\[-1ex]
12 hour & free convection             & $+4.8 \times 10^{-7}$ & 0                    & $+1000$ & $0$  \\[1ex] 
12 hour & weak wind strong cooling    & $+4.0 \times 10^{-7}$ & $4.0 \times 10^{-4}$ & $+833$  & $15$ \\[1ex] 
12 hour & mid wind mid cooling        & $+3.2 \times 10^{-7}$ & $6.0 \times 10^{-4}$ & $+667$  & $17$ \\[1ex] 
12 hour & strong wind weak cooling    & $+2.0 \times 10^{-7}$ & $8.0 \times 10^{-4}$ & $+417$  & $20$ \\[1ex] 
12 hour & strong wind                 & 0                     & $9.0 \times 10^{-4}$ & $0$     & $21$ \\[1ex] 
12 hour & strong wind no rotation     & 0                     & $6.0 \times 10^{-4}$ & $0$     & $17$ \\[1ex] 
12 hour & strong wind and sunny       & $-5.0 \times 10^{-7}$ & $9.0 \times 10^{-4}$ & $-1042$ & $21$ \\[1ex] 
24 hour & free convection             & $+2.4 \times 10^{-7}$ & 0                    & $+500$  & $0$  \\[1ex] 
24 hour & weak wind strong cooling    & $+2.0 \times 10^{-7}$ & $3.0 \times 10^{-4}$ & $+417$  & $13$ \\[1ex] 
24 hour & mid wind mid cooling        & $+1.6 \times 10^{-7}$ & $4.5 \times 10^{-4}$ & $+333$  & $16$ \\[1ex] 
24 hour & strong wind weak cooling    & $+1.0 \times 10^{-7}$ & $5.9 \times 10^{-4}$ & $+208$  & $17$ \\[1ex] 
24 hour & strong wind                 & 0                     & $6.8 \times 10^{-4}$ & $0$     & $18$ \\[1ex] 
24 hour & strong wind no rotation     & 0                     & $3.0 \times 10^{-4}$ & $0$     & $13$ \\[1ex] 
24 hour & strong wind and sunny       & $-3.0 \times 10^{-7}$ & $4.5 \times 10^{-4}$ & $-625$  & $16$ \\[1ex] 
48 hour & free convection             & $+1.2 \times 10^{-7}$ & 0                    & $+250$  & $0$  \\[1ex] 
48 hour & weak wind strong cooling    & $+1.0 \times 10^{-7}$ & $2.0 \times 10^{-4}$ & $+208$  & $11$ \\[1ex] 
48 hour & mid wind mid cooling        & $+8.0 \times 10^{-8}$ & $3.4 \times 10^{-4}$ & $+167$  & $14$ \\[1ex] 
48 hour & strong wind weak cooling    & $+5.0 \times 10^{-8}$ & $3.8 \times 10^{-4}$ & $+104$  & $15$ \\[1ex] 
48 hour & strong wind                 & 0                     & $4.5 \times 10^{-4}$ & $0$     & $16$ \\[1ex] 
48 hour & strong wind no rotation     & 0                     & $1.6 \times 10^{-4}$ & $0$     & $10$ \\[1ex] 
48 hour & strong wind and sunny       & $-1.0 \times 10^{-7}$ & $2.0 \times 10^{-4}$ & $-208$  & $11$ \\[1ex] 
\end{tabular}
\caption{Summary of surface boundary conditions for LES used to calibrate CATKE.
All LES are initialized with the buoyancy profile described in equation~\eqref{initial-buoyancy} and use the Coriolis parameter $f = 10^{-4} \, \mathrm{s^{-1}}$ except ``strong wind no rotation'' and ``strong wind and sunny'', which use $f=0$.
The ``suite'' indicates simulation duration.
$J_b$ is the surface buoyancy flux, $\tau_x$ is the kinematic momentum flux (momentum flux divided by ocean reference density), $Q \approx \rho_o c_p J_b / (\alpha g)$ is the heat flux associated with $J_b$, and $u_{10}$ is an estimate of the 10-meter wind speed associated with $\tau_x$ according to equation~\ref{u10-estimate} using
reference density $\rho_o = 1024 \, \mathrm{kg \, m^{-3}}$,
seawater heat capacity $c_p = 3991\, \mathrm{J \, {}^\circ C^{-1}}$,
thermal expansion coefficient $\alpha = 2 \times 10^{-4} \, \mathrm{{}^\circ C^{-1}}$,
gravitational acceleration $g = 9.81 \, \mathrm{m \, s^{-2}}$ are used for $Q$ and $u_{10}$.
When the surface buoyancy flux is negative ($J_b < 0$), $J_b$ represents $J_b = I(z=0)$, where $I(z)$ is the buoyancy flux associated with penetrating solar insolation in equation~\ref{solar-radiation}.
The forcing in equation~\eqref{tracers} is then defined as $F_b = - \d_z I$.
All fluxes use the convention that a positive flux carries quantities upwards, out of the ocean, which means a negative $\tau_x$ drives currents in the $+\bxh$ direction and a positive buoyancy flux cools the ocean by extracting buoyancy.
Additional LES used to validate CATKE are summarized in table~\ref{table:les-validate-summary}.
\label{table:les-summary}
}
\end{table}

\begin{table}  
\centering
\begin{tabular}{l l c c c c c}
\bf{Suite} & \bf{Case} & $J_b \, \mathrm{(m^2 \, s^{-3})}$ & $|\tau_x| \, \mathrm{(m^2 \, s^{-2})}$ & $Q$ $\mathrm{\left ( \tfrac{W}{m^2} \right )}$ & $u_{10} \, \mathrm{\left ( \tfrac{m}{s} \right )}$ \\[1ex] 
\hline \\[-1ex]
6 hour  & free convection             & $+9.6 \times 10^{-7}$ & 0                    & $+2000$ & $0$  \\[1ex] 
6 hour  & weak wind strong cooling    & $+8.0 \times 10^{-7}$ & $5.0 \times 10^{-4}$ & $+1666$ & $16$ \\[1ex] 
6 hour  & mid wind mid cooling        & $+6.4 \times 10^{-7}$ & $8.0 \times 10^{-4}$ & $+1333$ & $20$ \\[1ex] 
6 hour  & strong wind weak cooling    & $+4.0 \times 10^{-7}$ & $1.2 \times 10^{-3}$ & $+833$  & $23$ \\[1ex] 
6 hour  & strong wind                 & 0                     & $1.4 \times 10^{-3}$ & $0$     & $24$ \\[1ex] 
6 hour  & strong wind no rotation     & 0                     & $1.1 \times 10^{-3}$ & $0$     & $22$ \\[1ex] 
6 hour  & strong wind and sunny       & $-6.0 \times 10^{-7}$ & $1.5 \times 10^{-3}$ & $-1250$ & $25$ \\[1ex] 
72 hour & free convection             & $+8.7 \times 10^{-8}$ & 0                    & $+181$  & $0$  \\[1ex] 
72 hour & weak wind strong cooling    & $+7.5 \times 10^{-8}$ & $1.8 \times 10^{-4}$ & $+156$  & $11$ \\[1ex] 
72 hour & mid wind mid cooling        & $+6.0 \times 10^{-8}$ & $2.9 \times 10^{-4}$ & $+125$  & $13$ \\[1ex] 
72 hour & strong wind weak cooling    & $+3.8 \times 10^{-8}$ & $3.4 \times 10^{-4}$ & $+79$   & $14$ \\[1ex] 
72 hour & strong wind                 & 0                     & $4.1 \times 10^{-4}$ & $0$     & $15$ \\[1ex] 
72 hour & strong wind no rotation     & 0                     & $1.1 \times 10^{-4}$ & $0$     & $9$  \\[1ex] 
72 hour & strong wind and sunny       & $-5.0 \times 10^{-8}$ & $1.3 \times 10^{-4}$ & $-104$  & $9$  \\[1ex]
\end{tabular}
\caption{Summary of surface boundary conditions for LES used to validate CATKE.
See table~\ref{table:les-summary} for a description and a summary of the LES used to calibrate CATKE.
\label{table:les-validate-summary}
}
\end{table}

\subsection{The single column context}

\begin{figure}[ht]
    \includegraphics[width = 1\textwidth]{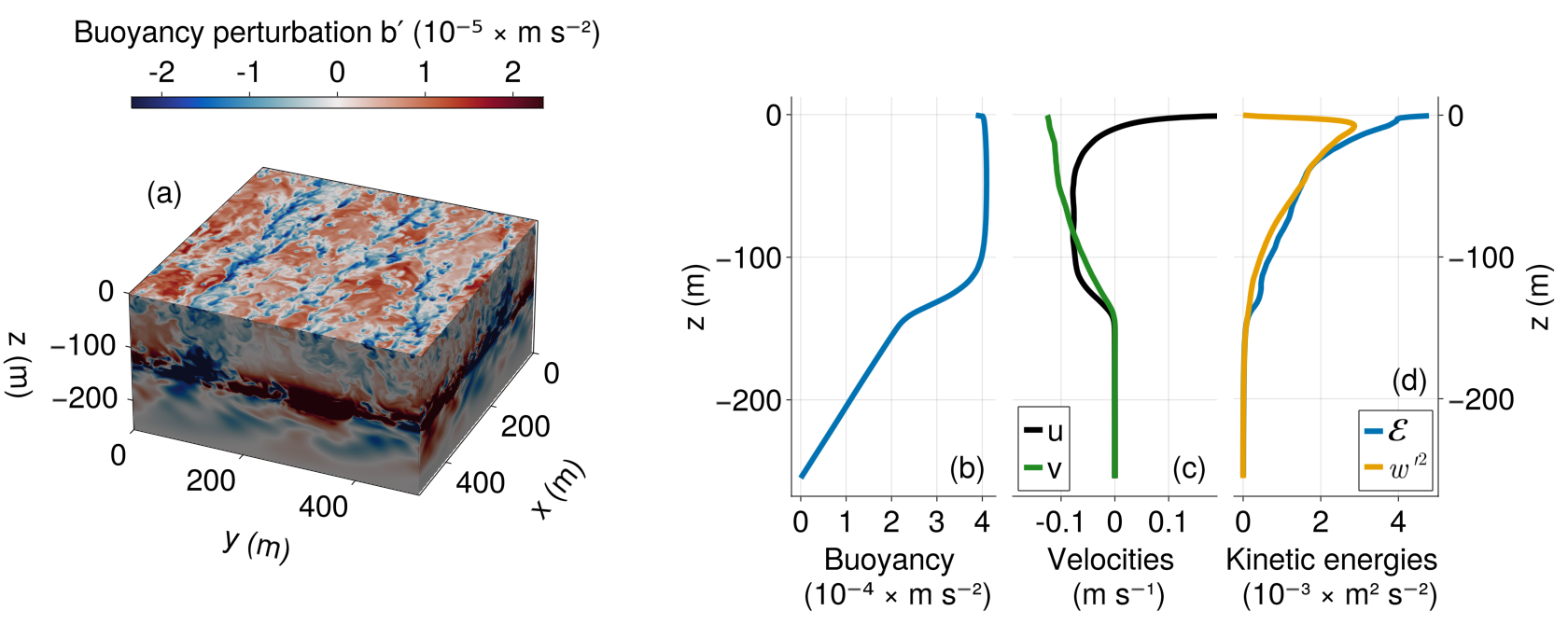}
    \caption{Illustration of horizontally-averaged data from the 12-hour strong wind, weak cooling LES. Panel (a) shows the buoyancy perturbation $b'$. Note the colorbar is strongly saturated to illustrate boundary layer structure; the buoyancy perturbation is particularly large at the base of the boundary layer, where the horizontally-averaged buoyancy gradient is also strong. (b) shows the horizontally-averaged buoyancy $b$, (c) shows the horizontally-averaged velocities $u, v$, and (d) shows the horizontally-averaged fluctuation kinetic energy, $\E \defn \left (\overline{u'^2} + \overline{v'^2} + \overline{w'^2} \right ) / 2$ and horizontally-averaged vertical velocity variance, $\overline{w'^2}$.}
    \label{fig:les-horizontal-averages}
\end{figure}

We would like to develop a model that can predict the horizontally-averaged momentum and buoyancy simulated by the LES.
We therefore decompose all three-dimensional variables~$\Psi$ in~\eqref{momentum}--\eqref{tracers} into a horizontally-averaged component $\psi \defn \bar \Psi$ and a fluctuation $\psi'$ such that,
\beq
\Psi(x, y, z, t) = \underbrace{\mystrut{1ex} \bar \Psi(z, t)}_{\defn \psi(z, t)} \, + \, \psi'(x, y, z, t) \com
\eeq
where the overline $\overline{()}$ denotes a horizontal average, and $\Psi \in (\UL, \VL, \WL, C)$ includes the velocity components $\UL$, $\VL$, $\WL$, and tracer concentrations $C$.
Note that the horizontal average of~\eqref{continuity} and the horizontal homogeneity of our LES implies that $\wL = 0$ and $\WL = w'$ and thus the vertical momentum equation reduces to a statement of wave-modified hydrostatic balance.
Figure~\ref{fig:les-horizontal-averages} shows horizontally-averaged buoyancy, velocity, and kinetic energy profiles alongside a three-dimensional visualization of the buoyancy perturbation $b'$ for the 12-hour strong wind, weak cooling case.

Next, we derive a set of equations that governs the horizontally-averaged zonal momentum $u(z, t)$, meridional momentum $v(z, t)$, and any tracer $c(z, t)$ by taking a horizontal average of~\eqref{momentum} and~\eqref{tracers} to obtain,
\begin{linenomath*}
\begin{gather}
    \d_t u - f v = - \d_z \overline{w' u'} + \bar F_u \com \label{u0} \\
    \d_t v + f u = - \d_z \overline{w' v'} + \bar F_v \com \\
    \d_t c = - \d_z \overline{w' c'} + \bar F_c \com \label{c0}
\end{gather}
\end{linenomath*}
where $u$, $v$ represent the horizontal average of the horizontal Lagrangian-mean velocities $\UL$, $\VL$, with superscript $\mathrm{L}$ is omitted to simplify the notation.
Lateral fluxes vanish from~\eqref{u0}--\eqref{c0} due to horizontal homogeneity.
No terms Stokes-drift-dependent terms enter into~\eqref{u0}--\eqref{c0} because $\bUS(z)$ is horizontally uniform.
Figure~\ref{fig:les-horizontal-averages} illustrates the horizontally-averaged buoyancy, velocity, and turbulent kinetic energy for the 12-hour strong wind, weak cooling case.

The parameterization problem may now be stated: we seek a parameterization that predicts the vertical fluxes $\overline{w' u'}$, $\overline{w' v'}$, and $\overline{w' c'}$ in terms of the resolved state $u, v, c$, boundary conditions, and potentially, additional auxiliary variables.
For example, the parameterization described in the next section uses a downgradient formulation $\overline{w' c'} \sim \d_z c$ to predict vertical tracer and momentum fluxes.

\subsection{Connection to the regional and global ocean modeling context}

Our LES, and the models that predict the horizontal average of the LES, may be described as ``single column models''. This nomenclature reflects the notion that the models simulate the vertical redistribution of momentum and tracers by turbulent motions in a single column of a three-dimensional ocean model. Indeed, we envision that the single column context is generalized to a large-scale ocean simulation merely by adding advection by motions somewhat larger than the scale of the LES domain.
This approach relies on two key assumptions.
First, the microscale turbulence must be horizontally homogeneous so as to ignore lateral flux divergences.
Second, there must be a scale separation between microscale turbulence and larger-scale motions so that interactions between the two can be ignored. 

For typical oceanic situations, the first assumption is likely satisfied because vertical gradients are much larger than horizontal ones on the scales of a ``single column model'' and thus the vertical flux divergences dominate over the horizontal ones.
In other words the ocean is more homogeneous in the horizontal than in the vertical on scales of O(100~m).
The second assumption is more problematic especially near the ocean surface and bottom boundaries.
While microscale turbulence does not significantly interact with mesoscale geostrophic eddies with scales of O(10--100~km), there is growing evidence of interactions between submesoscale frontal dynamics with scales of O(100 m -- 10 km) and microscale turbulence~\cite<see the reviews by>{thomas2008submesoscale, mcwilliams2016submesoscale, taylor2023submesoscale}.
Frontal instabilities are also effective at restratifying the ocean boundary layers during time of weak microscale turbulence~\cite<see for example>{boccaletti2007mixed}.
These interactions are presently ignored in the formulation of microscale turbulence parameterizations, but they are an obvious direction for future development of CATKE.
Following the approach outlined in this paper, this will require generating a library of simulations which resolve microscale turbulence in the presence of ocean fronts, extending CATKE to include those physics, and then calibrating the extended CATKE against the library of those simulations.

Similarly, microscale turbulent mixing in the ocean interior requires considering multiscale dynamics.
For example, internal waves generated by surface winds and tide-bathymetry interactions produce a direct cascade of internal wave energy to progressively smaller scales until wave breaking finally transfers energy to microscale turbulence.
Incorporating the physics of turbulent mixing driven by internal wave breaking is another area for future development.

\section{CATKE formulation}
\label{sec:CATKE-formulation}

CATKE models the horizontally-averaged vertical fluxes $\overline{w' \psi'}$ appearing on the right side of~\eqref{u0}--\eqref{c0} with a downgradient, mixing length formulation \cite{prandtl1925results},
\beq \label{eddy-diffusivity}
\overline{w' \psi'} \approx - \underbrace{\ell_\psi \sqrt{e}}_{\defn \K_\psi} \, \d_z \psi \com
\eeq
where $e$ is the turbulent kinetic energy, $\sqrt{e}$ is the turbulent velocity scale, and $\ell_\psi$ is the mixing length for the horizontally-averaged variable $\psi(z, t)$.
After choosing to parameterize turbulent transport with eddy diffusion that depends on the turbulent velocity $\sqrt{e}$ and mixing length $\ell_\psi$, the form $\K_\psi = \ell_\psi \sqrt{e}$ follows from dimensional analysis.
CATKE invokes three mixing lengths and three eddy diffusivities for horizontal velocities ($\ell_u$ and $\kappa_u$), tracers ($\ell_c$ and $\kappa_c$), and turbulent kinetic energy ($\ell_e$ and $\kappa_e$).

With~\eqref{eddy-diffusivity}, the single column equations become
\begin{linenomath*}
\begin{gather}
    \d_t u - f v = \d_z \left (\K_u \d_z u \right ) + \bar F_u \com \label{u1d} \\
    \d_t v + f u = \d_z \left (\K_u \d_z v \right ) + \bar F_v \label{v1d} \com \\
    \d_t c = \d_z \left (\K_c \d_z c \right ) + \bar F_c \label{c1d} \per
\end{gather}
\end{linenomath*}
In this paper we use a linear equation of state that relates density to a single thermodynamic constituent, such that the buoyancy $b$ is just another tracer,
\beq \label{b1d}
\d_t b = \d_z \left ( \K_c \d_z b \right ) + \bar F_b \com
\eeq
where $\bar F_b = - \d_z I$ corresponds to heating within the water column due to penetrating solar radiation, $I$.
The buoyancy gradient $N^2 \defn \d_z b$ appears in many of the scaling arguments central to CATKE's formulation, where $N$ is often referred to as the ``buoyancy frequency''.
Note that in more realistic simulations of seawater, $b$ and $N^2$ are functions of geopotential height, mean temperature, and mean salinity through the empirically-determined seawater equation of state~\cite{teos10-manual}.

Next we turn to the estimation of the turbulent kinetic energy $e$, and thus the turbulent velocity scale $\sqrt{e}$ in~\eqref{eddy-diffusivity}.
For this we first introduce the kinetic energy of the subgrid velocity field, $\E$, defined in terms of the velocity fluctuations $(u', v', w')$,
\beq \label{tke}
\E \defn \half \overline{| \bu' |^2} = \half \left ( \overline{u'^2} + \overline{v'^2} + \overline{w'^2} \right ) \per
\eeq
We postulate a close relationship between $e$ in~\eqref{eddy-diffusivity} and the subgrid kinetic energy, $\E$.
However, this is a relationship rather than an identity, because $\E$ has contributions from motions that are unrelated to the eddy diffusivity in~\eqref{eddy-diffusivity}.
For example, internal waves generated by convective plumes make a significant contribution to $\E$ below the base of boundary layer, despite that there is no mixing there.
We note further that if the kinetic energy and mixing length are actually known, the inexact relationship between $\E$ and $e$ manifests through a ``correlation coefficient'' \cite{taylor1922diffusion} that appears in formulations like~\eqref{eddy-diffusivity}.
We therefore define $e$ as a \textit{latent variable} which is linked to the averaged velocity and tracer fields via~\eqref{eddy-diffusivity}, rather than as corresponding directly to the observable, but less relevant quantity~\eqref{tke}.
This interpretation has important implications for calibration: rather than using the discrepancy between LES-derived $\E$ and $e$ to estimate free parameters, we only use the error in momentum and buoyancy profiles --- which are strongly affected by $e$ through~\eqref{eddy-diffusivity} --- to constrain the free parameters that govern the evolution of $e$.
In other words, $e$ can only be observed indirectly via the evolution of momentum and buoyancy.
Interpreting $e$ as a latent variable rather than as the actual subgrid kinetic energy $\E$ is also proposed by Kolmogorov \cite<see>{spalding1991kolmogorov} and \citeA{saffman1970model}.

Though we define $e$ as a latent variable that is linked to $u, v, c$ solely via~\eqref{eddy-diffusivity}, we nevertheless postulate a similarity between $e$ and $\E$ on physical grounds --- where there is turbulence, there will be mixing --- and following a litany of prior work \cite{saffman1970model, gaspar1990simple, spalding1991kolmogorov, umlauf2003generic}, use the evolution equation for $\E$ to derive a model for the evolution of $e$.
An equation describing the evolution of $\E$ can be derived from~\eqref{momentum}, including the molecular stress divergence $\nu \lap \left ( \bUL - \bUS \right )$ (we include the Stokes drift term here for completeness, though it does not contribute to the equation for $\E$).
The result is
\beq \label{e-exact}
\d_t \E =
    - \underbrace{\mystrut{1.5ex} \d_z \left ( \overline{w' \E'} + \overline{w'p'} - \nu \d_z \E \right )}_{\text{transport}}
\, \, \, -          \underbrace{\mystrut{1.5ex} \overline{\bu' w'} \bcdot \d_z \bu}_{\text{shear production}} 
+ \, \, \,          \underbrace{\mystrut{1.5ex} \overline{w' b'}}_{\text{buoyancy flux}} 
\, \, \, - \, \, \, \underbrace{\mystrut{1.5ex} \nu \overline{|\bnabla \bu'|^2}}_{\text{dissipation}} \com 
\eeq
where $\nu$ is the kinematic viscosity, $p$ is kinematic pressure (dynamic pressure divided by a reference density) and $\E' = \half | \bu'|^2 - \E$.
Note that because $\bu$ is the horizontally-averaged Lagrangian-mean velocity, the shear production term in~\eqref{e-exact} represents the total transfer of kinetic energy from the average $\bu$ to the fluctuations $\bu'$ --- including the so-called ``Stokes production'' term \cite{mcwilliams1997langmuir}.
Inspired by~\eqref{e-exact}, we formulate an equation for $e$ consisting of terms that mirror each term in equation~\eqref{e-exact}:
\beq \label{e1d}
\d_t e = 
               \underbrace{\mystrut{2.5ex} \d_z \left ( \kappa_e \d_z e \right )}_{\text{transport}}
    \, \, \, + \underbrace{\mystrut{2.5ex} \kappa_u | \d_z \b{u} |^2}_{\text{shear production}} 
          - \, \underbrace{\mystrut{2.5ex} \kappa_c N^2}_{\text{buoyancy flux}} 
    \,    - \, \underbrace{\mystrut{2.5ex} \frac{e^{3/2}}{\ell_D}}_{\text{dissipation}} \com 
\eeq
where $|\d_z \bu|^2 = (\d_z u)^2 + (\d_z v)^2$ is the square vertical shear of the horizontally-averaged velocity field $\bu$ (note that $w = 0$ because of horizontal homogeneity), $\K_e$ is the vertical diffusivity of $e$, $\ell_D$ is the ``dissipation length scale'', and we have labeled the corresponding terms in~\eqref{e-exact} and~\eqref{e1d}.
The shear production and buoyancy flux terms are formulated by applying the eddy diffusivity hypothesis~\eqref{eddy-diffusivity} to their corresponding expressions in equation~\eqref{e-exact}.
Like in the budget for $\E$, the shear production term in~\eqref{e1d} represents the total shear production including both ``Eulerian'' and ``Stokes'' production. 
We assume that the transport of~$e$, which helps to deepen boundary layers by modeling turbulence spreading away from turbulence-generating regions, can be modeled with an eddy diffusivity $\kappa_e = \ell_e \sqrt{e}$.
Finally, to model the dissipation of~$e$ we introduce the dissipation length scale $\ell_D$, which has a similar form to the mixing lengths $\ell_u$, $\ell_c$, and $\ell_e$.
The expression $e^{3/2} / \ell_D$ follows on dimensional grounds.

Equation~\eqref{e1d} requires boundary conditions.
We impose a no-flux condition on $e$ at the bottom.
(Extending CATKE to describe the bottom boundary layer in the future may require imposing a different bottom boundary condition.)
At $z=0$, we parameterize subgrid production of $e$ by wind stress and destabilizing buoyancy fluxes across the uppermost cell interface with
\beq \label{surface-TKE-flux}
J_e \defn - \kappa_e \d_z e \, \big |_{z=0} = - \C{\rm{shear}}{J} \, u_\star^3 - \C{\rm{conv}}{J} w_\Delta^3 \com
\quad \text{where} \quad w_\Delta^3 \defn \Delta z \max(J_b, 0) \com 
\eeq
and $\C{\rm{shear}}{J}$ and $\C{\rm{conv}}{J}$ are constant, non-dimensional free parameters, $J_b$ is the surface buoyancy flux defined such that $J_b > 0$ removes buoyancy and thus causes convection, $\Delta z$ is the distance between the top of the ocean domain and the first interior cell interface, 
and $w_\Delta^2$ is the convective TKE scale that follows from a balance between buoyant production and dissipation estimated using the grid spacing $\Delta z$ as a length scale.
$u_\star$ in~\eqref{surface-TKE-flux} is the ocean-side friction velocity,
\beq
u_\star \defn \left ( \tau_x^2 + \tau_y^2 \right )^{1/4} \com
\label{eq:ustar}
\eeq
defined in terms of the zonal and meridional kinematic momentum fluxes $\tau_x$ and $\tau_y$ (wind stresses divided by reference water density).
The boundary condition~\eqref{surface-TKE-flux} differs from boundary conditions used in the TKE-based models described by \citeA{blanke1993variability} and \citeA{madec2017nemo}, which prescribe TKE (rather than prescribing TKE flux), and depend only on the friction velocity $u_\star$. % in their surface boundary condition for $e$ \cite{blanke1993variability, madec2017nemo}.

The surface flux formulation in~\eqref{surface-TKE-flux} introduces the notation
\beq \label{free-parameter-notation}
\C{\rm{label}}{\rm{component}}
\eeq
for two free parameters $\C{\rm{shear}}{J}$ and $\C{\rm{conv}}{J}$, where ``label'' indicates the parameter's role and ``component'' refers to the variable or component to which the parameter associates.

\subsection{Turbulence length scale model}

We decompose the four length scales $\ell_\psi \in (\ell_u, \ell_c, \ell_e, \ell_D)$ into a shear-dominated length scale $\ell^{\rm{shear}}_\psi$ limited by density-stratification and boundaries, and a convection-dominated length scale $\ell^{\rm{conv}}_\psi$ limited by the depth of the convective boundary layer.
At any time and location, the maximum of these two length scales is chosen as the mixing length via
\beq \label{mixing-length}
\ell_\psi = \max \left (\ell^{\rm{conv}}_\psi, \ell^{\rm{shear}}_\psi \right ) \com
\eeq
encapsulating a sharp separation between turbulence regimes that exhibit distinct scaling laws.
We next describe a length scale formulation that can be calibrated to predict turbulent fluxes associated with the kinds of flows plotted in figure~\ref{fig:les-summary}.

\subsubsection{Shear turbulence length scale}
\label{sec:shear-turbulence-length-scale}

To represent shear dominated turbulence either in strong stratification or near the ocean surface, we use the length scale
\beq \label{variable-Pr-length}
\ell^{\rm{shear}}_\psi = \s_\psi(\Ri) \min \left (\frac{\sqrt{e}}{N_+}, \C{s}{} d \right ) \com
\quad \text{where} \quad N^2_+ \defn \max \left (0, \d_z b \right )
\eeq
with $d$ the distance to the ocean surface, $\C{s}{}$ a free parameter (``$s$'' for ``surface''), and $\s_\psi$ a ``stability function'' defined below.
$\sqrt{e} / N$ is the vertical distance traversed by a patch of turbulence expending all its kinetic energy $e$ to mix the uniform stratification $N$.
\citeA{blanke1993variability} point out that $\sqrt{e}/ N$ is a local or constant-stratification version of the more complete, but computationally expensive length scale proposed by \citeA{gaspar1990simple}.

We use~\eqref{variable-Pr-length} for $\ell^\text{shear}_c$, $\ell^\text{shear}_u$, and $\ell^\text{shear}_e$.
For the dissipation length scale $\ell^\text{shear}_D$, we use
\beq \label{dissipation-length}
\ell_D = \frac{1}{\s_D(\Ri)} \min \left ( \frac{\sqrt{e}}{N_+}, \C{s}{} d \right ) \com
\eeq
so that the stability function for the dissipation length scale is $1 / \s_D$
The alternative formulation in~\eqref{dissipation-length} yields a tight connection between $\s_D$'s free parameters and $e$ dissipation, and facilitates the physical interpretation of CATKE's parameters.

The stability functions $\s_\psi$ and $1/\s_D$ in~\eqref{variable-Pr-length}--\eqref{dissipation-length} modulate each length scale with the stably-stratified Richardson number
\beq
\Ri \defn \frac{\partial_z b}{| \d_z \b{u} |^2} \com
\eeq
which, among other meanings, indicates the role of shear production in turbulent mixing.
The stability functions give CATKE a turbulent Prandtl number,
\beq \label{prandtl}
\Pr(\Ri) \defn \frac{\kappa_u}{\kappa_c} = \frac{\s_u(\Ri)}{\s_c(\Ri)} \com
\eeq
that depends on $\Ri$.

\begin{figure}[htp]
    \includegraphics[width = 1\textwidth]{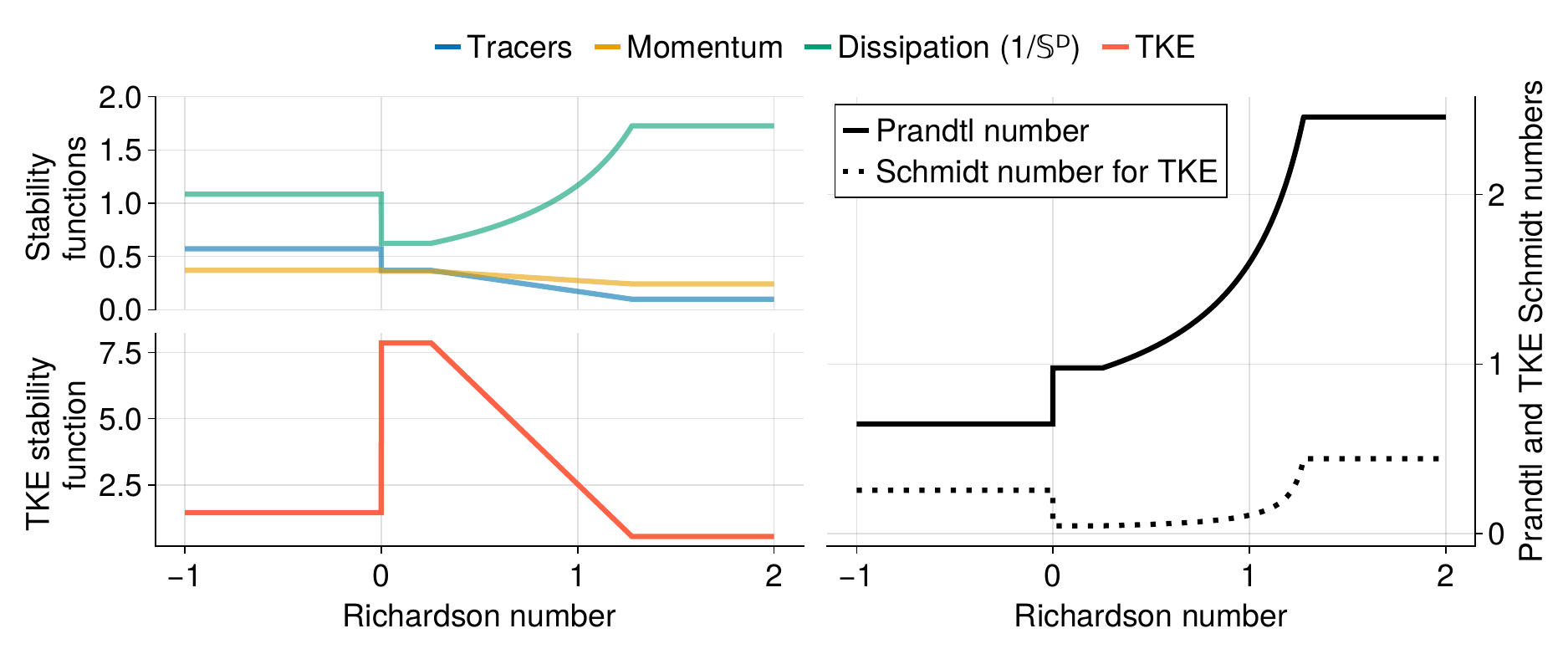}
    \caption{Stability functions (left panel), and Prandtl numbers and Schmidt numbers (right panel). The stability functions for tracers, momentum, and TKE are given by $\s_\psi$ in~\eqref{stability-function}. The stability function for dissipation length scale is $1 / \s_D$. The Prandtl number is $\s_u / \s_c$ and the Schmidt number for TKE is $\s_u / \s_e$.}
    \label{fig:stability-prandtl-schmidt}
\end{figure}

We propose a four-part functions $\s_\psi(\Ri)$,
\beq \label{stability-function}
\s_\psi(\Ri) = \left \{
\begin{matrix}
\C{-}{\psi}   & \quad \text{when} & \Ri < 0 \com \\[1ex]
\C{\lo}{\psi} & \quad \text{when} & 0 \le \Ri \le \C{0}{\Ri} \com \\[1ex]
\C{\lo}{\psi} + \left ( \C{\hi}{\psi} - \C{\lo}{\psi} \right ) \frac{\Ri - \C{0}{\Ri}}{\C{\delta}{\Ri}} & \quad \text{when} & \C{0}{\Ri} < \Ri < \C{0}{\Ri} + \C{\delta}{\Ri} \com \\[1ex]
\C{\hi}{\psi} & \quad \text{when} & \Ri \ge \C{0}{\Ri} + \C{\delta}{\Ri} \per
\end{matrix} \right .
\eeq
In~\eqref{stability-function}, the parameter $\C{0}{\Ri}$ is the ``transition $\Ri$''.
The four regions of the stability function are:
\begin{itemize}
\item Constant $\s_\psi = \C{-}{\psi}$ for unstably-stratified shear turbulence with $\Ri < 0$. 
\item Constant $\s_\psi = \C{\lo}{\psi}$ for near-neutral turbulence with $0 \le \Ri \le \C{0}{\Ri}$
\item Linearly-varying from $\C{\lo}{\psi}$ to $\C{\hi}{\psi}$ as $\Ri$ increases from $\C{0}{\Ri}$ to $\C{0}{\Ri} + \C{\delta}{\Ri}$.
\item Constant $\C{\hi}{\psi}$ when high $\Ri > \C{0}{\Ri} + \C{\delta}{\Ri}$.
\end{itemize}
The stability function~\eqref{stability-function} plays a similar role as the more elaborate stability functions used in two-equation models \cite{burchard2001comparative}, which are derived from an second-moment closure.
The stability functions in equation~\eqref{stability-function} are plotted in the left panel of figure~\ref{fig:stability-prandtl-schmidt} (see section~\ref{sec:calibration} for how the parameters are obtained via calibration to LES).

The four shear length scales introduce 15 free parameters: $\C{s}{}$, $\C{\delta}{\Ri}$, and $\C{0}{\Ri}$ used in all four length scales, along with 12 additional parameters associated with the coefficients $\C{-}{\psi}$, $\C{\lo}{\psi}$ and $\C{\hi}{\psi}$ for each length scale respectively.

\subsubsection{Turbulent Prandtl and Schmidt numbers in stably stratified shear turbulence}

Note that CATKE's $\Pr$ in~\eqref{prandtl} is a rational function of $\Ri$, slightly different from the piecewise linear formulation proposed by \citeA{blanke1993variability} and \citeA{madec2017nemo}.
In particular,
\beq
\Pr = \left \{ \,
\begin{matrix}
\C{-}{u} / \C{-}{c} & \Ri < 0 \\[1ex]
\C{\lo}{u} / \C{\lo}{c} & 0 \le \Ri \, \le \, \C{0}{\Ri} \\[1ex]
\frac{\C{\lo}{u} + \mu_u \left ( \Ri - \C{0}{\Ri} \right )}{\C{\lo}{c} + \mu_c \left ( \Ri - \C{0}{\Ri} \right )} & \C{0}{\Ri} < \Ri < \C{0}{\Ri} + \C{\delta}{\Ri} \\[2ex]
\C{\hi}{u} / \C{\hi}{c} & \Ri \, \ge \, \C{0}{\Ri} + \C{\delta}{\Ri} \\[1ex]
\end{matrix} \right . \com
\eeq
where $\mu_\psi \defn \left ( \C{\hi}{\psi} - \C{\lo}{\psi} \right ) / \C{\delta}{\Ri}$.
Similarly, the Schmidt number for TKE transport in stably-stratified shear turbulence is $\Sc \defn \kappa_u / \kappa_e$.
The Prandtl number and Schmidt number for calibrated parameters are visualized in the right panel figure~\ref{fig:stability-prandtl-schmidt}.

\subsubsection{Neutral, self-similar, wave-modulated, non-rotating, near-surface mixing}
\label{sec:von-karman-constant}

To interpret CATKE's near-surface mixing length $\ell_\psi \sim d$, we consider neutrally-stratified ($\d_z b = 0$), quasi-equilibrium ($\d_t u \approx \d_t e \approx 0$), non-rotating ($f=0$) near-surface turbulence driven by wind stress $\b{\tau} = \tau_x \bxh$.
We hypothesize that CATKE possesses a similarity solution in this scenario,
\beq \label{similarity-shear}
\d_z u \approx \frac{u_\star}{\varkappa \, d} \com
\eeq
where $u_\star$ is the friction velocity~\eqref{eq:ustar} (here simply $ \sqrt{|\tau_x|}$), $d = -z$ is the distance to the surface, and $\varkappa$ is a constant parameter.
If the ocean surface were rigid, $\varkappa$ could be interpreted as the celebrated von K\'arm\'an constant.
But because the LES we use in this paper include surface wave effects, $\varkappa$ has a slightly different interpretation --- as a ``wave-modified'' similarity layer constant, perhaps, as proposed by \citeA{samelson2022wind}.

To express $\varkappa$ in terms of CATKE's free parameters, we begin by assuming a balance between shear production and dissipation and neglecting diffusive turbulent transport to simplify~\eqref{e1d} to
\beq \label{von-karman-e}
\kappa_u \left (\d_z u \right )^2 \approx \frac{e^{3/2}}{\ell_D} \per
\eeq
Note that in neutral conditions,
\beq \label{neutral-momentum-diffusion}
\kappa_u = \C{0}{u} \C{s}{} d \sqrt{e} \com
\qquad \text{and} \qquad 
\ell_D = \frac{\C{s}{}}{\C{0}{D}} d\per
\eeq
Inserting~\eqref{similarity-shear} and~\eqref{neutral-momentum-diffusion} into~\eqref{von-karman-e} and rearranging, we find an expression that relates the constant $\varkappa$, $u_\star$, and $e$,
\beq \label{u-star-to-e}
\frac{u_\star^2}{e} \approx \varkappa^2 \frac{\C{0}{D}}{\C{0}{u} \left ( \C{s}{} \right )^2} \per
\eeq
Notice that $e$ is independent of $d$ in this expression.
This means that neglecting turbulent transport in~\eqref{von-karman-e} in the context of the similarity hypothesis~\eqref{similarity-shear} is at least self-consistent, though this assumption may fail when applied over significant portions of the boundary layer.
Next, integrating the quasi-equilibrium $x$-momentum equation $0 \approx \d_z \left ( \kappa_u \d_z u \right )$ from $z=0$ to $z=-d$ yields
\beq \label{catke-similarity-shear}
\d_z u \approx \frac{u_\star}{d} \underbrace{\frac{u_\star}{\C{0}{u} \C{s}{} \sqrt{e}}}_{= 1 / \varkappa} \com
\eeq
where we have used the neutral momentum diffusivity in~\eqref{neutral-momentum-diffusion} and the friction velocity definition $- \kappa_u \d_z u |_{z=0} = u_\star$.
Equation~\ref{catke-similarity-shear} identifies $\varkappa$ by comparison to~\eqref{similarity-shear}.
We next use~\eqref{u-star-to-e} to eliminate $u_\star / \sqrt{e}$ to obtain an expression for CATKE's wave-modified similarity layer constant $\varkappa$,
\beq \label{von-karman-def}
\varkappa \defn \C{s}{} \left [ \left ( \C{0}{u} \right )^3 \! \Big /\C{0}{D} \right ]^{1/4} \per
\eeq

\subsubsection{Steady-state gradient Richardson number for stably stratified shear turbulence}
\label{sec:critical-richardson-number}

CATKE's dependence on the stable length scale $\ell \sim \sqrt{e} / N$ is associated with a steady-state gradient Richardson number in stably-stratified shear turbulence \cite{blanke1993variability}.
To see this, we first note that in stable stratification and far from boundaries, the mixing and dissipation length scales become
\beq \label{stable-shear-turbulence-scales}
\ell_\psi = \s_\psi \frac{\sqrt{e}}{N} 
\quad \text{for } \quad \psi \in (u, c, e) \quad \text{ and} \qquad
\ell_D = \frac{1}{\s_D} \frac{\sqrt{e}}{N} \per
\eeq
Inserting~\eqref{stable-shear-turbulence-scales} into~\eqref{e1d} and neglecting turbulent transport (equivalently, assuming spatially-uniform $e$) yields
\beq \label{critical-Ri-tke}
\d_t e = \underbrace{\mystrut{3ex} N ( \s_c + \s_D) \Bigg ( \frac{\Ri^\dagger}{\Ri} - 1 \Bigg )}_{\defn r} e \com
\eeq
where $r$ is a rate and $\Ri^\dagger$ is the steady-state Richardson number,
\beq
\Ri^\dagger \defn \frac{\s_u}{\s_c + \s_D} %\approx 0.18 \per
\eeq
When the Richardson number $\Ri = \Ri^\dagger$ equals the steady-state value $\Ri^\dagger$, the shear production of TKE is perfectly balanced by TKE destruction via buoyancy flux and dissipation.
But if $\Ri < \Ri^\dagger$, then $r > 0$ --- and TKE will grow.
Conversely, if $\Ri > \Ri^\dagger$ then $r < 0$ and TKE will decay.
Finally we note that the functions $\s_\psi$, defined in~\eqref{stability-function}, depend on $\Ri$.
For example if $\Ri < \C{0}{\Ri}$, then $\Ri^\dagger = \C{\lo}{u} / \left ( \C{\lo}{c} + \C{\lo}{D} \right )$.
But if $\Ri^\dagger > \C{0}{\Ri} + \C{\delta}{\Ri}$, then $\Ri^\dagger = \C{\hi}{u} / \left ( \C{\hi}{c} + \C{\hi}{D} \right )$.

\subsubsection{Convective turbulence length scale}
\label{sec:convective-mixing-length}

To formulate a length scale for free convection, we divide the freely convecting boundary layer into two regions: a ``convecting layer'' with unstable $N^2 < 0$, and a ``penetration layer'' with thickness $\delta$.
In the penetration layer, $N^2(z) > 0$ but $N^2(z + \delta) < 0$, where we note that the vertical coordinate $z$ increases upwards and is defined such that $z < 0$.
(We use ``penetration layer'' rather than ``entrainment layer'' used by \citeA{deardorff1970convective} because it is less likely to be confused with other types of ``entrainment''.)
Our formulation for the convective length scale models both rapid mixing in the convective layer as well as entrainment into the boundary layer from below by plumes plunging through the convecting layer into the stably-stratified penetration layer below.

Our dynamic length scale for mixing in the convective layer is based on a dimensional analysis first proposed by \citeA{deardorff1970convective} that links the turbulent velocity $\sqrt{e}$ ($\mathrm{m \, s^{-1}}$), surface buoyancy flux $J_b$ (m$^2$/s$^3$), and convective layer depth, $h$ (m),
\beq \label{turbulent-convection-scaling}
\sqrt{e} \sim \left (h \, J_b \right )^{1/3} \per
\eeq
Recasting~\eqref{turbulent-convection-scaling} in terms of a time-scale $t_\text{mix} \sim h / \sqrt{e}$ for convective mixing over the depth $h$ yields
\beq \label{convective-mixing-time}
t_\text{mix} \sim \left ( \frac{h^2}{J_b} \right )^{1/3} \per
\eeq
But if we represent convection as a diffusive process with diffusivity $\kappa_c$, then we also have that
\beq \label{diffusivity-mixing-time}
t_\text{mix} \sim \frac{h^2}{\kappa_c} \per
\eeq
Equating~\eqref{convective-mixing-time} and~\eqref{diffusivity-mixing-time} yields a scaling relation for the convective diffusivity $\kappa_c$.

Now consider convection driven by constant destabilizing buoyancy fluxes $J_b$ and increasing $h(t)$: according to~\eqref{convective-mixing-time}, the mixing time then evolves according to $t_\text{mix} \sim h^{2/3}$.
On the other hand, if we instead we impose a \textit{constant} $\kappa_c$ --- a commonly used parameterization when $N^2 < 0$ \cite{madec2017nemo, kuhlbrodt2018low, gutjahr2021comparison, jungclaus2022icon} --- then~\eqref{diffusivity-mixing-time} implies that, spuriously, $t_\text{mix} \sim h^2$. 
Thus, constant convective adjustment diffusivities inaccurately exhibit $t_\text{mix} \sim h^2$ and may produce bias when convection competes with other processes such as lateral restratification, or biogeochemical production and destruction.

To capture $t_\text{mix}$ consistently between~\eqref{convective-mixing-time} and~\eqref{diffusivity-mixing-time} over the convective region where $N^2 < 0$, we introduce a dynamic convective mixing length scale $\ell^h_\psi$ that scales with $h$,
\beq \label{unstable-convective-mixing-length}
\ell^h_\psi \defn \C{h}{\psi} \frac{e^{3/2}}{\widetilde J_b + J_b^\text{min}} \sim h \com
\eeq
where the regularizer $J_b^\text{min}$ is a minimum convective buoyancy flux parameter chosen small enough to have no impact on CATKE-parameterized solutions, and
$\widetilde J_b$ is an estimate of the slowly-evolving part of the buoyancy flux $J_b$ averaged over time-scales $t \sim t_\text{mix}$.
%In~\eqref{unstable-convective-mixing-length}, $\widetilde J_b$ is an estimate of the slowly-evolving part of the buoyancy flux averaged over time-scales $t \sim t_\text{mix}$.
We compute $\widetilde J_b$ by integrating
\beq \label{time-average-q}
\d_t \widetilde J_b = \underbrace{\mystrut{3ex} \left ( \frac{J_b}{\ell_D^2(z=0)} \right )^{1/3}}_{\mystrut{4ex} \sim t_\text{mix}^{-1}} \left ( J_b - \widetilde J_b \right ) \com
\eeq
where $\ell_D$ is the dissipation length scale and $(\ell_D^2 / J_b)^{1/3} \sim t_\text{mix}$ scales with the instantaneous convective mixing time.
Equation~\eqref{time-average-q} relaxes $\widetilde J_b$ to $J_b$ over the time-scale $t_\mathrm{mix}$ as defined by~\eqref{convective-mixing-time}, and therefore effectively acts to average $J_b$ in time.
We use the dissipation length scale $\ell_D$ in~\eqref{time-average-q} rather than the tracer mixing length $\ell_c$ because we hypothesize that convective turbulence evolution time-scale is most closely related to the time-scale for turbulent kinetic energy dissipation rather than the time-scale for tracer mixing.
In quasi-equilibrium, $\widetilde J_b \approx J_b$.
Because $\ell^h_\psi \sim h$, CATKE's convective tracer diffusivity scales with $\kappa_c \sim h \sqrt{e}$.

The second objective of our convective mixing length formulation is to correctly predict the evolution of $h$.
For this we introduce a model for ``penetrative mixing'' \textit{below} the convective mixed layer associated with convective plumes that plunge through the mixed layer and penetrate into the strongly stratified region below.
The ``empirical law of convection'' \cite{large1994oceanic, siebesma2007combined, van2018kpp, souza2020uncertainty, souza2022flux} is the observation, robust across a wide range of convective conditions, that penetrative fluxes at the penetration level  $z_p$ scale with
%the entrainment flux at the entrainment level $z_e$ scales with
\beq \label{empirical-law-convection}
\overline{w' b'} \, |_{z=z_p} \sim - J_b \qquad \text{such that} \qquad h^2 \sim \frac{J_b t}{N^2} \com
\eeq
for initially-constant buoyancy gradient $N^2$ and constant buoyancy flux $J_b$.

To ensure that CATKE reproduces~\eqref{empirical-law-convection}, we introduce a ``penetrative mixing length'',
\beq \label{free-entrainment-length}
\ell^p_{\psi} \defn \C{p}{c} \frac{\widetilde J_b}{N^2 \sqrt{e} + J_b^\text{min}} \com
\eeq
which is applied at the height $z_p < 0$ defined via
\beq
N^2(z_p) > 0
\qquad \text{and} \qquad
N^2(z_p + \delta) < 0 \com
\eeq
where $\delta$ is the thickness of the penetration layer.
At $z=z_p$,~\eqref{free-entrainment-length} produces $\overline{w' b'} = - \ell^p_c \sqrt{e} N^2 \approx - \C{p}{c} J_b$ in accordance with the empirical law in~\eqref{empirical-law-convection}.
Our numerical implementation of the convective mixing length uses $\delta = \Delta z$ where $\Delta z$ is the grid spacing at $z_p$.
This assumes that the entrainment layer is thinner than the grid spacing: when $\delta > \Delta z$, CATKE solutions may exhibit a ``thin entrainment layer bias'' even if the boundary layer deepening rate is correct.

Finally, because $e$ is much larger in shear turbulence than in convective turbulence with similar mixing rates, the scaling~\eqref{unstable-convective-mixing-length} will greatly overestimate the mixing length when $e$ is produced by both convection and shear.
%\textcolor{red}{Can we find a reference that points this out too?}
To limit the impact of the convective mixing length in the presence of shear, we use an estimate of the flux Richardson number,
\beq \label{sheared-convection-number}
\widetilde{Ri_f} \defn \frac{d \sqrt {e} | \d_z \bu |^2}{\widetilde J_b + J^\text{min}_b} \com
\eeq
where $d = -z$ is depth, which measures the relative contribution of shear production (the numerator) versus buoyancy flux (the denominator) to the TKE budget in unstable stratification.
We then use this estimate to reduce the convective mixing length by
\beq \label{shear-reduction-factor}
\epsilon_{sp} \defn \max \big (0,  1 - \C{sp}{} \, \widetilde{Ri_f} \big ) \com
\eeq
where $\C{sp}{}$ is a free parameter.
The reduction factor~\eqref{shear-reduction-factor} is used in lieu of more detailed understanding of how shear acts to limit turbulence correlation scales during convection.
Note that the numerator in~\eqref{sheared-convection-number} estimates shear production using the mixing length $d$, which is appropriate for shear-driven turbulent mixing.
This formulation means that the free convection length scale is more limited at depth, where convective plumes are less connected to destabilizing surface buoyancy fluxes.

\begin{figure}[htp]
    \includegraphics[width = 1\textwidth]{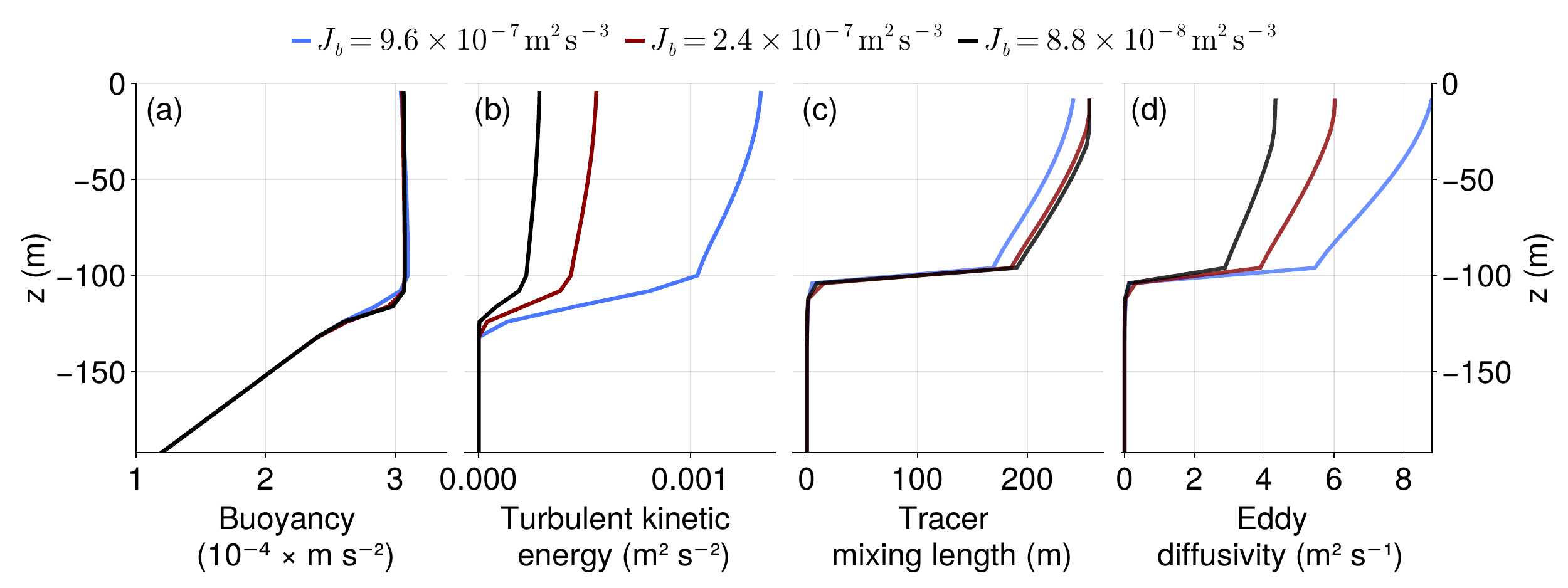}
    \caption{CATKE mixing length and eddy diffusivity during free convection for three cases with boundary layer depth $h \approx 100$ m.
    (a) CATKE-predicted buoyancy profiles for the three cases, (b) profiles turbulent kinetic energy, $e$, (c) tracer mixing lengths $\ell_c$, (d) tracer eddy diffusivities $\kappa_c$.
    %Blue shades the unstable ``convecting layer'' where $N^2 < 0$ and thus $\ell_c \sim h \sim e^{3/2} / J_b$.
    %Red shades the ``penetration layer'' just below the unstable convecting layer, where $N^2(z) > 0$ but $N^2(z + \Delta z)<0$.
    The buoyancy fluxes $J_b$ correspond to heat fluxes $Q \approx 2000$, 500, and 183 $\mathrm{W \, m^{-2}}$ using $Q \approx \rho_o c_p J_b / \alpha g$ and $\rho_o = 1024 \, \mathrm{kg \, m^{-3}}$, $c_p = 3991 \, \mathrm{J \, {}^\circ C^{-1}}$, $\alpha = 2 \times 10^{-4} \, \mathrm{\, {}^\circ C^{-1}}$, and $g = 9.81 \, \mathrm{m \, s^{-2}}$.}
    \label{fig:convective-length-scale}
\end{figure}

Putting~\eqref{unstable-convective-mixing-length},~\eqref{free-entrainment-length}, and~\eqref{shear-reduction-factor} together yields the piecewise parameterization
\beq \label{convective-length}
\ell^{\rm{conv}}_\psi(z) = \epsilon_{sp} \left \{
\begin{matrix}
\ell^h_\psi & \text{if } N^2 < 0 \text{ and } J_b > 0 \com \\[1ex]
\ell^p_\psi & \text{if } N^2 > 0 \com N^2(z + \Delta z) < 0 \com \text{ and } J_b > 0 \com \\[1ex]
0 & \text{otherwise} \per
\end{matrix} \right .
\eeq
%The formulation~\eqref{convective-length} defines the CATKE-predicted boundary layer structure during free convection.
%The CATKE-predicted convective boundary layer structure therefore differs --- slightly --- from \textit{observed} boundary layer structure wherein $N^2 |_\text{data}$ is close to zero or slightly positive \textit{above} the entrainment layer.
Figure~\ref{fig:convective-length-scale} illustrates the behavior of the convective length scale predicted by CATKE in~\eqref{convective-length} for three free convection cases with surface buoyancy fluxes $J_b = 9.6 \times 10^{-7}$, $2.4 \times 10^{-7}$, and $8.8 \times 10^{-8} \, \mathrm{m^2 \, s^{-3}}$ integrated for 6, 24, and 72 hours respectively, using the initial buoyancy profile in equation~\eqref{initial-buoyancy}, which is also used for all our LES.
Figure~\ref{fig:convective-length-scale}(a) shows CATKE-simulated buoyancy profiles after integrating for 6, 24, and 72 hours.
Figure~\ref{fig:convective-length-scale}(b) shows that stronger forcing cases have greater levels of turbulent kinetic energy.
Figure~\ref{fig:convective-length-scale}(c) shows the tracer mixing length, which above $z=-100$ meters is dominated by the convective mixing length.
Though each case has different TKE and different surface buoyancy flux, they nevertheless predict similar tracer mixing lengths which are $O(100)$ meters and thus similar to the boundary layer depth, corroborating the dimensional analysis in equation~\eqref{turbulent-convection-scaling}.
Figure~\ref{fig:convective-length-scale}(d) shows the eddy diffusivity for the three cases --- unlike a typical constant-diffusivity convective adjustment model, CATKE's ``convective adjustment diffusivity'' varies depending on the strength of the surface buoyancy flux.
Because the predicted mixing length is similar for all three cases, the tracer diffusivity varies with the surface buoyancy flux due to variation in the turbulent kinetic energy.

\section{\textit{A posteriori} calibration against large eddy simulations}
\label{sec:calibration}

We calibrate CATKE's 23 free parameters in an \textit{a posteriori} \cite{duraisamy2021perspectives, frezatposteriori2022} single-column context using horizontally-averaged data from 21 LES described in section~\ref{sec:les-data} and~\ref{les-description}.
\textit{A posteriori} calibration estimates free parameters by minimizing the error between LES data --- $b(z, t)$, $u(z, t)$, $v(z, t)$, and the forced passive tracer $c(z, t)$ extracted from solutions of~\eqref{momentum}--\eqref{tracers} --- and single column simulations of $b$, $u$, $v$, and $c$ in~\eqref{u1d}--\eqref{c1d} that use CATKE as a parameterization.
The minimization is computed over the whole time series and thus in \textit{a posteriori} calibration free parameters are determined by directly minimizing simulation bias.
In this way, \textit{a posteriori} calibration incorporates numerical and other errors that accumulate during a simulation.
Moreover, \textit{a posteriori} calibration can leverage any observational data computable from the predicted solution, even only indirectly informative data.
For example, in this work we calibrate elements of the TKE equation using only horizontally-averaged momentum and buoyancy profiles derived from LES.  

\subsection{The importance of \textit{a posteriori} calibration}

Explicitly minimizing simulation bias distinguishes \textit{a posteriori} calibration from other methods that minimize other biases that are only indirectly related to simulation bias --- for example by attempting to compute free parameters directly from data, usually by considering subcomponents of the parameterization in isolation \cite<examples may be found in>{umlauf2003generic, reichl2019parameterization}.
These latter methods are called ``\textit{a priori}'' \cite{duraisamy2021perspectives}, because they hinge critically on additional and often problematically strong hypotheses --- such as an assumption of structurally perfect, unbiased parameterization (permitting a direct computation of free parameters from limited data), or an assumption that free parameters are uncorrelated with one another (permitting free parameters to be determined in isolated contexts, rather than leveraging all data simultaneously).

To illustrate the pitfalls of \textit{a priori} calibration, we consider integrating a parameterized single column equation for buoyancy $b$,
\beq \label{parameterization}
\d_t b = - \d_z \underbrace{\mystrut{1.5ex} \mathcal{J}(b; \c)}_{\text{parameterization}} + \underbrace{\mystrut{1.5ex} \xi}_{\text{noisy error}} \per
\eeq
In~\eqref{parameterization}, we include two terms: \textit{(i)} the divergence of a parameterized flux $\mathcal{J}$ that depends on both the simulated buoyancy $b$ (omitting here for simplicity other aspects of the state such as~$u$ or~$v$) and a set of free parameters $\c$, and \textit{(ii)} an explicit ``error'' term $\xi$ that represents spatial and temporal discretization errors.
We additionally define the ideal or ``perfect'' solution as $b^\dagger$.
When equation~\eqref{parameterization} is integrated forward to predict the evolution of~$b$, fluctuations away from the perfect solution~$b^\dagger$ inevitably develop due both to structural errors in $\mathcal{J}$ and because of the discretization error $\xi$, leading to an error $\varepsilon = b - b^\dagger$ that grows as $\sqrt{t}$ \cite<see, for example>{gardiner2021elements}. %(optimistically, if the errors are random, cf Langevin).

This error accumulation is potentially fatal for \textit{a-priori}-calibrated parameterizations: because the parameters $\mathbb{C}$ are determined by evaluating $\mathcal{J}(b^\dagger)$ in terms of the \textit{perfect} $b^\dagger$, while the predictions $\mathcal{J}(b)$ made in terms of the noisy $b$ are unconstrained by the calibration procedure.
At best, the unconstrained predictions $\mathcal{J}(b)$ are inaccurate.
At worst, however, the errors $\mathcal{J}(b) - \mathcal{J}(b^\dagger)$ self-amplify without bound, thwarting prediction altogether \cite{rasp2018deep, brenowitz2019spatially, rasp2020coupled}.

\textit{A posteriori} calibration avoids all of these pitfalls by definition, since $\mathcal{J}(b, \c_\star)$ computed in terms of the simulated $b$ and optimal parameters $\c_\star$ is explicitly constrained by minimizing the discrepancy between $\mathcal{J}(b, \c)$ and data.
Put differently: \textit{a posteriori} calibration ``teaches'' $\mathcal{J}$ how to make accurate, stable predictions in terms of potentially noisy inputs $b$.
We leverage this feature to realize a key innovation of this work: we explicitly minimize spatial discretization error by including single-column simulations with 2-, 4-, and 8-meter resolution in our loss function.

\subsection{Ensemble Kalman Inversion for \textit{a posteriori} calibration}

The downside of \textit{a posteriori} calibration is  that nonlinear inverse problems are difficult to solve.
In this work we use an ensemble-based, gradient-free method called Ensemble Kalman Inversion \cite<EKI; >{iglesias2013ensemble}.
A major advantage of EKI is that it does not require a gradient or adjoint of the CATKE-parameterized single column model. Instead, EKI only requires the ability to evaluate the loss functions for an ensemble of free parameters.
The EKI algorithm can be construed either as the integration of a dynamical system or as an iterative scheme for repeatedly refining an initial distribution of free parameter values.

EKI minimizes the ``EKI objective function'' $\Phi$, defined as
\beq \label{eki-objective}
\Phi(\G, \y; \c) \defn \big \| \M^{-1/2} \left [ \G(\c) - \y \right ] \big \|^2 \com
\eeq
where $\y$ denotes observations, $\G(\c)$ denotes a parameterized prediction of the observations made with a set of free parameters $\c$, and $\M$ is a covariance matrix that represents the uncertainty of $\y$.
$\Phi$ measures the discrepancy between $\G(\c)$ and $\y$ given uncertainty $\M$.
The data $\y$ is extracted from 21 of the LES described in table~\ref{table:les-summary} that have intermediate surface forcing, each coarse-grained three times to 2-, 4-, and 8-meter vertical resolution.
$\G$ is constructed by assembling $21 \times 3 = 63$ single column simulations, representing a prediction of each of the 21 LES cases at the three vertical resolutions.

We note that the near-surface dynamics in the LES seems uncertain.
For example, the LES profiles exhibit strong unstable near-surface buoyancy gradients for strongly-forced convective cases.
Though these features are robust to changes in LES resolution (see ~\ref{les-description}), we are unsure whether the simple implicit LES turbulence closure is missing crucial turbulent mixing processes important near a wavy, bubbly, broken ocean surface.
We therefore omit the top 8 meters of the LES domain from $\y$ to avoid overconstraining parameters based on the most uncertain elements of the LES data.

EKI finds a set of optimal parameters $\c = \c_\star$ that minimize $\Phi(\G, \y, \c)$ in~\eqref{eki-objective} by evolving an ensemble of parameter sets using the algorithm described in~\ref{calibration-details}.
In this work we use relatively large ensembles with 1000 members.
This means that every EKI iteration requires performing up to $21 \times 3 \times 1000 = 63,000$ single column simulations, corresponding to 21 LES cases and 3 vertical resolutions for every ensemble member.
To make the calibration as efficient as possible, we implement CATKE in Oceananigans and leverage a feature that permits us to integrate an ensemble of single column models in parallel in the configuration of a single three-dimensional simulation on a GPU. 
As a result, each EKI iteration requires evolving 9 effectively three-dimensional simulations (3 resolutions for each of the 12-, 24- and 48-hour suites).
On an Nvidia Titan V GPU and with 1,000 ensemble members, a single EKI iteration takes 40-50 seconds, and the entire calibration takes 4-6 hours.
In the course of this work we have performed complete calibrations of CATKE's parameters hundreds of times --- to experiment with new formulations, new numerical schemes, and to tweak the calibration setup.
This workflow represents a new ``calibration-based'' paradigm in parameterization development, where physical formulation or numerical implementation changes are tested against the baseline by comparing predictions for independently calibrated parameterizations.
The 23 calibrated free parameters that correspond to the version of CATKE described in this paper and the previously described LES are listed in table~\ref{favorite-parameters}.

\begin{table}  

\centering
\begin{tabular}{c c c c}
 \bf{Symbol} & \bf{Description} & \bf{Optimal value} & \bf{Bounds} \\[1ex] 
 \hline \\[-1ex]
 $\C{\rm{shear}}{J}$    & Wind stress TKE surface flux              & 3.18   & $(0, 2)$      \\[1ex] 
 $\C{\rm{conv}}{J}$     & Convective TKE surface flux               & 0.38   & $(0, 2)$     \\[1ex] 
 $\C{s}{}$              & Near-surface mixing scale                 & 1.13   & $(0, 2)$      \\[1ex] 
 $\C{h}{c}$             & Tracer free convection scale              & 4.79   & $(0, 8)$     \\[1ex] 
 $\C{-}{c}$             & Tracer mixing for negative $\Ri$          & 0.57   & $(0, 2)$      \\[1ex] 
 $\C{\lo}{c}$           & Tracer mixing for near-neutral $\Ri$      & 0.37   & $(0, 2)$      \\[1ex] 
 $\C{\hi}{c}$           & Tracer mixing for high $\Ri$              & 0.098  & $(0, 2)$      \\[1ex] 
 $\C{p}{c}$             & Tracer free entrainment scale             & 0.11   & $(0, 2)$      \\[1ex] 
 $\C{h}{u}$             & Momentum free convection scale            & 3.71   & $(0, 8)$     \\[1ex] 
 $\C{-}{u}$             & Velocity mixing for negative $\Ri$        & 0.37   & $(0, 2)$      \\[1ex] 
 $\C{\lo}{u}$           & Velocity mixing for near-neutral $\Ri$    & 0.36   & $(0, 2)$      \\[1ex] 
 $\C{\hi}{u}$           & Velocity mixing for high $\Ri$            & 0.24   & $(0, 2)$      \\[1ex] 
 $\C{h}{e}$             & TKE free convection scale                 & 3.64   & $(0, 8)$     \\[1ex] 
 $\C{-}{e}$             & TKE transport for negative $\Ri$          & 1.44   & $(0, 8)$      \\[1ex] 
 $\C{\lo}{e}$           & TKE transport for near-neutral $\Ri$      & 7.86   & $(0, 8)$      \\[1ex] 
 $\C{\hi}{e}$           & TKE transport for high $\Ri$              & 0.55   & $(0, 8)$      \\[1ex] 
 $\C{h}{D}$             & Dissipation free convection scale         & 3.25   & $(0, 8)$     \\[1ex] 
 $\C{-}{D}$             & Dissipation scale for negative $\Ri$      & 0.92   & $(0, 8)$      \\[1ex] 
 $\C{\lo}{D}$           & Dissipation scale for near-neutral $\Ri$  & 1.60   & $(0, 8)$      \\[1ex] 
 $\C{\hi}{D}$           & Dissipation scale for high $\Ri$          & 0.58   & $(0, 8)$      \\[1ex] 
 $\C{0}{\Ri}$           & Stability function transitional $\Ri$     & 0.25   & $(0, 2)$      \\[1ex] 
 $\C{\delta}{\Ri}$      & Stability function $\Ri$ width            & 1.02   & $(0, 2)$      \\[1ex] 
 $\C{sp}{}$             & Sheared plume scale                       & 0.50   & $(0, 2)$     \\[4ex] 
\end{tabular}
\caption{A summary of CATKE's free parameters.
Note that ``near-neutral $\Ri$'' means $\Ri \le \C{0}{\Ri}$, while ``high $\Ri$'' means $\Ri \ge \C{0}{\Ri} + \C{\delta}{\Ri}$.
The bounds limit the values a parameter can take during calibration, using the method described in~\ref{appendix:bounds}.
The prior distributions for each parameter span the range between the bounds.
\label{favorite-parameters}
}
\end{table}

\section{Validation}
\label{sec:validation}

We next assess CATKE's ability to make accurate predictions in a single column context with the free parameters listed in table~\ref{favorite-parameters}.
First, we derive quantities with well-understood physical interpretations from CATKE's free parameters, and evaluate whether their calibrated values are close to expected or directly measured values reported in the literature.
Second, we compare CATKE-parameterized simulations both to the 21 constant-forcing LES used for calibration and to an additional 12 constant-forcing LES that are both more strongly and more weakly forced than the calibration LES.
Third, we conduct a 34-day CATKE-parameterized simulation of equatorial deep-cycle turbulence using the dataset provided by \citeauthor{whitt2022simulation}~\citeyear{whitt2022simulation}, and then compare the results to the LES used therein.
This third validation context is useful because it involves both time-dependent surface forcing, solar insolation, and lateral flux divergences derived from a high resolution tropical GCM.
Finally, we evaluate CATKE's sensitivity to vertical resolution and time-step size.
These all provide a measure of confidence in CATKE's ability to not only represent the LES data used for calibration but also to extrapolate to differently-forced conditions, time-dependent surface forcing, and GCM-like contexts that include lateral flux divergences from for example, the advection of momentum, temperature, and salinity.
All of this said, we maintain a caveat that CATKE should still be assessed, and likely recalibrated, in a regional or global context that is more similar to the context in which CATKE is intended to be used.

\subsection{Derived quantities}

Table~\ref{derived-parameteres} shows several quantities that can be derived or computed in terms of CATKE's calibrated free parameters.
Note that there is unknown uncertainty in these estimates, so the precise values must be taken with a grain of salt. 
Uncertainty quantification, using the methodology proposed by \citeA{cleary2021calibrate} for example, is left for future work.

\subsubsection{Steady-state Richardson number}

Section~\ref{sec:critical-richardson-number} shows how a steady-state $\Ri$ may be derived from CATKE's TKE equation.
From the parameters in table~\ref{favorite-parameters}, we find that
\beq
\Ri^\dagger \defn \frac{\C{0}{u}}{\C{0}{c} + \C{0}{D}} \approx 0.18 \com
\eeq
which lies in the ``near-neutral'' stability function regime, since $\C{\lo}{\Ri} = 0.25 > \Ri^\dagger$.
$\Ri^\dagger = 0.18$ is somewhat less than the 0.23 used by \citeauthor{blanke1993variability}~\citeyear{blanke1993variability}, or the celebrated value $\Ri = 1/4$ that determines the stability of a laminar stratified shear layer.
In section~\ref{sec:tropical-turbulence}, we find that $\Ri^\dagger$ is a crucial parameter controlling mixing in forced stably-stratified turbulence, and that LES tend to exhibit $\Ri$ in the range 0.2--0.23.

\subsubsection{Near-surface similarity constant}

Section~\ref{sec:von-karman-constant} shows how a near-surface similarity constant --- analogous to the von K\'arm\'an constant for turbulence near rigid non-wavy walls --- may be computed from the near-wall and momentum stability function parameters.
In terms of the parameters in table~\ref{favorite-parameters} from~\eqref{von-karman-def} we find that
\beq
\varkappa = \C{s}{} \left [ \left ( \C{0}{u} \right )^3 \! \Big /\C{0}{D} \right ]^{1/4} \approx 0.47 \com
\eeq
which is slightly higher than the celebrated rigid-wall von K\'arm\'an constant value of $0.4$.
A slightly higher similarity constant is consistent with the notion that surface waves act to increase the coherence of turbulent motions, which increases mixing lengths and suppresses turbulent kinetic energy dissipation.

A similar wave-induced enhancement to the similarity constant is proposed by \citeA{samelson2022wind}.
However, \citeA{samelson2022wind} models the enhancement as a function of wind at ten meters height, $u_{10}$.
In our case, the LES are forced with varying $u_{10}$, but constant Langmuir number $\La \approx 0.3$ (see table~\ref{table:les-summary} for a summary of the LES cases).
Thus we must either hypothesize that surface waves can be modeled with a $\La$-dependent enhancement of $\varkappa$, or that CATKE is missing physics.
Either way, we are unable to proceed further in determining wave-induced enhancements to $\varkappa$ without LES that vary both $u_{10}$ and $\La$, so we save such considerations for future work.

\subsubsection{The turbulent Prandtl number}

The turbulent Prandtl number is defined as
\beq
\Pr \defn \frac{\kappa_u}{\kappa_c} \com
\eeq
which is derived for CATKE in section~\ref{sec:shear-turbulence-length-scale}.
For various regimes of turbulence we obtain
\begin{itemize}
\item $\Pr_{c} \approx 0.77$ for weakly-sheared convection,
\item $\Pr_{-} \approx 0.65$ for unstably-stratified shear turbulence,
\item $\Pr_0 \approx 0.98$ for near-neutral shear turbulence,
\item $\Pr_\infty \approx 2.46$ for strongly-stratified shear turbulence.
\end{itemize}
A turbulent $\Pr$ that increases from less than unity to above unity as $\Ri$ crosses zero is consistent with laboratory and DNS studies \cite<for example,>{li2019turbulent}, as well as what is typically used in two-equation models \cite<for example,>{burchard2001comparative}.
On the other hand, one-equation models \cite{blanke1993variability, madec2017nemo} typically prescribe $\Pr$ to a value of 10 or higher as $\Ri$ tends to infinity.
It is unlikely that our boundary layer LES are informative for such high $\Ri$ mixing, so more LES are needed to assess and perhaps refine CATKE's stability function to capture very high $\Ri$ regimes.

\subsubsection{The turbulent Schmidt number}

Calibration determines that $Sc = 0.26$ for unstably-stratified shear turbulence with $\Ri < 0$, and then varies between $0.046 < \Sc < 0.44$ as $\Ri$ increases from 0 to $\C{0}{\Ri} + \C{\delta}{\Ri}$.
As a result, TKE is transported much more rapidly than momentum or tracers in shear-dominated turbulence, and similarly to momentum or tracers in convective or weakly-sheared stratified turbulence.
Rapid TKE diffusion relative to momentum or tracer diffusion introduces an ``implicitly non-local'' element to CATKE's mixing predictions, because TKE transport can generate mixing in a region that is displaced from the region of TKE generation.

\subsubsection{Stratified turbulence mixing coefficient}

The ``mixing coefficient'' --- the ratio between buoyancy flux and dissipation in stably-stratified turbulence \cite{gregg2018mixing, colm2020open} --- measures the relative TKE converted to potential energy in the process of mixing buoyancy vs TKE dissipation.
Using~\eqref{e1d} and assuming stably-stratified turbulence far from boundaries such that $\ell_c = \s_c \sqrt{e} / N$, $\ell_D = \sqrt{e} / (\s_D N)$, and $\kappa_c = \s_c e / N$, we find that
\beq \label{mixing-coefficient}
\Gamma \defn - \frac{\text{buoyancy flux}}{\text{dissipation}} = \frac{\s_c}{\s_D} \per
\eeq
The free parameters in table~\ref{favorite-parameters} imply that the mixing coefficient~$\Gamma$ varies between $\Gamma_0 \approx 0.26$ for near-neutral turbulence and $\Gamma_\infty \approx 0.17$ for strongly-stratified (shear-free) turbulence.
The latter is applicable to internal wave breaking, where an extensive literature suggests that $\Gamma_\infty \approx 0.2$ \cite{gregg2018mixing}.

\begin{table}  
\centering
\begin{tabular}{c c c c}
 \bf{Symbol} & \bf{Value} & \bf{Description} \\[1ex] 
 \hline \\[-1ex]
 $\Ri^\dagger$      & 0.18  & Steady-state gradient Richardson number \\[1ex] 
 $\varkappa$      & 0.47  & Near-neutral near-surface similarity constant \\[1ex] 
 $\Pr_0$          & 0.98   & Near-neutral turbulent Prandtl number ($\Ri \to 0$)              \\[1ex] 
 $\Pr_\infty$     & 2.46   & Strongly-stratified turbulent Prandtl number ($\Ri \to \infty$)  \\[1ex] 
 $\Pr_{-}$        & 0.65  & Unstably-stratified shear turbulence Prandtl number ($\Ri < 0$)   \\[1ex] 
 $\Pr_{c}$        & 0.77  & Free convection turbulent Prandtl number ($\Ri \to -\infty$)      \\[1ex] 
 $\Gamma_0$       & 0.23  & Near-neutral mixing coefficient ($\Ri \to 0$) \\[1ex]
 $\Gamma_\infty$  & 0.17  & Strongly-stratified mixing coefficient ($\Ri \to \infty$) \\[1ex]
 $\Sc_0$          & 0.046 & Near-neutral turbulent TKE Schmidt number ($\Ri \to 0$)              \\[1ex] 
 $\Sc_\infty$     & 0.44  & Strongly-stratified turbulent TKE Schmidt number ($\Ri \to \infty$)  \\[1ex] 
 $\Sc_{-}$        & 0.26  & Unstably-stratified shear turbulence TKE Schmidt number ($\Ri < 0$)   \\[1ex] 
 $\Sc_{c}$        & 1.02  & Free convection turbulent TKE Schmidt number ($\Ri \to -\infty$)      \\[2ex] 
\end{tabular}
\caption{A summary of parameters and non-dimensional numbers derived from CATKE's calibrated free parameters.
\label{derived-parameteres}
}
\end{table}

\subsection{Validation against constant-forcing LES and comparison with other parameterizations}

In this section, we validate CATKE's ability to make predictions both for within and outside the range of surface forcings used for calibration.
To add context to this validation exercise and connect with other studies, we include a comparison with predictions from the $K$-profile parameterization \cite<KPP; >{large1994oceanic}, and the ``Langmuir turbulence'' second-moment closure (SMC-LT) described by \citeauthor{harcourt2015improved}~\citeyear{harcourt2015improved}, whose results depend additionally on the Stokes drift profile we used for LES.
All simulations, including those with KPP and SMC-LT, use staggered vertical grids with 128 points, in a 256-meter deep domain and thus with 2-meter vertical resolution.
We use a 5-minute time step for CATKE, a 2-minute time step for KPP, and a 1-second time-step for SMC-LT.
Such a short time-step was used for SMC-LT because we observed that the results were sensitive to time steps 20 seconds and longer for the strong forcing cases.

We should treat these comparisons with some caution, because KPP or SMC-LT were calibrated to somewhat different datasets than what we use for CATKE.
That said, we find that for every constant-forcing case, CATKE predicts the boundary layer depth simulated by LES --- both inside and outside the training dataset --- more accurately than either KPP or SMC-LT.
This is an important result because boundary layer is a key metric determining the short-term sensitivity of climate predictions \cite{gregory2000vertical, held2010probing}.
Moreover and by design of the calibration problem (because we omit the upper 4 meters of the LES profiles from the error estimate), CATKE predicts more well-mixed near-surface profiles during convection, and thus warmer sea surface temperatures, than either KPP or SMC-LT.
With this broad summary of CATKE's main successes stated, we focus the subsequent discussion for each case on CATKE's biases and areas to focus on for future improvements.

\subsubsection{Constant forcing validation: free convection}

\begin{figure}
    \includegraphics[width = 1\textwidth]{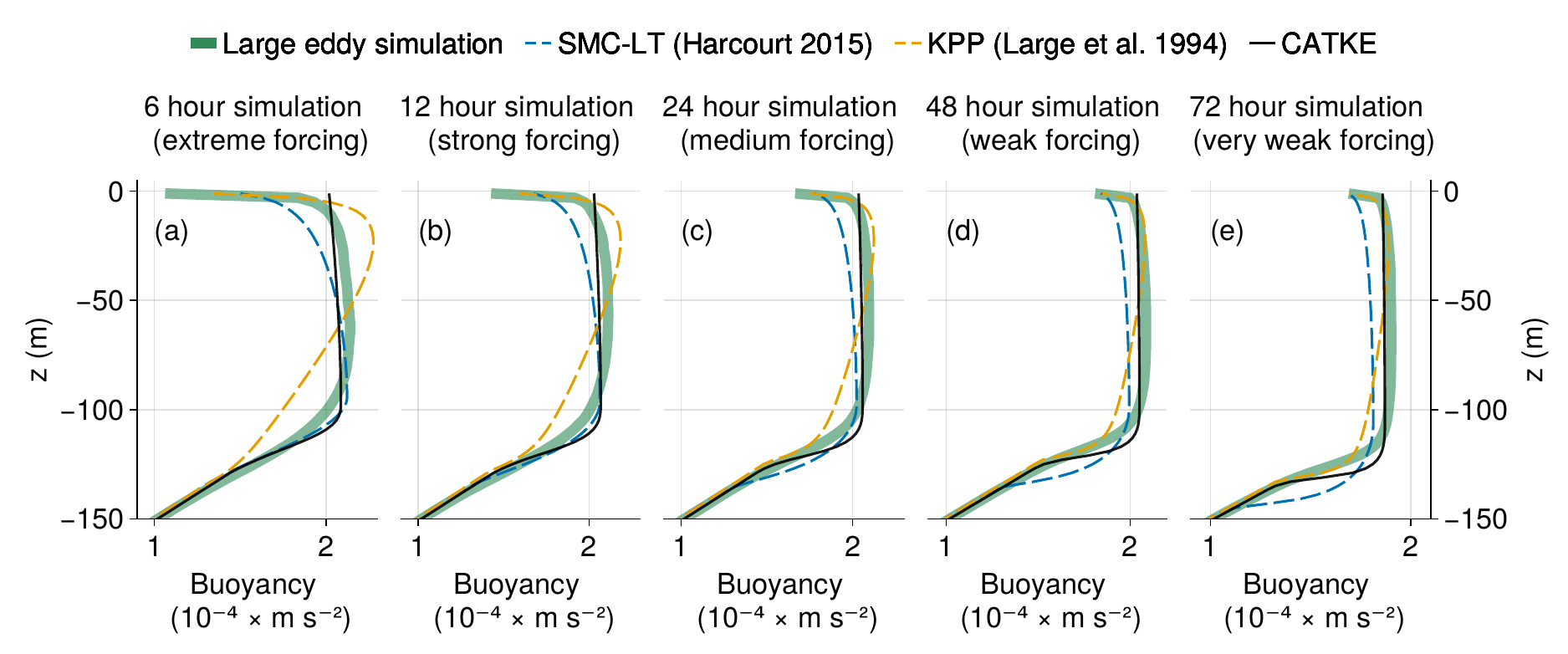}
    \caption{A four-way comparison for the ``free convection'' constant forcing cases described in~\ref{table:les-summary} and~\ref{les-description} between LES, CATKE, the $K$-profile parameterization \cite<KPP>{large1994oceanic}, and the Langmuir turbulence second moment closure described by \citeA{harcourt2015improved} (SMC-LT). KPP and SMC-LT are implemented in the General Ocean Turbulence Model \cite<GOTM,>{umlauf2005second}. Panels (a)--(e) show free convection for forcing of decreasing strength, corresponding to the 6-, 12-, 24-, 48-, and 72-hour suites, respectively. The free convection cases have no wind forcing and destabilizing buoyancy fluxes that correspond, roughly, to heat fluxes between 181 and $2000 \, \mathrm{W \, m^{-2}}$. The initial condition is density stratified with a depth-varying buoyancy gradient that varies between $10^{-6} \, \mathrm{s^{-2}}$ and $2 \times 10^{-5} \, \mathrm{s^{-2}}$.
    \label{free-convection-validation}
    }
%\end{figure}
%\begin{figure}
\includegraphics[width = 1\textwidth]{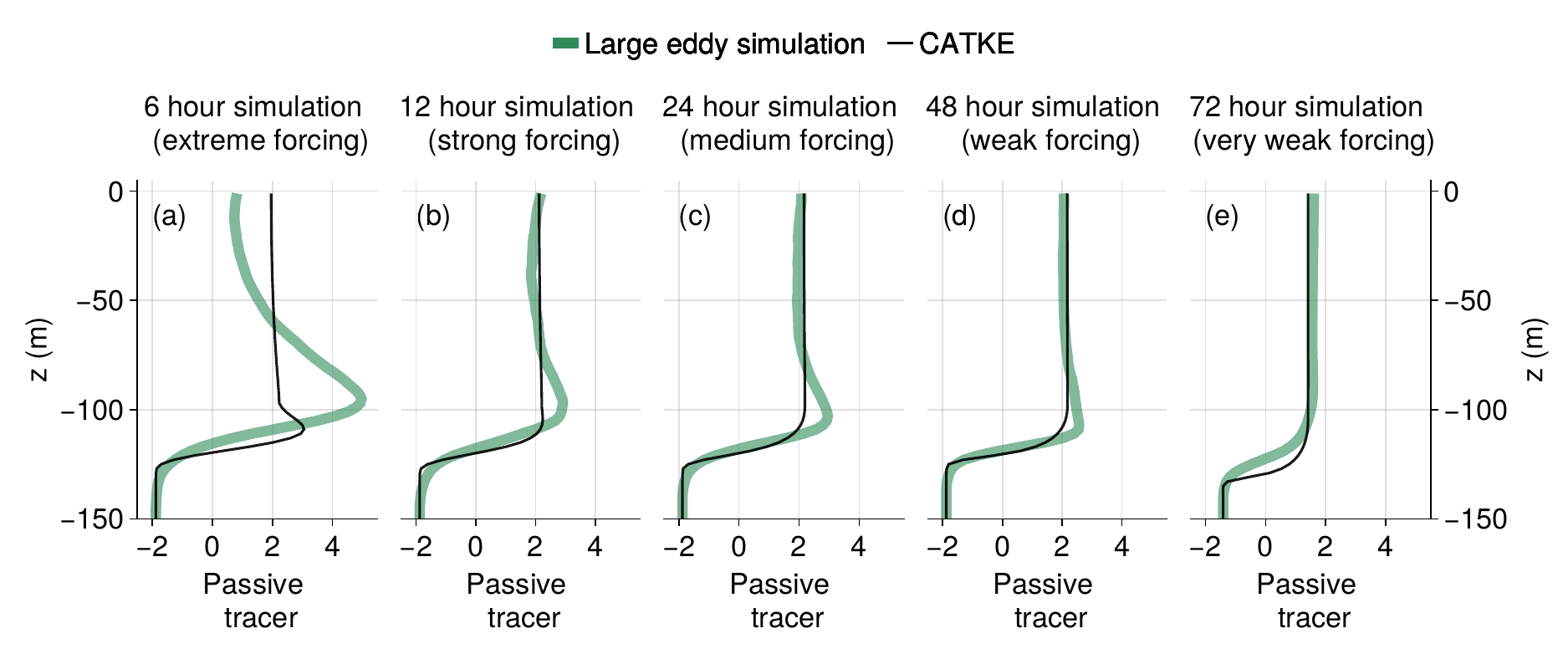}
    \caption{Comparison between the forced passive tracer profile simulated by LES and CATKE for free convection. The passive tracer forcing, which is described in appendix~\ref{sec:passive-tracer-forcing}, is a Gaussian centered on $z=-96$ m and $8$ m wide. The strength of the forcing depends on the suite: the 6-, 12-, 24-, 48-, and 72-hour suites use 15 minute, 30 minute, 1 hour, 2 hour, and 4 hour forcing time scales, respectively.
    \label{tracer-free-convection-validation}}
\end{figure}

We begin with the free convection cases plotted in figure~\ref{free-convection-validation}.
The free convection cases represent some of the best predictions of KPP and SMC-LT.
Boundary layer depth is well-predicted by all parameterizations to within 10 meters, with perhaps the greatest bias coming from SMC-LT in the weakly-forced 72-hour case --- despite that KPP has known structural biases for representing free convection \cite{souza2020uncertainty}. 
Oddly, for the more strongly forced cases, a large portion of the KPP profiles are stably stratified within the boundary layer, and capped by a very strong unstable stratification near the surface.
Of the three, CATKE's convective mixing length most capably keeps the boundary layer nearly-neutrally stratified during strong free convection.

For near-surface buoyancy (and equivalently sea surface temperature, or SST) the three parameterizations make somewhat different predictions.
For example, CATKE predicts a nearly-mixed boundary layer due to its convective mixing length, which means that it predicts a warmer SST.
On the other hand KPP, SMC-LT, and the LES all predict layers (of varying thickness) of unstable stratification next to the surface, and thereby also predict substantially colder SST than CATKE.
Caution is probably warranted when interpreting the LES results, however: our LES may exhibit spuriously reduced mixing near the upper boundary where the simulated scale of turbulent eddies shrinks significantly below the grid scale.
Addressing this uncertainty in the LES data requires the use of observations of the near-surface temperature profiles to inform modifications to the LES, which is left for future work.

The buoyancy profiles in figure~\ref{free-convection-validation} reveal bias in CATKE's predictions of the detailed structure of the lower half of the convecting boundary layer.
One contribution to this bias is well-known: in free convection, buoyancy fluxes in the lower half of the boundary layer are upgradient.
In order to accurately capture the boundary layer depth, CATKE must accurately predict the buoyancy flux --- and therefore cannot avoid erroneously predicting a slightly unstably stratified buoyancy profile where in the LES the profile is either nearly mixed or actually slightly stably stratified.
No amount of calibration or additional free parameters can fix this bias given CATKE's downgradient formulation --- the only recourse is to introduce a non-downgradient, and therefore non-local, contribution to CATKE's fluxes.
For example, CATKE could be augmented with a mass flux scheme in the manner of \citeA{siebesma2007combined, giordani2020eddy}.
Other alternatives include evolving fluxes directly as in \citeauthor{garanaik2024new}~\citeyear{garanaik2024new}, or adding additional tracer variance equations and computing non-gradient fluxes in terms of those \cite{legay2024derivation}.
But even this may not be sufficient --- for example, even though KPP has a non-local model for fluxes, it still has significant biases in convective boundary layer buoyancy structure.

To investigate CATKE's free convection bias further, figure~\ref{tracer-free-convection-validation} compares CATKE's predictions of the forced passive tracer profile with LES.
This comparison reveals that while CATKE generally models the tracer profile well (except for the extreme, extrapolating, 6-hour case in panel a), CATKE tends to overmix especially in the lower part of the boundary layer, where the LES profiles exhibit a slight peak and a bit more shape.
Thus in addition to lacking a non-local contribution to fluxes, CATKE also overpredicts mixing to some degree, especially near the base of the boundary layer.
Solving this bias could simultaneously motivate adding non-local contributions to convective fluxes as well as modifying the depth structure of the convective mixing length.

\subsubsection{Constant forcing validation: shear-driven turbulence}

\begin{figure}
    \includegraphics[width = 1\textwidth]{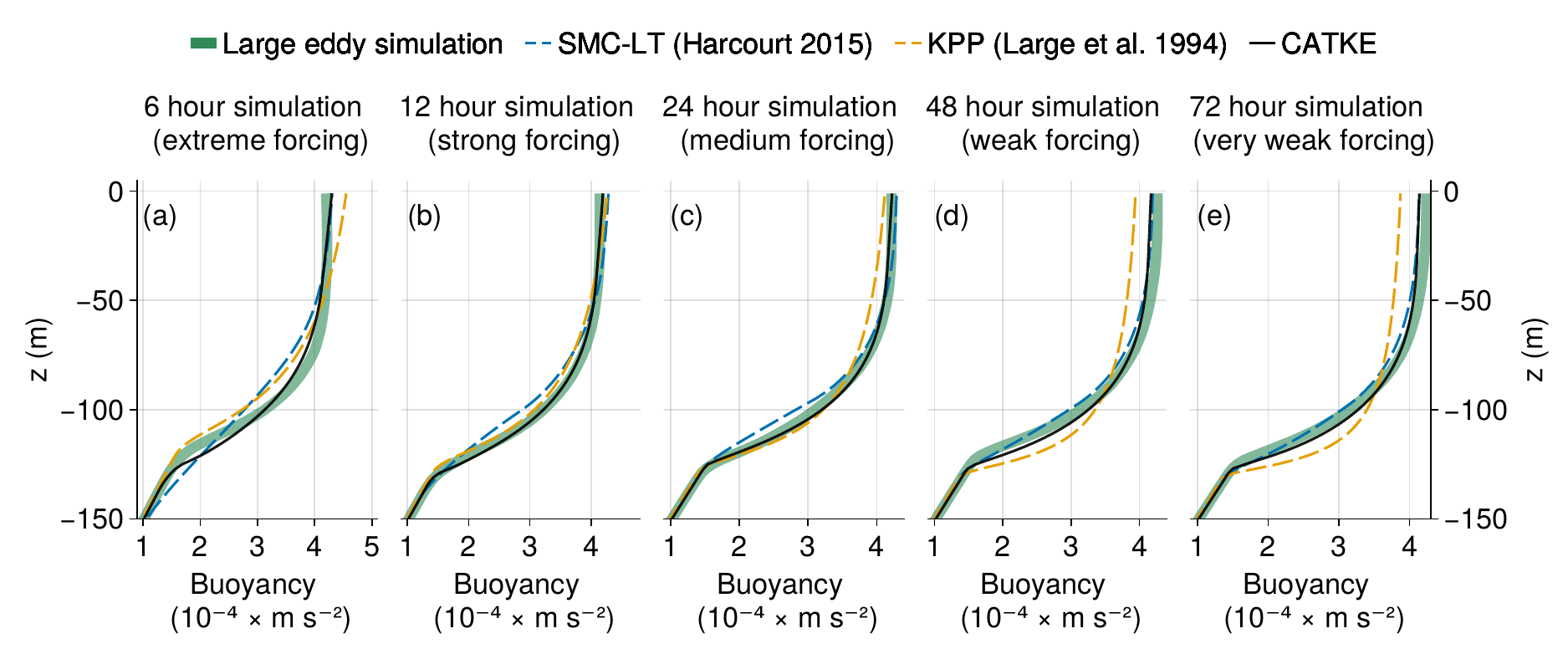}
    \caption{A four-way comparison between LES and three turbulence closures (CATKE, KPP, and SMC-LT) for the ``strong wind, no rotation'' constant forcing cases described in table~\ref{table:les-summary} and~\ref{les-description}. The strong wind, no rotation cases are forced by surface stresses that correspond roughly to 9--22 $\mathrm{m \, s^{-1}}$ atmospheric wind at a height of 10 $\mathrm{m}$. See figure~\ref{free-convection-validation}.
    \label{strong-wind-no-rotation-validation}
    }
% \end{figure}
% \begin{figure}[!h]
    \includegraphics[width = 1\textwidth]{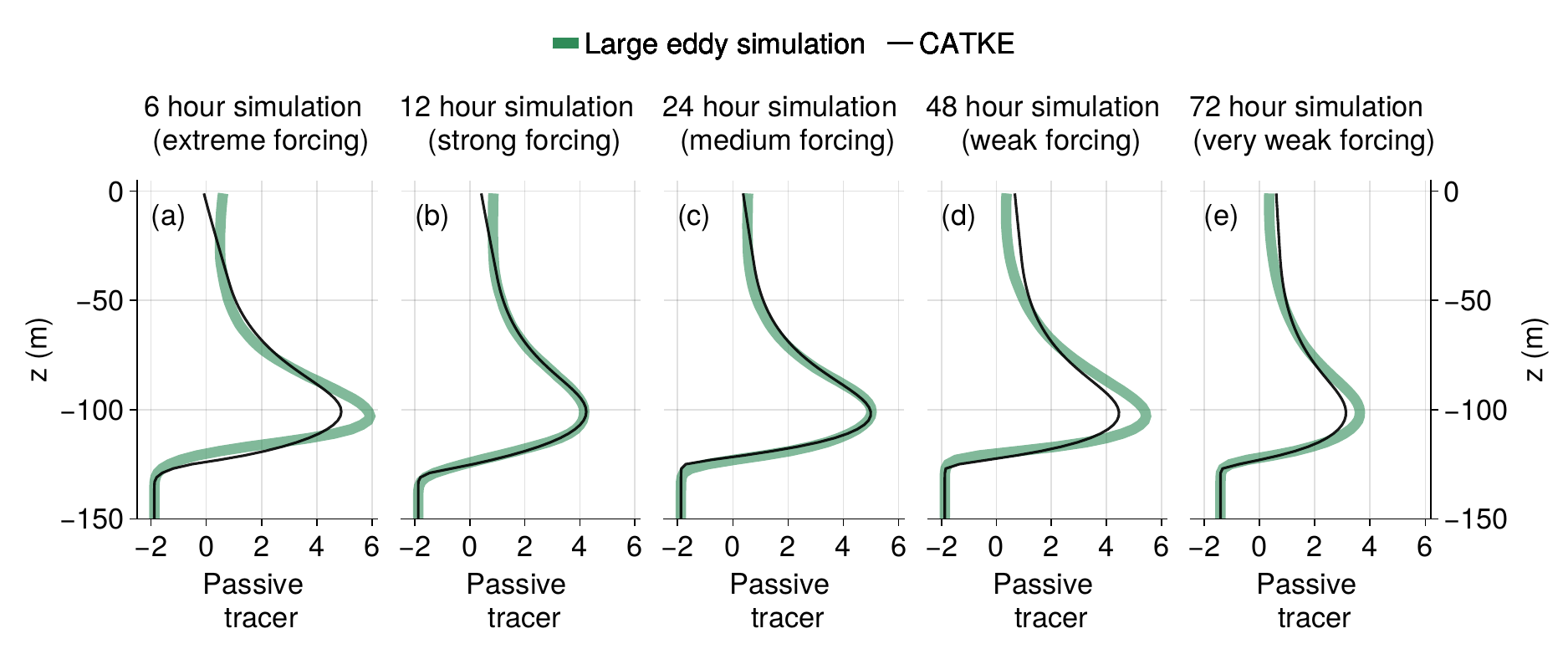}
    \caption{Comparison between the forced passive tracer profile simulated by LES and CATKE for strong wind, no rotation. See figure~\ref{tracer-free-convection-validation}.
    \label{tracer-strong-wind-no-rotation-validation}
    }
\end{figure}

We next turn to pure shear- or wind-driven turbulence.
We have two such cases, one without rotation and thus representing near-equatorial mixing, and a second with a Coriolis parameter of $f = 10^{-4} \, \mathrm{s^{-1}}$ corresponding to a latitude of about $43^\circ$.
The wind forcing that would produce the momentum flux applied to the strong wind, no rotation cases spans from 9--22 $\mathrm{m \, s^{-1}}$.
The wind forcing in the strong wind (and rotating) cases spans 15--24 $\mathrm{m \, s^{-1}}$.
%These cases therefore represent relatively high wind situations, but do not extend all the way to the strongest winds observed over the ocean (cite).
%\textcolor{red}{So there is future work to conduct additional validation for cases forced by even stronger winds...}

A comparison between LES, SMC-LT, KPP, and CATKE for the strong wind, no rotation case is shown in figure~\ref{strong-wind-no-rotation-validation}.
All parameterizations make similar and good predictions for boundary layer depth and surface temperature, except for SMC-LT in the 6-hour case, where it overmixes slightly.
A comparison between CATKE and LES simulations of the forced passive tracer for the strong wind, no rotation case is shown in figure~\ref{tracer-strong-wind-no-rotation-validation}, revealing that CATKE fares far better for this case than for free convection, and more specifically exhibits a slight tendency to overmix near the base of the boundary layer and to undermix near the surface.

\begin{figure}
    \includegraphics[width = 1\textwidth]{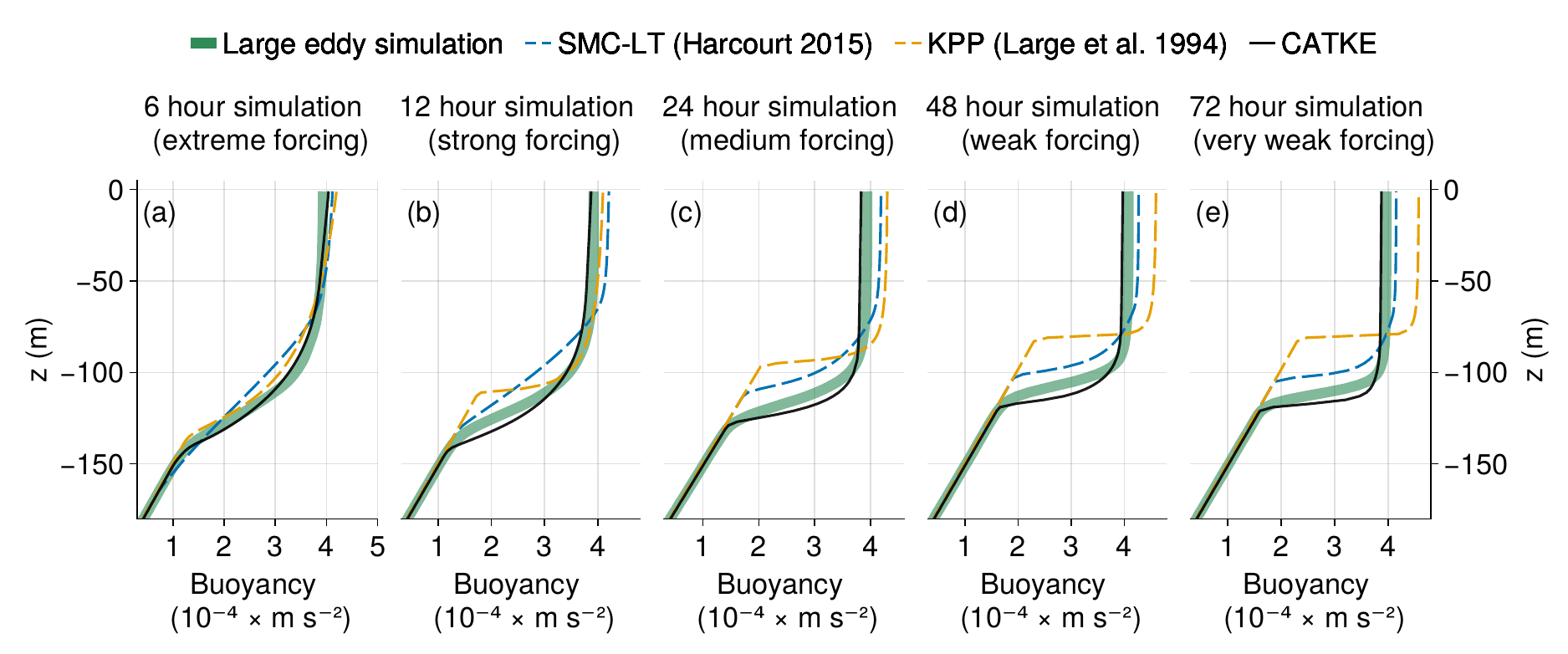}
    \caption{A four-way comparison between LES and three turbulence closures (CATKE, KPP, and SMC-LT) for the strong wind constant forcing cases described in table~\ref{table:les-summary} and~\ref{les-description}. The strong wind cases are rotating with Coriolis parameter $f = 10^{-4} \, \mathrm{s^{-1}}$ and forced by surface stresses that correspond roughly to 15--24 $\mathrm{m \, s^{-1}}$ atmospheric wind at 10 meters height. See figure~\ref{free-convection-validation}.
    \label{strong-wind-validation}
    }
% \end{figure}
% \begin{figure}[!h]
    \includegraphics[width = 1\textwidth]{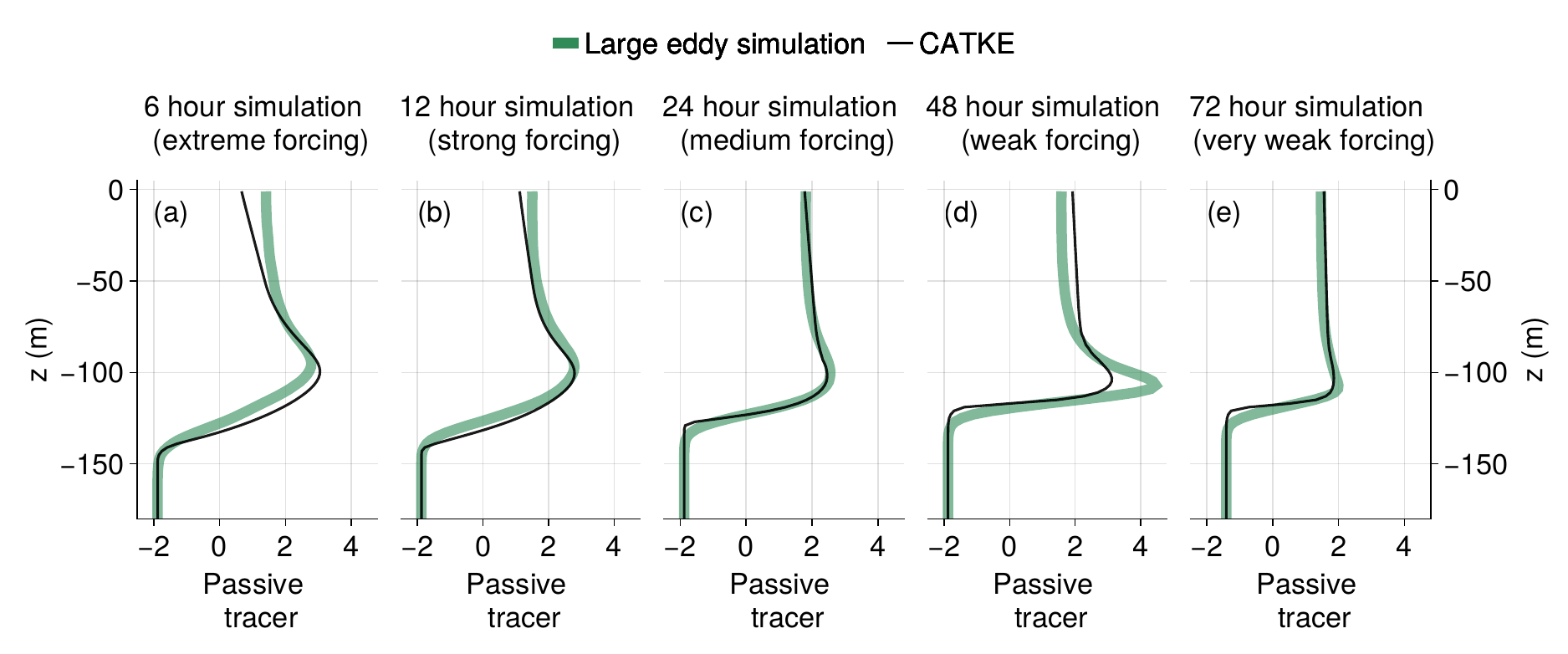}
    \caption{Comparison between the forced passive tracer profile simulated by LES and CATKE for strong wind. See figure~\ref{tracer-free-convection-validation}
    \label{tracer-strong-wind-validation}
    }
\end{figure}

The strong wind case with rotation plotted in figure~\ref{strong-wind-validation} proves more challenging for CATKE and extremely challenging for SMC-LT and KPP.
For all forcing strength, SMC-LT and KPP exhibit serious shallow bias and warm SST bias.
CATKE simulations, on the other hand, are good but exhibit a tendency to overmix slightly, resulting in boundary layers that are approximately 5\% too deep.
Figure~\ref{tracer-strong-wind-validation} compares CATKE and LES predictions of the forced passive tracer for the strong wind case, corroborating the ``overmixing bias'' especially for the 6- and 48-hour suites, while additionally revealing undermixing near the surface.
%In particular, errors in the precise structure of the entrainment layer at the bottom of the boundary layer simulated by CATKE can be seen in the more strongly forced 6- and 12-hour cases.
%Bias in rotating situations could motivate adding a cross-gradient contribution to the eddy diffusivity model in~\eqref{eddy-diffusivity} along the lines suggested by Svensson (1979, equations 16-17 expanded for weak rotation),
%\beq
%\overline{\bu' w'} = - \ell_u \sqrt{e} \, \d_z \bu - \C{f}{} \ell_u^2 f \bzh \times \d_z \bu \com
%\eeq
%where $\C{f}{}$ would be an additional free parameter.
%\nocite{svensson1979structure}

\subsubsection{Constant forcing validation: mixed shear and convective turbulence}

\begin{figure}
    \includegraphics[width = 1\textwidth]{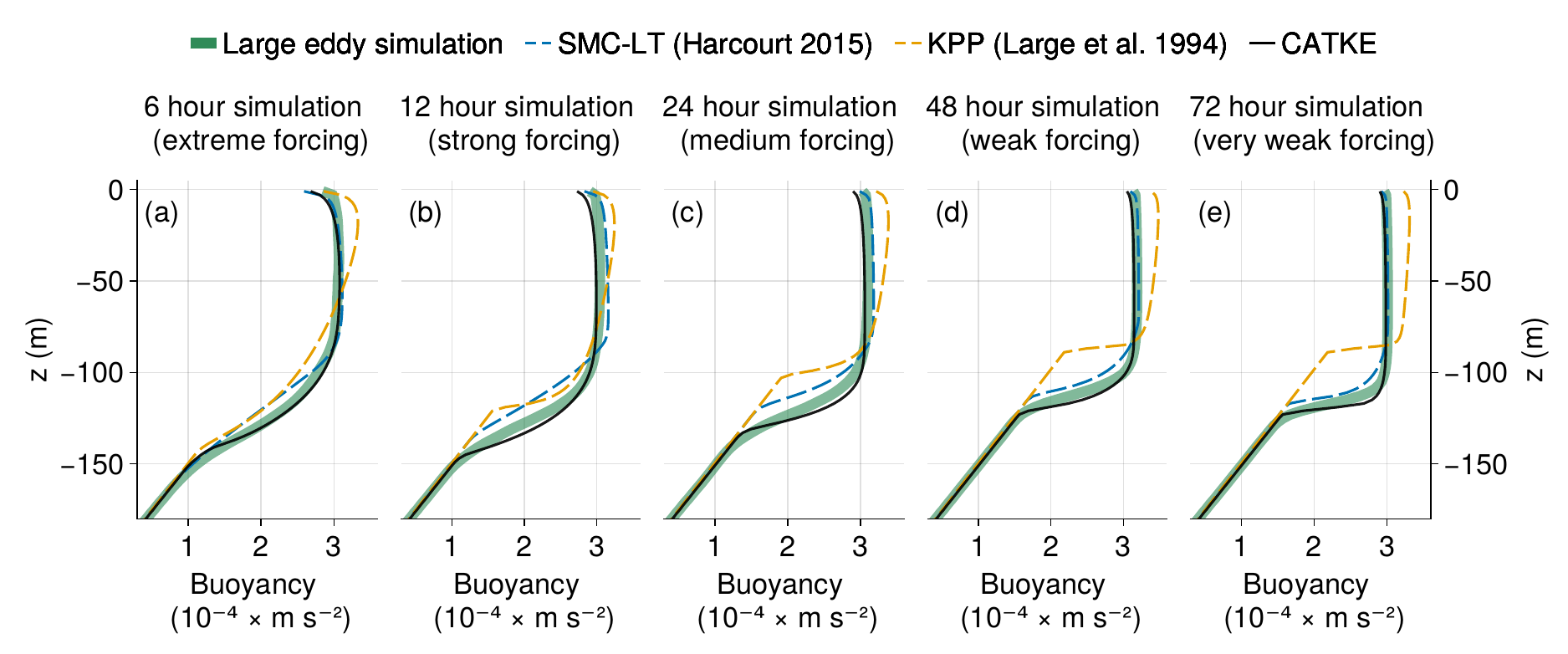}
    \caption{A four-way comparison between LES and three turbulence closures (CATKE, KPP, and SMC-LT) for the ``strong wind, weak cooling'' constant forcing cases described in table~\ref{table:les-summary} and~\ref{les-description}. The strong wind weak cooling cases are rotating with Coriolis parameter $f = 10^{-4} \, \mathrm{s^{-1}}$, forced by surface stresses that correspond roughly to 14--23~$\mathrm{m \, s^{-1}}$ atmospheric wind at 10 meters height, and destabilizing buoyancy fluxes that correspond roughly to heat fluxes between \mbox{79--833}~$\mathrm{W \, m^{-2}}$. See figure~\ref{free-convection-validation}.}
    \label{strong-wind-weak-cooling-validation}
% \end{figure}
% \begin{figure}[!h]
    \includegraphics[width = 1\textwidth]{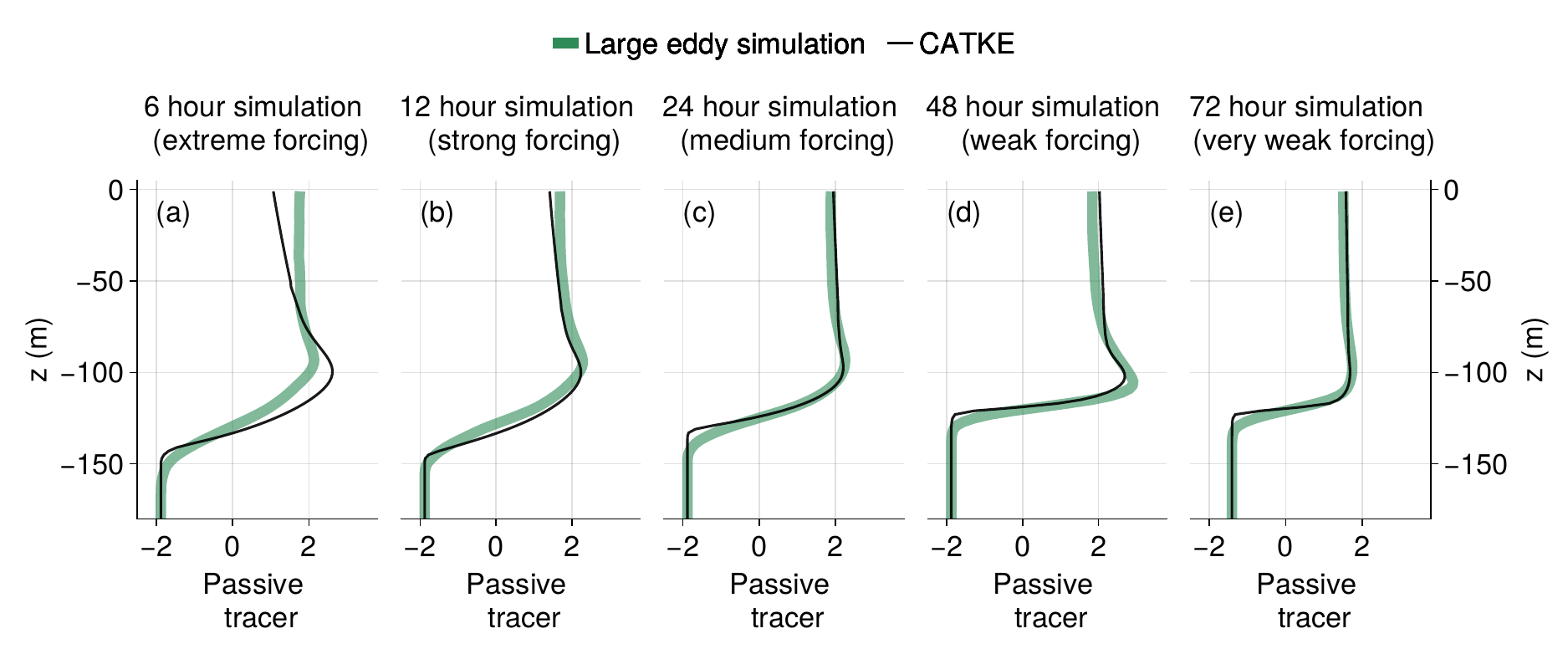}
    \caption{Comparison between the forced passive tracer profile simulated by LES and CATKE for strong wind, weak cooling. See figure~\ref{tracer-free-convection-validation}.}
    \label{tracer-strong-wind-weak-cooling-validation}
\end{figure}

\begin{figure}
    \includegraphics[width = 1\textwidth]{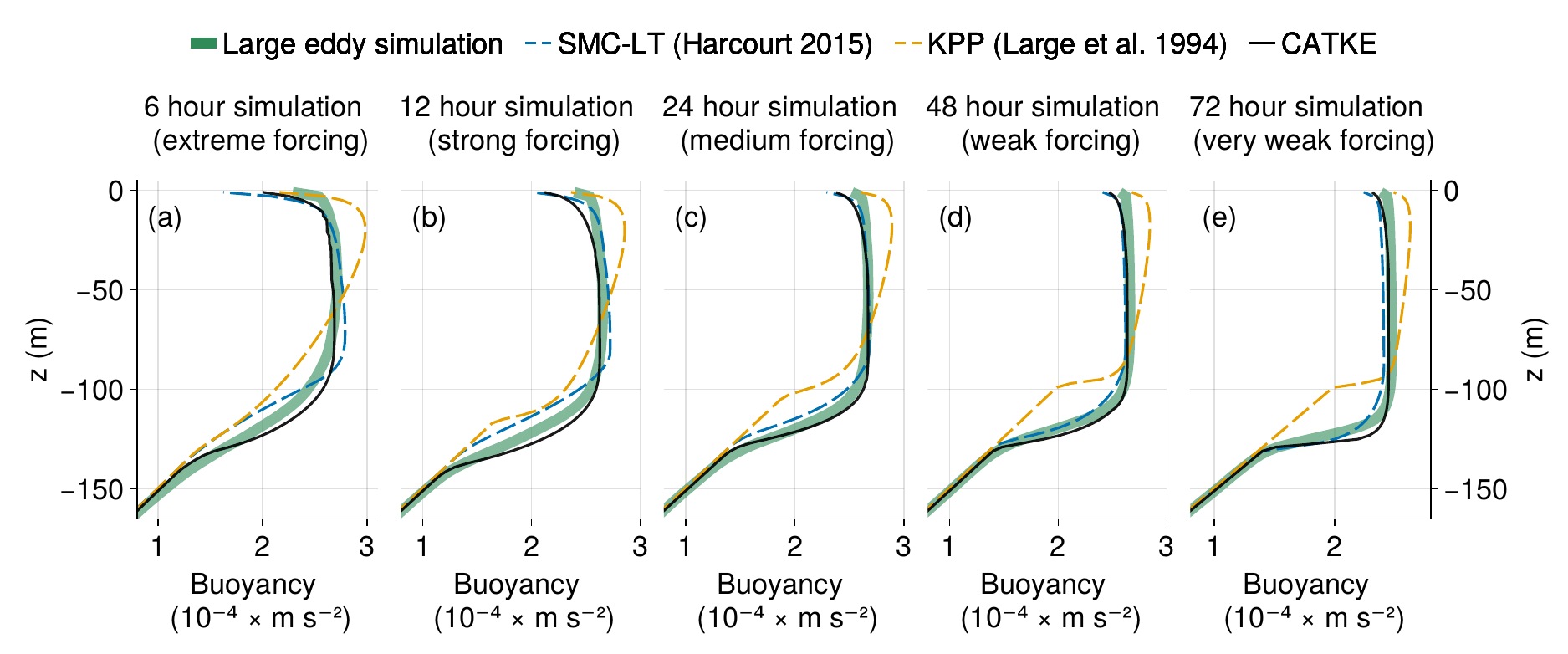}
    \caption{A four-way comparison between LES and three turbulence closures (CATKE, KPP, and SMC-LT) for the ``mid wind, mid cooling'' constant forcing cases described in table~\ref{table:les-summary} and~\ref{les-description}. The mid wind mid cooling cases are rotating with Coriolis parameter $f = 10^{-4} \, \mathrm{s^{-1}}$, forced by surface stresses that correspond roughly to 13--20 $\mathrm{m \, s^{-1}}$ atmospheric wind at 10 meters height, and destabilizing buoyancy fluxes that correspond roughly to heat fluxes between 125--1333 $\mathrm{W \, m^{-2}}$. See figure~\ref{free-convection-validation}.
    \label{med-wind-med-cooling-validation}
    }
    
% \end{figure}
% \begin{figure}
    \includegraphics[width = 1\textwidth]{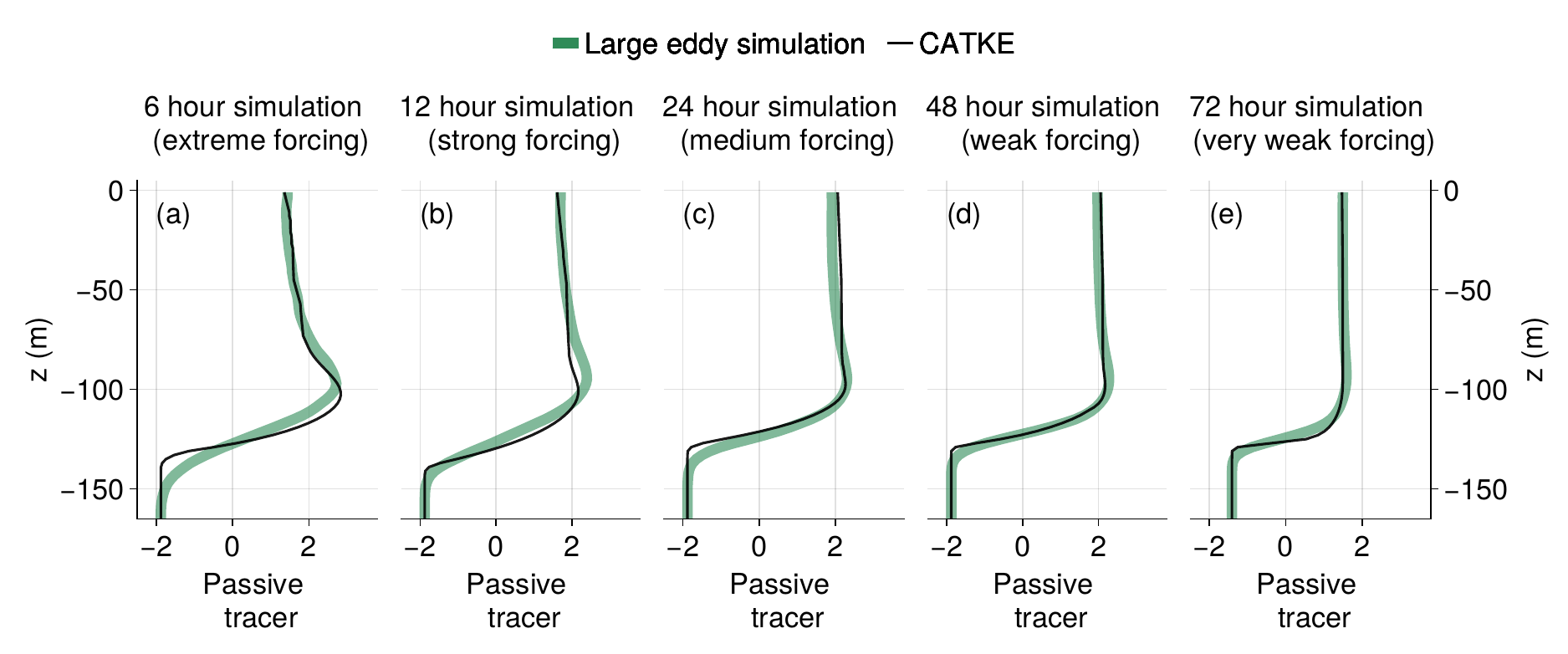}
    \caption{Comparison between the forced passive tracer profile simulated by LES and CATKE for mid wind, mid cooling. See figure~\ref{tracer-free-convection-validation}.
    \label{tracer-med-wind-med-cooling-validation}}
\end{figure}

\begin{figure}
    \includegraphics[width = 1\textwidth]{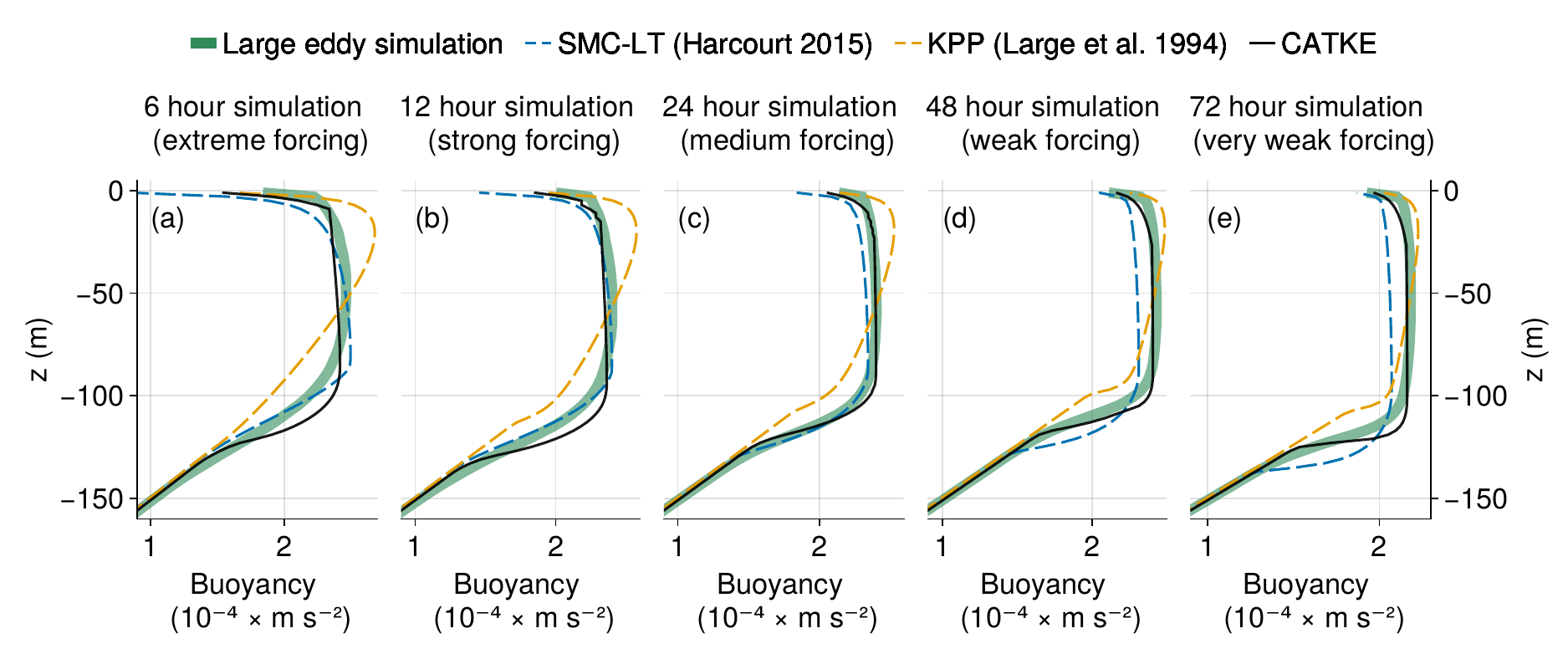}
    \caption{A four-way comparison between LES and three turbulence closures (CATKE, KPP, and SMC-LT) for the ``weak wind, strong cooling'' constant forcing cases described in table~\ref{table:les-summary} and~\ref{les-description}. The weak wind strong cooling cases are rotating with Coriolis parameter $f = 10^{-4} \, \mathrm{s^{-1}}$, forced by surface stresses that correspond roughly to 11--16 $\mathrm{m \, s^{-1}}$ atmospheric wind at 10 meters height, and destabilizing buoyancy fluxes that correspond roughly to heat fluxes between 156--1666 $\mathrm{W \, m^{-2}}$. See figure~\ref{free-convection-validation}.
    \label{weak-wind-strong-cooling-validation}}
% \end{figure}
% \begin{figure}
    \includegraphics[width = 1\textwidth]{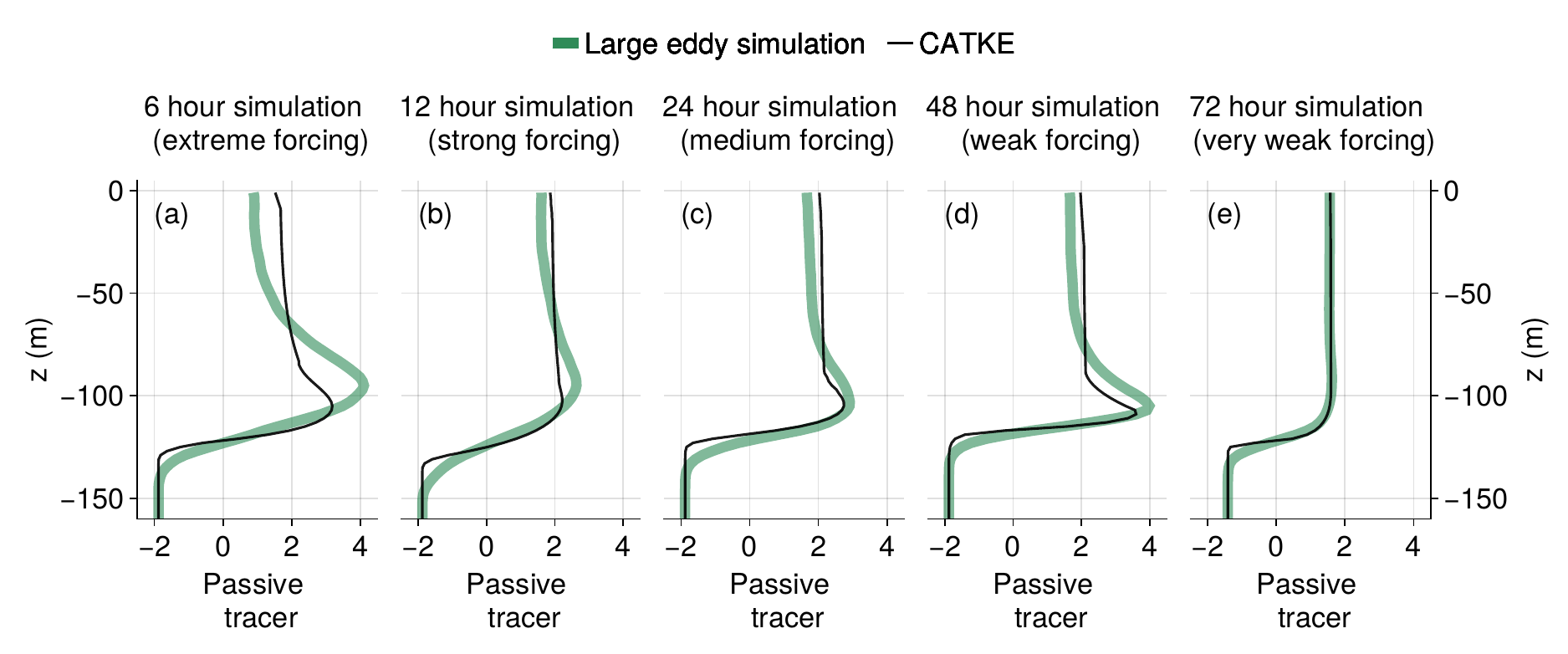}
    \caption{Comparison between the forced passive tracer profile simulated by LES and CATKE for weak wind, strong cooling. See figure~\ref{tracer-free-convection-validation}. See figure~\ref{tracer-free-convection-validation}.
    \label{tracer-weak-wind-strong-cooling-validation}}
\end{figure}

CATKE simulations are also accurate for cases involving both wind and destabilizing buoyancy forcing, which produces a mixed regime of turbulence with both shear and buoyant production of TKE.
We have three mixed cases comprising a total of 15 LES with both wind and buoyancy forcing: strong wind, weak cooling, medium wind, weak cooling, and weak wind, strong cooling.
Results for these 15 cases are shown in figures~\ref{strong-wind-weak-cooling-validation},~\ref{med-wind-med-cooling-validation}, and~\ref{weak-wind-strong-cooling-validation}.
KPP exhibits significant shallow bias for all cases.
SMC-LT exhibits less shallow bias than KPP, but still more than CATKE.
Because KPP and SMC-LT also predict a spuriously strong unstable buoyancy gradient near the surface (compared to the present LES), the SST biases are more variable.
CATKE, on the other hand, makes good predictions for all cases except in the weak wind, strong cooling cases where it overmixes.

Figures~\ref{tracer-strong-wind-weak-cooling-validation},~\ref{tracer-med-wind-med-cooling-validation}, and~\ref{tracer-weak-wind-strong-cooling-validation} compare CATKE and LES predictions of the forced passive tracer for strong wind, weak cooling, mid wind mid cooling, and weak wind weak cooling cases.
The most bias is exhibited in the weak wind strong cooling case, where it tends to overmix as exhibits in both the boundary layer depth in figure~\ref{strong-wind-weak-cooling-validation} and the tracer profiles in figure~\ref{tracer-strong-wind-weak-cooling-validation}.
This shows that the most difficult cases are free convection and ``weak wind, strong cooling'' --- the cases where convective dynamics dominate.

The ``weak winds, strong cooling'' case is the most challenging for CATKE.
For this case, the 72-hour LES is forced by 156 $\mathrm{W \, m^{-2}}$ equivalent heat flux and 11 $\mathrm{m \, s^{-1}}$ 10-meter atmospheric winds, while the 6-hour LES is forced by 1666 $\mathrm{W \, m^{-s}}$ and 16 $\mathrm{m \ s^{-1}}$ 10-meter winds.
In the 6- and 12-hour cases, KPP exhibits a similar ``stable stratification bias'' as seen in free convection in figure~\ref{free-convection-validation}.
SMC-LT exhibits a shallow bias for the strongly forced cases and a deep biased for the weakly forced cases (and quite accurate predictions for the 24-hour case).
%Arguably, CATKE exhibits it's greatest boundary layer depth bias of the entire constant-forcing dataset in the 6-hour weak winds, strong cooling case.
CATKE also predicts a too-sharp entrainment layer that is much thinner than the broad entrainment layer observed in the LES in the 6- and 12-hour weak winds, strong cooling cases.
These simulations are farthest from quasi-equilibrium in time and may exhibit strong non-locality.
Despite CATKE's errors for the 6-hour case, however, CATKE's boundary layer depth predictions for the 24-, 48-, and 72-hour case are accurate.
%Fixing CATKE's biases for these cases could motivate generalizing CATKE's stability function for $\Ri < 0$ from the present piecewise constant formulation to, for example, a piecewise linear formulation.
%However, because this embellishment would add at least 6 new free parameters, we do not pursue it here.

\subsubsection{Constant forcing validation: summary}

CATKE exhibits less bias than either KPP or SMC-LT across all cases, even when making predictions ``outside'' its training dataset.
In particular, CATKE generates good predictions of boundary layer depth, even in convective dominated cases where an analysis of tracer profiles suggests that CATKE tends to overmix.
%However, the forced passive tracer results especially reveal significant biases in the predicted mixing rate during free convection.
Fixing CATKE's convecitve biases will likely require additional work with both the convective mixing length, and CATKE's stability function formulation for $\Ri < 0$.

CATKE makes good predictions relative to KPP or SMC-LT in part because its formulation expresses reasonable physical hypotheses, but also because its parameters have been calibrated comprehensively to minimize bias across a wide range of physical scenarios and vertical resolutions.
In particular, the simulations that CATKE has been trained on are more similar to the extrapolation test cases (the 6- and 72-hour cases) than the datasets that either KPP or SMC-LT have been trained on.
This generates an ambiguity in comparing the three: do KPP and SMC-LT exhibit greater bias because of structural issues with their formulation, or do they need to be recalibrated in a similar manner as CATKE?
We cannot answer this question conclusively.
While KPP has known structural biases \cite<see, for example,>{souza2020uncertainty}, the formulation of SMC-LT is seemingly more general than CATKE.
We therefore suspect that \textit{a posteriori} calibration of SMC-LT will allow it to make predictions that are as or more accurate than CATKE.
And until this calibration is performed, any judgments about the biases of SMC-LT must be taken with a grain of salt.

\subsection{Deep cycle turbulence in the tropics}
\label{sec:tropical-turbulence}

We next turn to a validation case that requires significant extrapolation outside of the constant-forcing dataset: 34 days of deep cycle turbulence in the tropics forced by time-varying winds, surface heat fluxes, and surface freshwater fluxes, as well as lateral flux divergences derived from a regional ocean model.
The scenario and LES that we use to validate the single column model simulations are described by \citeA{whitt2022simulation}.

\begin{figure}
    \includegraphics[width = 1\textwidth]{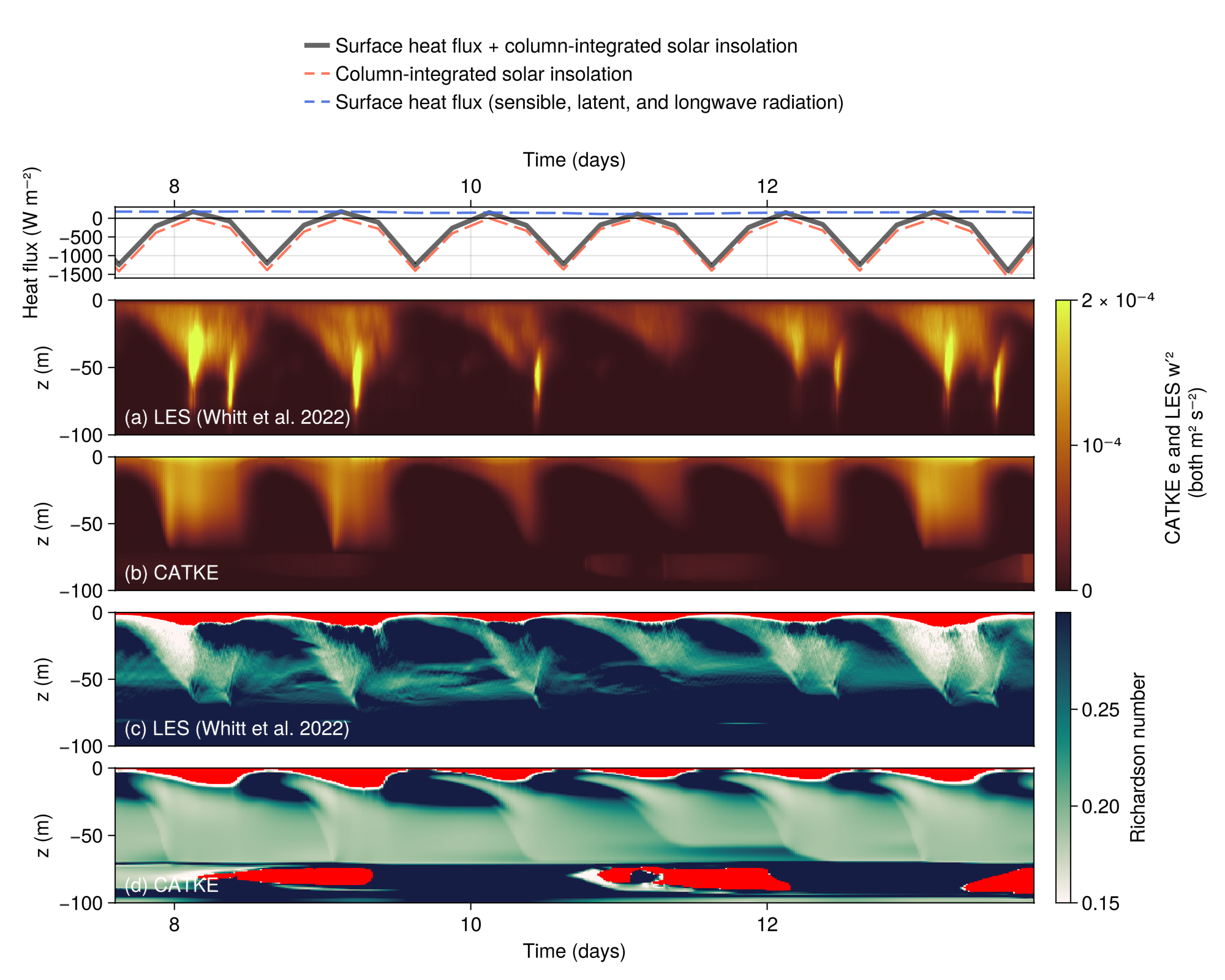}
    \caption{Overview of the tropical turbulence validation case. Panels show: (a)~the forcing and heat fluxes, (b)~vertical kinetic energy $\overline{w'^2}$ from the LES described by \citeA{whitt2022simulation}, (c)~CATKE's TKE variable, (d)~the Richardson number computed from the horizontally-averaged LES momentum and buoyancy profiles, and (e)~the Richardson number predicted by CATKE. The shaded red areas in panels~(d) and~(e) indicate a negative Richardson number. Shown here are days 8--13 out of the entire 34-day time-series. The heat fluxes are negative during the day (heat going downwards, into the ocean) and positive at night (heat going up, out of the ocean). The LES vertical kinetic energy and CATKE turbulent kinetic energy exhibit intermittent bursting. In the deep region below the boundary layer where turbulent bursting occurs, LES-derived Richardson numbers get as low as 0.15. In the CATKE solution and in the same region, the Richardson number reaches a minimum of about 0.18.}
    \label{tropical-turbulence-overview}
\end{figure}

Figure~\ref{tropical-turbulence-overview} illustrates the complex dynamics of this situation by showing vertical kinetic energy from the LES, TKE from CATKE, and $\Ri$ from days 8 to 13 of the time-series.
In this topical turbulence scenario, a combination of wind stress and stabilizing solar insolation in daytime produces a shallow, stably-stratified jet in the upper $\sim$10 meters of the water column.
As day turns to night, outgoing radiation starts to dominate the incoming solar insolation to reduce and eventually destabilize the upper part of the water column, producing turbulent mixing driven by a combination of convective buoyancy flux and shear.
Momentum is thereby mixed downwards and injected into the stably stratified region below the base of the boundary layer.
Remarkably, because the region below the boundary layer is close to marginally stable \cite{smyth2013marginal}, this nocturnal injection of momentum from the boundary layer eventually leads to shear instability which spans the entire, roughly 100 m depth of the region below the mixed layer.
More often then not, the turbulence ``pulsates'' --- initial bursts of turbulence mix momentum and buoyancy and thus decay rather quickly, only to precipitate a second, and even a third burst of turbulence later on the evening \cite{smyth2017pulsating}.
The process, which is called ``deep cycle turbulence'', repeats itself the next day.

\begin{figure}
    \includegraphics[width = 1\textwidth]{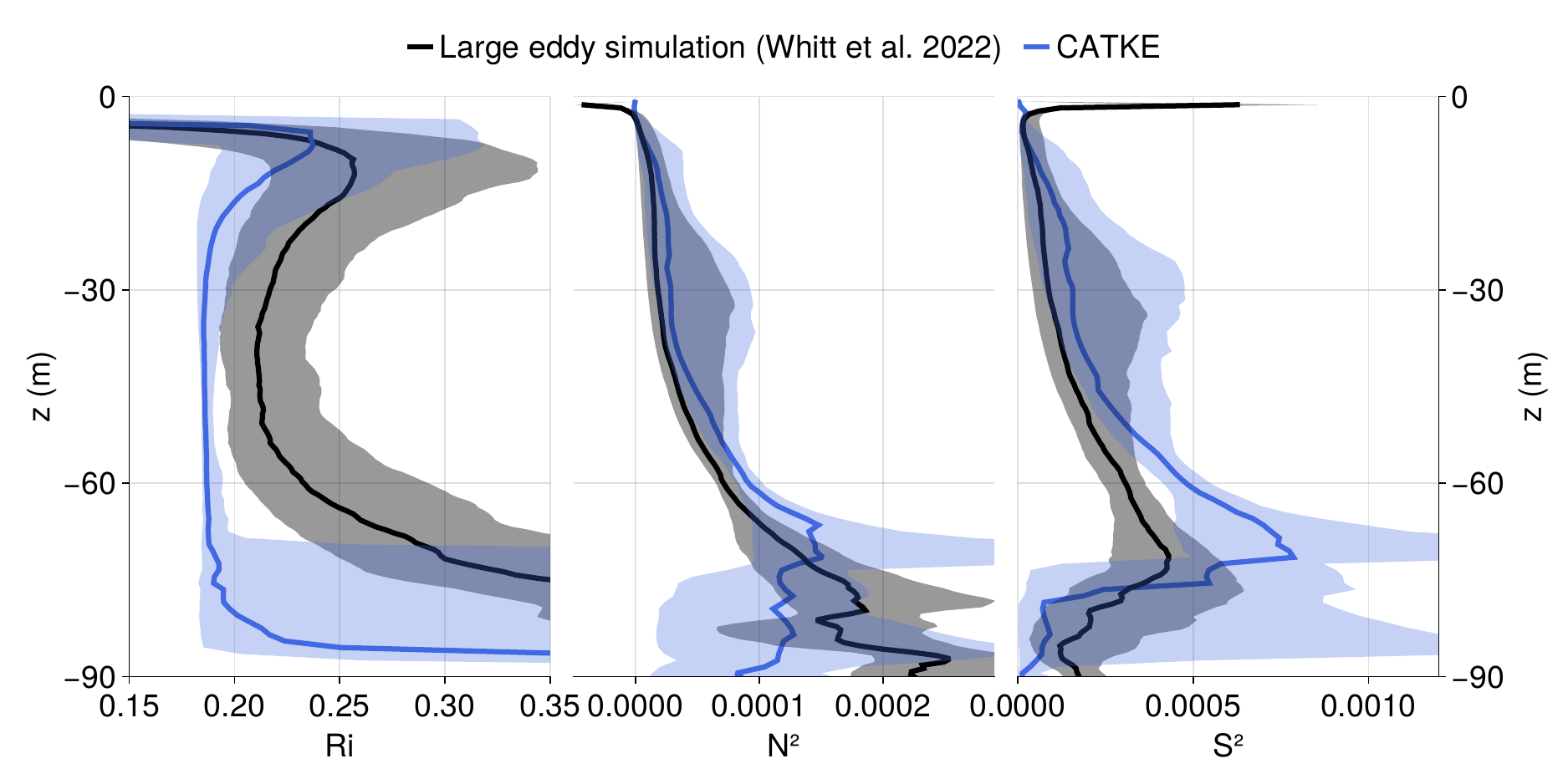}
    \caption{Median $\Ri$, shear ($S^2$), and buoyancy frequency $N^2$ at each depth computed from 34 days of realistic equatorial turbulence simulated by LES and CATKE. The LES $\Ri$ is computed in terms of the horizontally-averaged shear and buoyancy. Shading shows the range between the 25\% and 75\% quantiles. CATKE's prediction of $\Ri$ is narrowly peaked around its steady-state Richardson number, $\Ri^\dagger = 0.18$. This reveals a bias in CATKE: the median $\Ri$ in the LES is more variable and in particular, does not reach values as low as 0.18. Turning to the buoyancy gradient and shear, it seems that CATKE overpredicts both. This reflects complexity: apparently CATKE undermixes both momentum and buoyancy, but exhibits \textit{more bias} for momentum mixing, which permits the development of lower $\Ri$ than observed in the LES.
    Given that CATKE has already been calibrated to cases that presumably exhibit similar stratified shear mixing physics as this tropical turbulence case, fixing the $\Ri$, $N^2$, and $S^2$ biases may require changing the formulation of CATKE's stability functions.}
    \label{tropical-turbulence-statistics}
\end{figure}

The slow growth and intermittent bursting of turbulence at night is prominent in LES vertical kinetic energy shown in figure~\ref{tropical-turbulence-overview}a.
Figure~\ref{tropical-turbulence-overview}b shows that CATKE exhibits a qualitatively similar bursting behavior, though the timing of the bursts are sometimes misrepresented.
Moreover, inspection of the Richardson number plotted in figures~\ref{tropical-turbulence-overview}c and d reveals that CATKE sometimes underpredicts, and sometimes overpredicts the Richardson number.
Figure~\ref{tropical-turbulence-statistics} investigates this further by plotting the median $\Ri$, $N^2$, and $S^2$ and shading the range of values between the 25\% and 75\% quantiles.
The $\Ri$ statistics in the left panel are striking: while the $\Ri$ in the LES is relatively variable with a broad peak around $\Ri \approx 0.21$, CATKE's $\Ri$ are narrowly concentrated around its steady state value $0.18$.
Turning to $N^2$ (middle panel) and $S^2$ (right panel), we see that the $\Ri$ bias is not straightforwardly associated with a bias in either $N^2$ or $S^2$ --- both are slightly overpredicted (indicating undermixing), but nevertheless exhibit similar medians and ranges compared to the LES.

\begin{figure}
    \includegraphics[width = 1\textwidth]{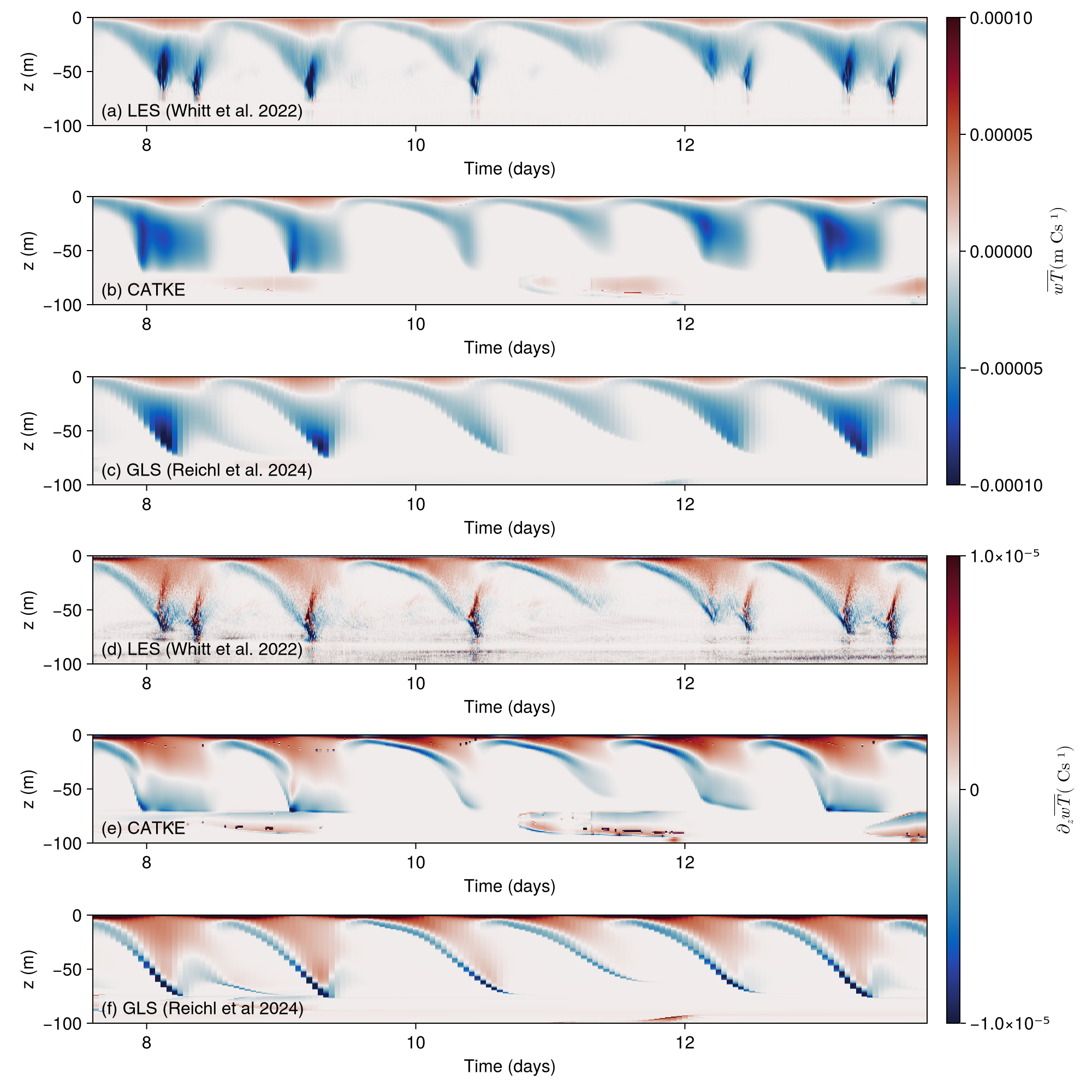}
    \caption{A comparison of the vertical temperature flux and vertical temperature flux divergence in tropical turbulence between LES \cite{whitt2022simulation}, CATKE, and the Generic Length Scale (GLS) turbulence closure as reported by \citeA{reichl2024improved}.}
    \label{tropical-turbulence-fluxes}
\end{figure}

Despite the errors in burst timing and Richardson number, we argue that CATKE's predictions should be interpreted as relatively accurate.
To make this point, figure~\ref{tropical-turbulence-fluxes} compares the vertical temperature flux and flux divergence between the LES, CATKE, and a third single column GOTM run that uses the ``Generic Length Scale'' (GLS) closure reported by \citeA{reichl2024improved}. 
GLS is a second-moment closure similar to SMC-LT \cite{umlauf2003generic}, which is used to facilitate a comparison with \citeA{reichl2024improved}.
For some reason, the bursting behavior observed in both the LES and CATKE solutions is absent from GLS --- suggesting that CATKE may hold an edge over GLS (at least, the GLS implemented in GOTM with default free parameter choices) in modeling intermittent forced stratified shear turbulence.
The vertical structure of the flux divergences in CATKE are also more similar to the LES than the GLS solution.

\subsection{Sensitivity to vertical resolution and time-step}

\begin{figure}
    \includegraphics[width = 1\textwidth]{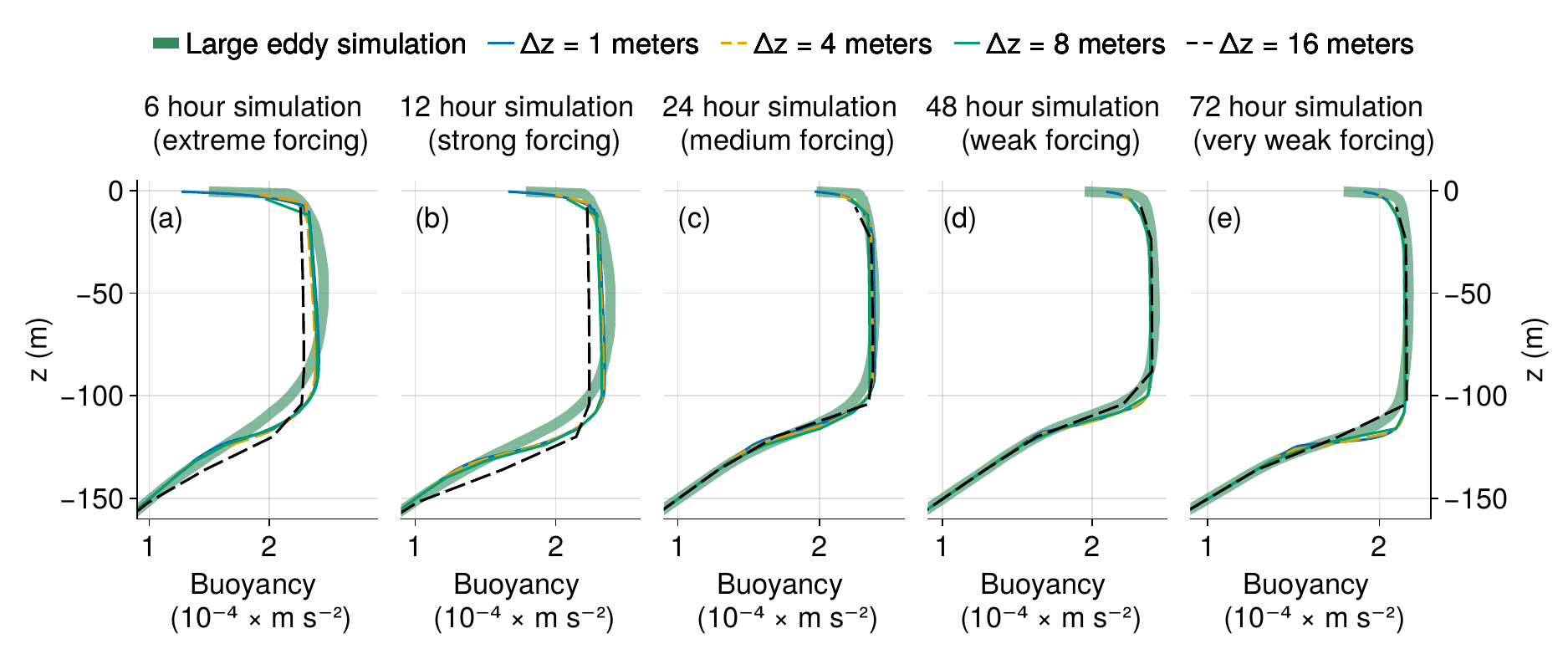}
    \caption{Illustration of sensitivity of CATKE predictions to vertical resolution for the weak wind, strong cooling case. Four vertical resolutions are shown: 1, 4, 8, and 16 meters. CATKE's calibration explicitly minimized errors between LES and CATKE simulations at 2, 4, and 8 meter resolution, such that the 1 and 16 meter cases represent ``extrapolation in resolution.'' The predictions are converged for resolutions 8 meters and finer, but the 16 meter resolution results exhibit small discrepancies from the converged solutions.}
    \label{vertical-resolution-sensitivity}
\end{figure}

Next we investigate the sensitivity of CATKE's predictions to numerical parameters like vertical resolution and time-step size --- a well-appreciated concern with ocean microscale parameterizations \cite{reffray2015modelling, van2018kpp}.
The sensitivity of CATKE's predictions to vertical resolutions ranging from 1 to 16 meters is shown in figure~\ref{vertical-resolution-sensitivity} for the weak wind, strong cooling case (the case for which CATKE exhibits the most bias).
Recall that CATKE was calibrated using simulations with 2-, 4-, and 8-meter vertical resolution, such that 1 and 16 meters represent ``extrapolation''.
Based off the results in figure~\ref{vertical-resolution-sensitivity}, we preliminarily conclude that CATKE is insensitive to vertical resolutions 8 meters and finer.
At 16 meter resolution, CATKE's predictions are still good compared to the biases observed for KPP and SMC-LT, but nevertheless start to deviate from the higher-resolution solutions and, in particular, tend to overmix.
It may be that with such a coarse resolution, it simply is not possible to resolve the structure of strongly-stratified entrainment layers at the base of the boundary layer.

\begin{figure}
    \includegraphics[width = 1\textwidth]{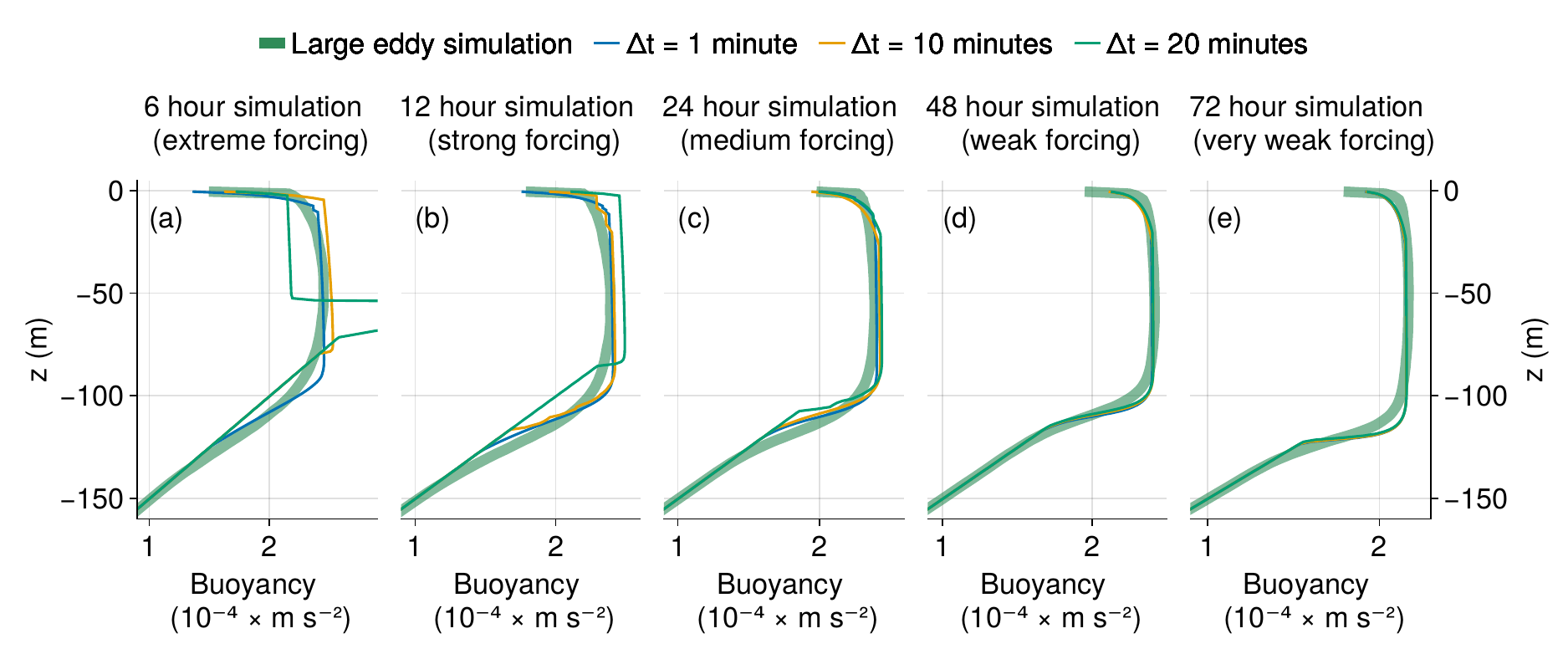}
    \caption{Sensitivity of CATKE predictions to time step for 1 meter vertical resolution for the weak wind, strong cooling case. At 1 meter resolution and in the strong forcing conditions of the 12- and 6-hour suites, CATKE solutions show time-step dependence for time steps longer than 1 minute. To enable longer time steps for high vertical resolutions in the presence of strong forcing, the substepping scheme described in~\ref{CATKE-numerics} is used and demonstrated in figure~\ref{substepping}.}
    \label{time-step-sensitivity}
\end{figure}

The sensitivity of CATKE's predictions to time-step size --- at a vertical resolution of 1 meter --- are shown in figure~\ref{vertical-resolution-sensitivity}.
Note that CATKE requires a smaller time step for finer vertical resolution.
Put differently, smaller time-steps are required to resolve the evolution of TKE, momentum, and tracers, and associated vertical transmission of information, on finer grids.
More strongly forced cases also require smaller time steps.
Figure~\ref{time-step-sensitivity}, and additional tests, show that with 1 meter vertical resolution, CATKE requires time-steps 2 minutes or shorter to resolve the dynamics associated with surface forcing as strong as that encountered in the 6-hour-suite.
(A 5-minute time step is adequately converged for the 12-, 24-, 48-, and 72-hour suite, however.)

\begin{figure}
    \includegraphics[width = 1\textwidth]{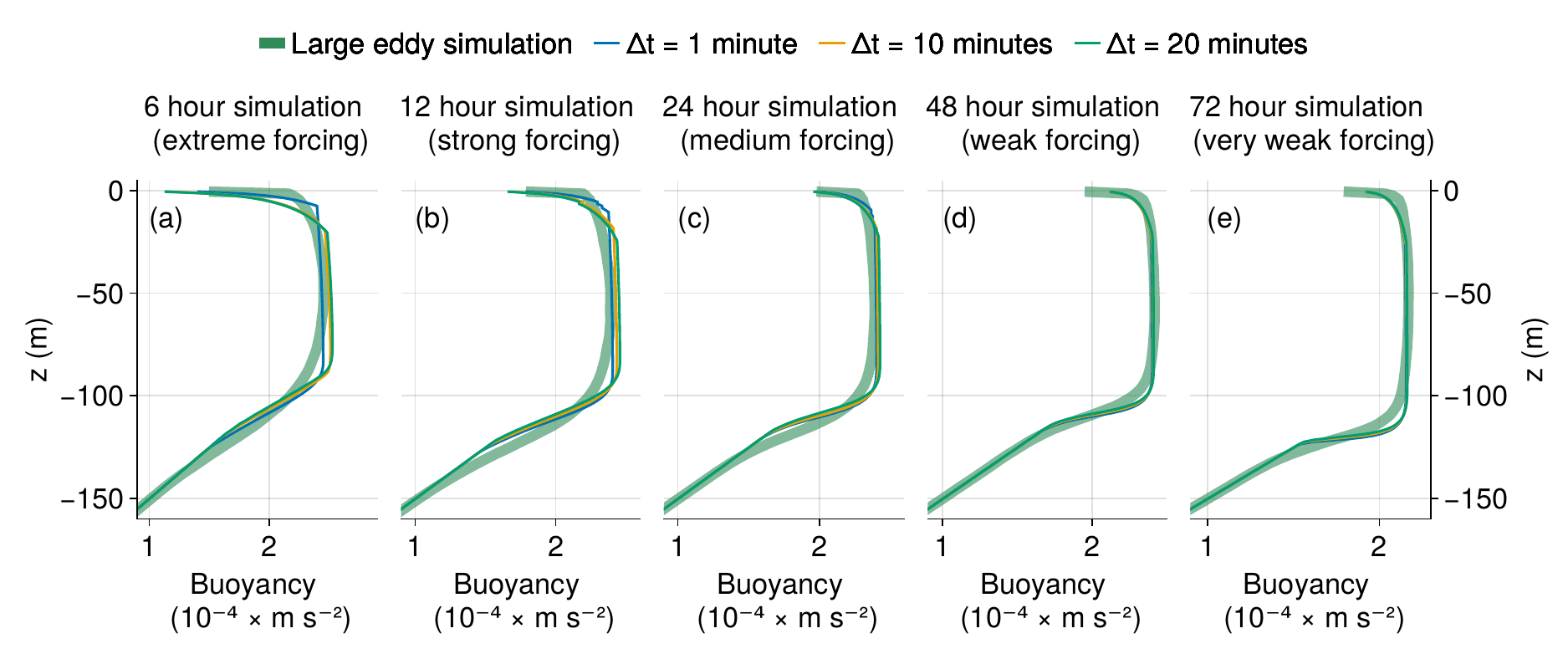}
    \caption{A comparison between LES and CATKE-parameterized single column simulations at 1 meter vertical resolution and three different momentum and tracer time-steps, when turbulent kinetic energy is substepped with a 30 second time step using the scheme described in~\ref{CATKE-numerics}. For the 6-hour suite, the time-step dependence is greatly reduced compared to the non-substepped case shown in figure~\ref{time-step-sensitivity}, but is not entirely converged. We suspect this is because even momentum and tracers require a time step shorter than 20 minutes for such strong forcing at high vertical resolution.}
    \label{substepping}
\end{figure}

We address this sensitivity of CATKE's predictions to time-step by implementing a novel split-explicit scheme that substeps the TKE using a short time-steps, while evolving momentum and tracers with a longer time-step.
The details are given in~\ref{CATKE-numerics}.
The results are shown in figure~\ref{substepping}, showing that CATKE generates converged predictions for momentum and tracer time-steps between 1 and 20 minutes when the TKE is substepped with a short 30 second time step.

\section{Discussion}
\label{sec:discussion}

This paper describes a novel one-equation parameterization for vertical fluxes by ocean microscale turbulence called ``CATKE''.
CATKE extends existing one-equation parameterizations \cite{blanke1993variability, madec2017nemo} with a dynamic model for convective adjustment capable of describing the wide range of convective mixing rates observed in the ocean surface boundary layer.
CATKE's 23 free parameters are calibrated against large eddy simulations accounting for discretization errors. We use \textit{a posteriori} calibration, meaning that the CATKE parameters are calibrated to capture the full temporal evolution of the coarse-grained variables rather than, for example, matching the unresolved eddy fluxes.
This approach improves both the accuracy and the stability of the calibrated parameterization. 

Our decision to develop a one-equation TKE-based parameterization rather than a $K$-profile parameterization \cite<KPP, see>{large1994oceanic, mcwilliams2009buoyancy, van2018kpp, reichl2018simplified, reichl2019parameterization} merits some discussion.
KPPs have a major advantage over TKE-based parameterizations in coarse resolution ocean models (especially with different time-steps for momentum and tracer variables) because they admit time-steps as long as 2 hours \cite{reichl2018simplified}.
In part, we are interested in one-equation parameterization because our focus is higher resolution, mesoscale-permitting and mesoscale-resolving simulations that require 1--10 minute time-steps to satisfy the advective numerical stability constraints of energetic solutions on relatively high-resolution grids.
CATKE adds no additional time step constraints to such simulations, while offering some significant benefits:
\textit{(i)} dynamic prediction of diffusivity vertical structure versus prescription via ``shape functions'';
\textit{(ii)} turbulent intensity growth and relaxation time scales or ``memory'', and
\textit{(iii)} better computational performance on hardware with fine-grained parallelism such as Graphics Processing Units (GPUs) used for example by Oceananigans \cite{OceananigansJOSS, silvestri2024mesoscaleGPU} and Veros \cite{hafner2021fast}, which are ill-suited for the nonlinear solvers for boundary layer depth common to KPP-type models \cite{zhang2020optimizing}.

Our calibration to a relatively limited range of LES cases reported in this paper is just the first step towards using CATKE for global ocean modeling and climate projection.
In particular, our ultimate objective is more accurate climate predictions with quantified uncertainties.
Addressing this ultimate goal requires first quantifying the uncertainty of CATKE's free parameters relative to high-resolution data, using the calibration context presented in this work.
Next, with prior parameter distributions in hand, CATKE's free parameters must then be recalibrated concomitant with other climate model free parameters against global climate observations to account for physics missing from the limited LES context used in this work, and to account for interactions between CATKE and other components of the climate model.

A second future step is to further calibrate CATKE to a more comprehensive suite of LES forced with temporally-varying surface fluxes, surface wave fields with $\La \ne 0.3$, and horizontal flux divergences \cite<for example following>{whitt2022simulation}.
These calibrations against more comprehensive LES will provide robust prior estimates of CATKE's parameters in preparation of the final goal of calibrating CATKE in a global context, by minimizing the mismatch between predictions of the ocean climate state and relevant observations with global or near-global coverage. 
More comprehensive calibration to more LES and to observations in a global context will likely reveal deficiencies to be addressed by further development of CATKE's formulation, such as accounting for the effect of surface waves on CATKE's mixing and dissipation length scales.

\appendix

\section{A synthetic dataset generated by large eddy simulations}
\label{les-description}

We use a synthetic dataset to calibrate and assess CATKE consisting of 35 idealized large eddy simulations (LES) of the ocean surface boundary layer with imposed constant surface fluxes of temperature and momentum and a simple surface wave field.

\subsection{Initial conditions}

The LES are initialized from rest with zero velocity and the piecewise-linear buoyancy stratification
\beq \label{initial-buoyancy}
b(z, t=0) = \left \{
\begin{matrix}
N^2_1 \, z & \quad \text{for} \, \, z > -h_1 \com \\[1ex]
N^2_2 \, z + \left (N^2_2 - N^2_1 \right ) h_1  & \quad \text{for} \, \, -h_2 > z > -h_1 \com \\[1ex]
N^2_3 \, z + \left (N^2_3 - N^2_2 \right ) h_2 +  \left (N^2_2 - N^2_1 \right ) h_1 & \quad \text{for} \, \, z < -h_2 \com
\end{matrix} \right .
\eeq
with $N^2_1 = N^2_3 = 2 \times 10^{-6} \, \rm{s^{-2}}$, $N^2_2 = 10^{-5} \, \rm{s^{-2}}$, $h_1 = 48 \, \rm{m}$, and $h_2 = 72 \, \rm{m}$.

% \subsection{Penetrating solar radiation}
% \label{appendix:solar-radiation}
% In the case of $J_b < 0$ and thus downward heat fluxes that represent heating by solar insolation, we distribute the buoyancy flux divergence throughout the interior of the model using the interior forcing term $F_b$ in~\eqref{tracers}, where $F_b = -\d_z I$ is the divergence of the solar insolation flux $I$, which is modeled with 
% \beq
% I(z) = J_b \left [ \epsilon_1 \ee^{z / \lambda_1} + \left (1 - \epsilon_1 \right ) \ee^{z / \lambda_2} \right ] \com
% \eeq
% where $\epsilon_1$ is the fraction of penetrating radiation absorbed over the vertical scale $\lambda_1$, and $(1 - \epsilon_1)$ is the remaining fraction of the radiation absorbed over $\lambda_2$.
% All simulations use $\epsilon_1 = 0.6$, $\lambda_1 = 1 \, \mathrm{m}$, and $\lambda_2 = 16 \, \mathrm{m}$ \cite<for example, see the solar insolation used by>{whitt2022simulation}.

\subsection{Passive tracer forcing}
\label{sec:passive-tracer-forcing}

We additionally simulate the evolution of a passive tracer $c$ which is forced by
\beq
F_c(z) = \omega_+ \ee^{- \left ( z - z_c \right )^2 / 2 \lambda_c^2} - \omega_- \com
\eeq
where $z_c$ is the depth of the forcing, $\lambda_c$ is the width of the forcing, $\omega_+$ is an inverse forcing time-scale that varies between each suite, and $\omega_-$ is chosen so that $F_c$ has zero mean, that is
\begin{linenomath*}
\begin{align}
\omega_- & \defn \frac{\omega_+}{L_z} \int_{-L_z}^0 \ee^{- \left ( z - z_c \right )^2 / 2 \lambda_c^2} \di z \nonumber\\
& \approx \omega_+ \frac{\lambda_c \sqrt{2 \pi}}{L_z} \com
\end{align}
\end{linenomath*}
where $L_z$ is the depth of the domain.
The approximation of the integral holds when the forcing is far from boundaries, or when $-L_z \ll z_c - \lambda_c$ and $0 \gg z_c + \lambda_c$.
We use $z_c = -96$~m and $\lambda_c = 8$~m for all cases.
For the forcing time scale $\omega_+^{-1}$, we use 15 minutes, 30~minutes, 1~hour, 2~hours, and 4~hours for the 6, 12, 24, 48, and 72~hour suites, respectively.

\subsection{Constant-flux boundary conditions}

The 35 simulations differ in their boundary conditions and Stokes drift.
The 35 simulations, which have different boundary conditions and S are organized into 5 ``suites'', each of which has 7 cases:
free convection, weak wind strong cooling, medium wind medium cooling,
strong wind weak cooling, strong wind, strong wind no rotation, and strong wind and sunny.
The suites differ by both forcing strength and duration, simulating 6, 12, 24, 48, and 72 hours of boundary layer turbulence respectively.
The forcing strength is chosen for each suite and case so that the boundary layer deepens to roughly half the depth of the domain;
for example, the ``6-hour suite'' has the strongest forcing, and the ``72-hour suite'' has the weakest forcing.
``Strong wind no rotation'' and ``strong wind and sunny'' use $f=0$, while the rest use the Coriolis parameter $f = 10^{-4} \, \rm{s^{-1}}$.
The surface fluxes for the 35 LES are summarized in tables~\ref{table:les-summary} and~\ref{table:les-validate-summary}.
To draw a connection between the LES suites and real air-sea flux conditions, tables~\ref{table:les-summary} and~\ref{table:les-validate-summary} provide an estimate of heat fluxes $Q$ for each case,
as well as an estimate of the atmospheric wind at 10 meters height using similarity theory (reduced to the case of neutral buoyancy fluxes for simplicity, see \citeA{large2009global}),
\beq \label{u10-estimate}
u_{10} = \sqrt{\frac{|\tau_a|}{c_{10}}} \com
\quad \text{where} \quad
c_{10} = \left ( \frac{\varkappa}{\log{\left (10 / \ell_r \right )}} \right )^2 \com
\quad \text{and} \quad
\ell_r = 0.011 \frac{|\tau_a|}{g} \com
\eeq
where $\ell_r$ is the Charnock roughness length given gravitational acceleration $g = 9.81 \, \mathrm{m \, s^{-2}}$, $\varkappa = 0.4$ is the von K\'arm\'an constant,
and $\tau_a = \rho_o \tau_x / \rho_a$ is the atmospheric kinematic momentum flux given ocean reference density $\rho_o = 1024 \, \mathrm{kg \, m^{-3}}$ and atmosphere density
$\rho_a = 1.2 \, \mathrm{kg \, m^{-3}}$.

\subsection{Stokes drift model}
\label{stokes-drift}

For all wind-forced cases, we additionally impose a surface wave field with a surface Stokes drift amounting to a constant ``Langmuir number'' $\La = \sqrt{u_\star / \US(z=0)} \approx 0.3$.
Our Stokes drift prescription models a surface wave field with the friction-number-dependent peak wavenumber
\beq
k_p = C_k \frac{g}{u_\star^2} \com
\eeq
where $u_\star = \sqrt{|\tau_x|}$ is the water-side friction velocity, $g$ is gravitational acceleration, and we use $C_k = 10^{-6}$.

We follow \citeA{lenain2020contribution} to estimate the depth-profiles of Stokes drift and Stokes drift shear.
The Stokes drift beneath a spectrum of deep-water waves is
\beq
\US(z) = 2 \int_{k_p}^{k_i} \ee^{2 k z} k \sqrt{g k} \, \chi(k) \di k \com
\eeq
where $\chi(k)$ is a one-dimensional wave spectrum that neglects ``directional spreading''.
The spectrum $\chi(k)$ is divided into an ``equilibrium range'' just above the spectral peak $k_p$, and a ``saturation range'' at even higher wavenumbers:
\beq
\chi(k) = \Bigg \{
\begin{matrix}
    \frac{C_\beta}{2 \sqrt{g}} a_\star k^{-5/2} \quad & \text{for} \quad k_p < k < k_n \quad \text{(equilibrium)} \com \\
    C_B k^{-3} \quad & \text{for} \quad k_n < k < k_i \quad \text{(saturation)} \com
\end{matrix}
\eeq
where $k_n$ is a transition wavenumber between equilibrium and saturation ranges, $k_i$ is an upper wavenumber cutoff above which waves are assumed to be isotropic and there do not contribute to Stokes drift.
$a_\star = u_\star \sqrt{\rho_o / \rho_a}$ is the air-side friction velocity defined in terms of the water-side friction velocity $u_\star$, a reference air density $\rho_a = 1.2 \, \rm{kg \, m^{-3}}$ and ocean density $\rho_o = 1024 \, \rm{kg \, m^{-3}}$.
Wavenumbers \textit{below} the spectral peak $k_p$ are assumed too weak to contribute appreciably to Stokes drift.

Both the transition wavenumber $k_n$ and the isotropic wavenumber $k_i$ decrease with increasing $u_\star$:
\begin{gather}
k_n \defn C_r g a_\star^{-2} \com \\
k_i \defn C_i g a_\star^{-2} \com
\end{gather}
where $C_r = 9.7 \times 10^{-3}$ and $C_i = 0.072$.

The Stokes drift is
\beq
\US(z) = C_\beta a_\star \int_{k_p}^{k_n} \frac{\ee^{2 k z}}{k} \di k + 2 C_B \sqrt{g} \int_{k_n}^{k_i} k^{-3/2} \ee^{2 k z} \di k \per
\eeq 
Noting that $\int_{k_p}^{k_n} k^{-1} \ee^{2 k z} \di k = \Ei(2 k_n z) - \Ei(2 k_p z)$, where $\Ei$ is the exponential integral function, we find
%\beq
%\int_{k_n}^{k_i} k^{-3/2} \ee^{2kz} \di z = \upsilon(k_n) - \upsilon(k_i) \com \quad \text{where} \quad \upsilon(k) = \frac{2}{\sqrt{k}} \left [ \ee^{2 k z} + \sqrt{2 \pi k |z|} \, \erf \left( \sqrt{2 k |z|} \right) \right ] \per 
%\eeq
%Thus
\beq
\US(z) = C_\beta a_\star \left [ \Ei(2 k_n z) - \Ei(2 k_p z) \right ] + 2 C_B \sqrt{g} \left [ \upsilon(k_n) - \upsilon(k_i) \right ] \com
\eeq
and 
\begin{linenomath*}
\begin{align}
\d_z \US &= 2 C_\beta a_\star \int_{k_p}^{k_n} \ee^{2 k z} \di k + 4 C_B \sqrt{g} \int_n^I \frac{\ee^{2kz}}{\sqrt{k}} \di k \com \\
&= C_\beta a_\star \frac{\ee^{2k_p z} - \ee^{2 k_n z}}{|z|} + 2 C_B \sqrt{\frac{2 \pi g}{|z|}} \left [ \erf \left (\sqrt{2 k_n |z|} \right ) - \erf \left (\sqrt{2 k_i |z|} \right ) \right ] \com
\end{align}
\end{linenomath*}
for the Stokes shear.

\subsection{LES uncertainty: effects of resolution and Stokes drift}

%Explain that we incorporate the uncertainty of the LES results into the calibration, which means that the LES do not have to be ``perfect'' but rather need to have well-characterized uncertainty.
%The characterization of LES uncertainty becomes important, however, during validation, because we want to assess whether the remaining bias exhibited by single column results lies within the LES uncertainty (and therefore is expected) or outside (therefore representing a real bias).

\begin{figure}[ht]
    \label{12-hour-resolution-dependence}
    \includegraphics[width = 1\textwidth]{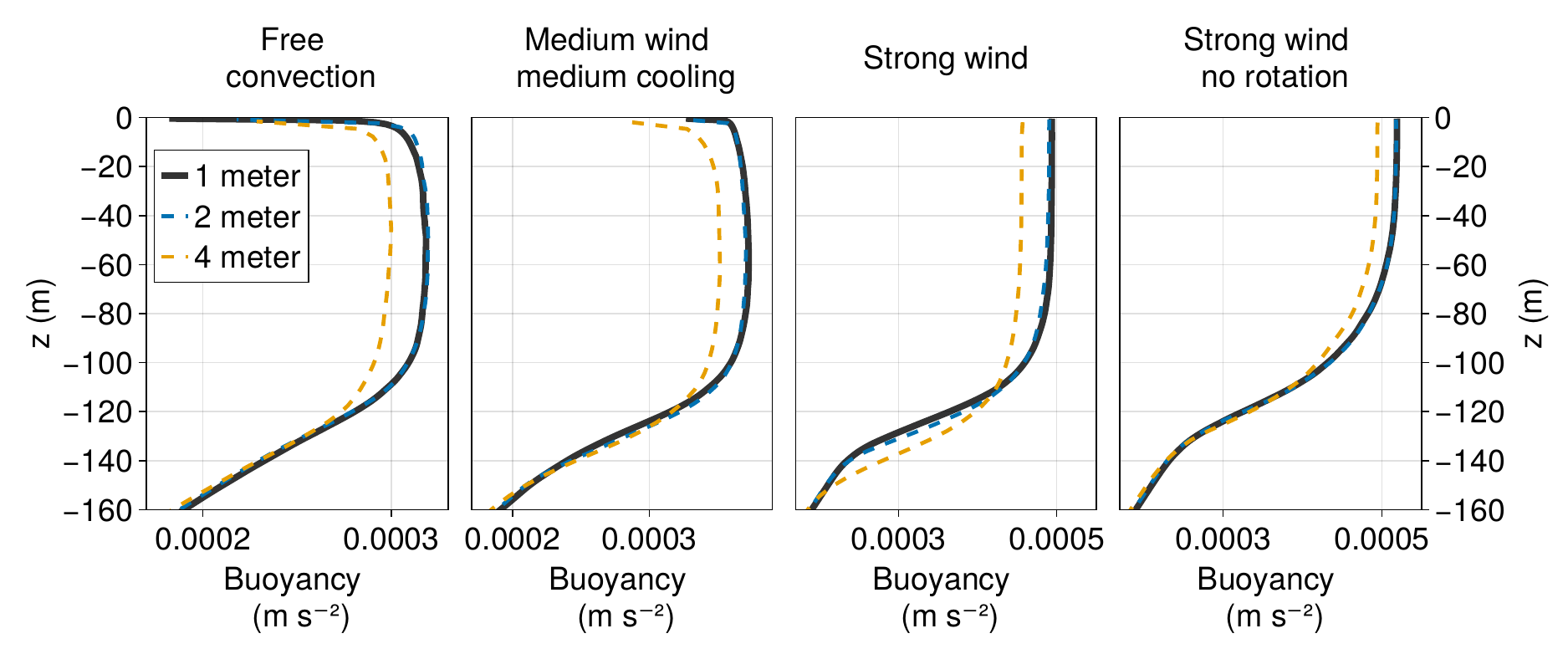}
    \caption{Resolution dependence of 12-hour LES.}
\end{figure}

All LES use 2 meter horizontal resolution and a stretched vertical resolution that varies from 0.8 meters in the upper half of domain to 2.3 meters at the bottom.
We refer to this as ``1 meter'' vertical resolution.
To account for the effects of resolution on the 35 LES used as synthetic observations in this paper, we run 70 additional LES on coarser grids with double (``2 meter'') and quadruple (``4 meter'') resolution, and use these to estimate the observational uncertainty used in calibration (see~\ref{sec:calibration} for more details).
The effect of resolution depends on forcing strength: for the 6 and 12 hour suite, the results are nearly identical for 1- and 2-meter vertical resolution.
Figure~\ref{12-hour-resolution-dependence} shows the results for 4 cases in the 12 hour suite.
Note that in the free convection case, the first two grid points exhibit a strong unstable stratification in the 12 hour suite.
We attribute this to an artificial reduction of mixing near the top boundary of the LES.
It might be possible to address this artificially-strong unstable mean stratification by introducing, for example, a surface-concentrated eddy diffusivity.
However, because the LES are used only for training CATKE and thus matter mostly in their predicted boundary layer depth, we choose instead to ignore the top 4 m when computing the LES--CATKE discrepancy during calibration.

\begin{figure}[ht]
    \label{24-hour-resolution-dependence}
    \includegraphics[width = 1\textwidth]{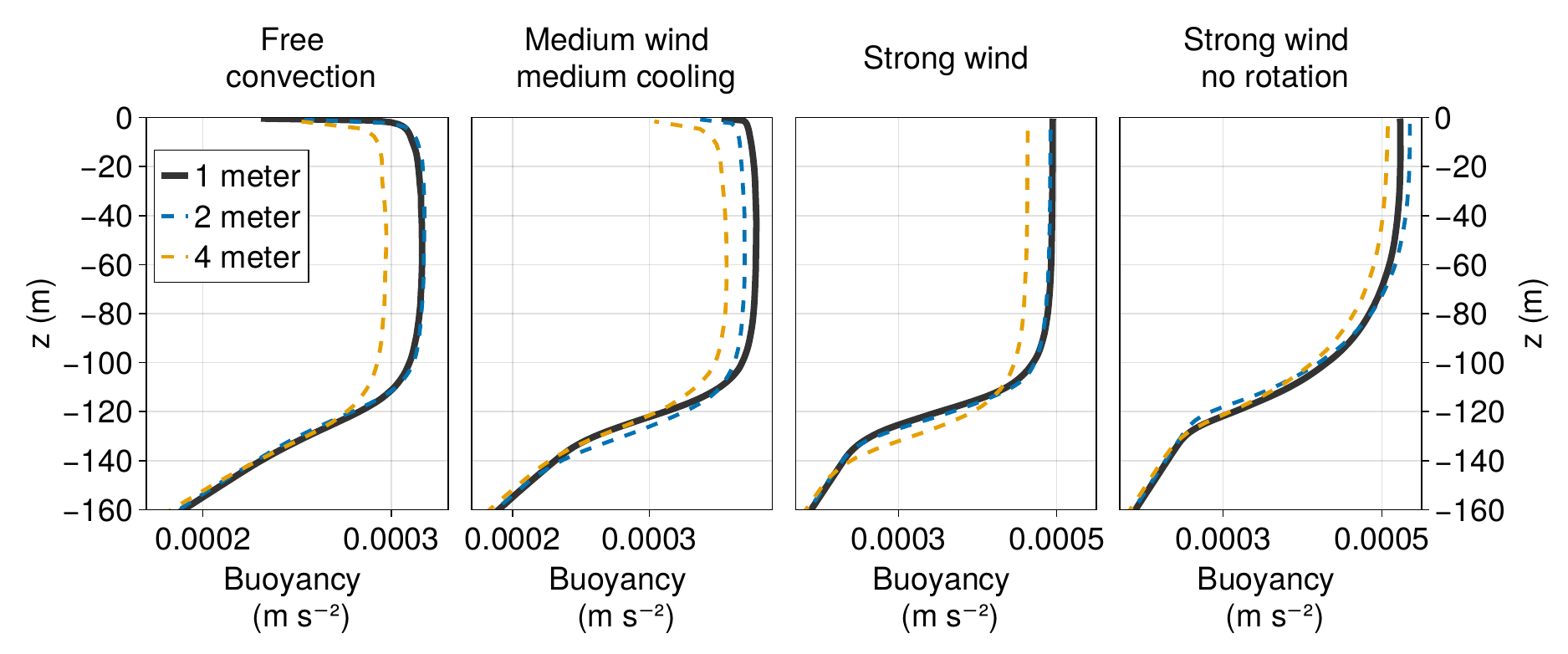}
    \caption{Resolution dependence of 24-hour LES.}
\end{figure}

Figure~\ref{24-hour-resolution-dependence} shows the resolution dependence of the 24-hour suite.
These LES show slightly more resolution dependence than the 12-hour suite, especially for cases forced by a combination of wind and cooling.
This indicates that our LES data for more weakly forced cases are \textit{less certain} than the strongly forced cases.
Interestingly, we find that CATKE exhibits the least bias for the weakly forced cases than for the strongly forced cases.
This means that the bias exhibited in the strongly-forced cases is real bias, while the weakly forced cases may be interpreted as exhibiting essentially no bias.

The LES also use an ``implicit closure'' technique whereby advection is discretized with a 9th order weighted essentially non-oscillatory scheme (or WENO for short) and no explicit subgrid-scale closure is added.

\subsection{Effect of Stokes drift on LES results}

Next we turn to the effect that including the Stokes drift profile described in section~\ref{stokes-drift} has on our LES results.
The inclusion of Stokes drift in our LES is an attempt to make them slightly more realistic.
In other words, we hypothesize that calibrating CATKE to LES without surface waves would generally lead to a shallow bias in mixed layer depth prediction with CATKE --- since surface waves are always present above real wind-forced ocean surface boundary layers.

\begin{figure}[ht]
    \label{12-hour-stokes-dependence}
    \includegraphics[width = 1\textwidth]{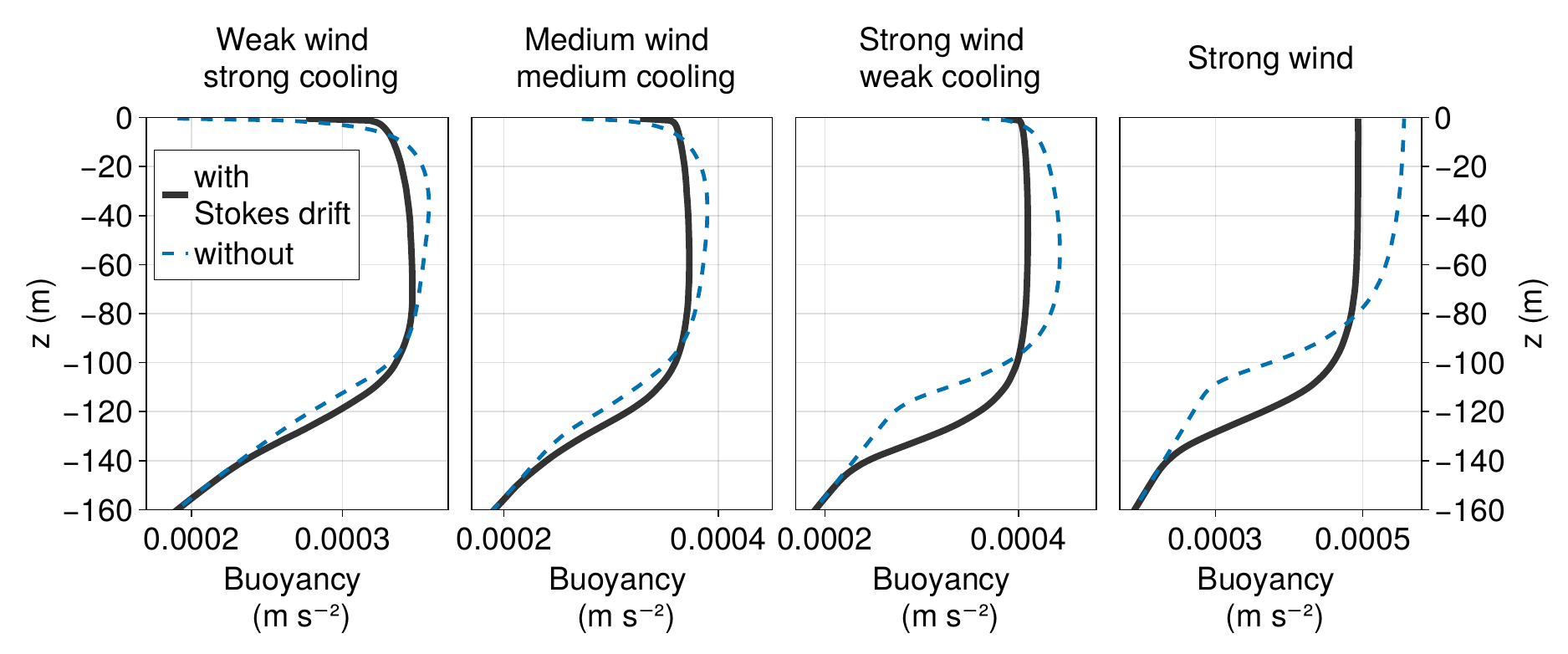}
    \caption{Stokes drift dependence of 12-hour LES.}
\end{figure}

This notion is corroborated by figure~\ref{12-hour-stokes-dependence}, which shows the horizontally-averaged buoyancy profiles for 4 LES in the 12 hour suite, with and without Stokes drift.
As expected, the inclusion of Stokes drift produces more mixing and makes the boundary layer deeper.
The effect of Stokes drift is minor in the case of weak and medium winds (leftmost and second from left panels).
In the strong wind (and rotating) case, the inclusion of Stokes drift makes the boundary layer 20 meters deeper, or around 20\% of the total.
In the strong wind, no rotation case, the case without Stokes drift completely fails to transition to the turbulence. 
(A small amount of cooling would probably be required to produce turbulence in the strong wind, no rotation case without Stokes drift.)

\section{Split-explicit turbulent kinetic energy time stepping and vertical discretization
\label{CATKE-numerics}}

The time discretization is a little non-trivial since we step forward velocity and tracers first, then step forward TKE and also use substepping/split--explicit scheme for TKE.
In the single column case, we integrate equations~\eqref{u1d}--\eqref{c1d} with the backward Euler scheme
\begin{linenomath*}
\begin{gather}
\frac{u^{n+1} - u^n}{\Delta t} = \d_z \left ( \kappa_u^n \d_z u^{n+1} \right ) \label{substep-u} \com \\
\frac{v^{n+1} - v^n}{\Delta t} = \d_z \left ( \kappa_u^n \d_z v^{n+1} \right ) \label{substep-v} \com \\
\frac{c^{n+1} - c^n}{\Delta t} = \d_z \left ( \kappa_c^n \d_z c^{n+1} \right ) \label{substep-c} \com
\end{gather}
\end{linenomath*}
where $\Delta t$ is the time step and the superscripts $n$ or $n+1$ indicate the time-level at which the quantity is evaluated.
For the TKE equation~\eqref{e1d}, we introduce a substepping scheme that uses $M$ short time-steps $\Delta t / M$ to integrate $e$ between $n$ to $n+1$,
\beq \label{substep-e}
\frac{e^{m+1} - e^m}{\Delta t / M} =
\underbrace{\mystrut{2.7ex} \d_z \left ( \kappa_e^m \d_z e^{m+1} \right )}_\text{transport}
+ \underbrace{\mystrut{2.7ex} \kappa_u^n \half \left ( \d_z \bu^n + \d_z \bu^{n+1} \right ) \cdot \d_z \bu^{n+1}}_\text{shear production}
+ \overline{w'b'}^m
- \underbrace{\mystrut{2.7ex} \frac{\sqrt{e^m}}{\ell_D^m} e^{m+1}}_\text{dissipation} \com
\eeq
where the superscripts $m$ and $m+1$ denote the substep level.
The buoyancy flux $\overline{w'b'}^m$ in~\eqref{substep-e} is discretized in time using the conditionally-implicit ``Patankar trick'' \cite{burchard2002energy}, such that
\beq
\overline{w'b'}^m = \left \{
\begin{matrix}
- \kappa_c^n \d_z b^{n+1} & \text{when} & \d_z b^{n+1} \le 0 \\[1ex]
- \kappa_c^n \d_z b^{n+1} \frac{e^{m+1}}{e^m} & \text{when} & \d_z b^{n+1} > 0
\end{matrix} \right .
\eeq
which improves the stability of~\eqref{substep-e} and keeps $e$ from becoming too negative.
Note that shear production is not updated during substepping.
The time discretization of the shear production term in~\eqref{substep-e}, which incorporates shear measured at the time step $n$ and $n+1$, also follows \citeA{burchard2002energy} and requires an algorithm that stores the velocity field at time step $n$, stepping forward momentum and tracers, and then substepping forward $e$.

We discretize $u$, $v$, $c$, and $e$ on a staggered vertical grid (not shown), with all variables vertically located at cell centers --- a deviation from \citeA{blanke1993variability}, \citeA{burchard2002energy}, or \citeA{madec2017nemo} who place $u, v, c$ at vertical cell centers but TKE at vertical cell interfaces where the diffusivity is computed (otherwise known as ``$w$-locations'').
Because $\kappa_c$, $\kappa_c$, and $\kappa_e$ are located at vertical cell interfaces, this discretization means that $e$ must be reconstructed from cell centers to cell interfaces to compute $\kappa_u$, $\kappa_c$, and $\kappa_e$ according to~\eqref{eddy-diffusivity}.
The vertical spatial discretization of the shear production term is derived from the mean kinetic energy equation following \citeA{burchard2002energy}, but adapted to our cell-centered placement of $e$. 
We use a tridiagonal solve to advance $u, v, c, e$ in~\eqref{substep-u}--\eqref{substep-e} over each time step of substep, treating both diffusion and linear terms in~\eqref{substep-e} implicitly.

\section{A posteriori calibration}
\label{calibration-details}

We use Ensemble Kalman Inversion \cite<EKI; >{iglesias2013ensemble} to calibrate CATKE.
EKI is a gradient-free and computationally inexpensive method for solving nonlinear inverse problems.
EKI supposes that a forward map $\G(\c)$ can predict uncertain observations $\y$ given a set of free parameters $\c$,
\begin{linenomath*}
\beq \label{model-data-relationship}
\y = \G(\c) + \eta \com
\eeq
\end{linenomath*}  
where $\eta \sim \mathcal{N} \left (0, \M \right )$ is normally-distributed random uncertainty with covariance $\M$.
Four objects appear in the model-data relation~\eqref{model-data-relationship},
\begin{enumerate}
    \item \textit{Observations} $\y$ with $M$ discrete elements $\y_m$. In this paper, each $\y_m$ represents a state variable like velocity $U$ or temperature $\Theta$ at a particular depth and time, computed from LES data by horizontal averaging and vertical coarse-graining, and then normalized and shifted to have zero mean and unit variance.
    \item A \textit{parameter set} $\c$ with $P$ free parameter values $\c_p$.
    \item A \textit{forward map} $\G(\c)$ whose elements $\G_m(\c)$ predict the observation $\y_m$.
    $\G(\c)$ represents a \textit{model} that maps a parameter set $\c$ to the space of observations $\y$.
    In our case, constructing $\G(\c)$ requires forward evaluations of 63 single column models parameterized by $\c$, each predicting the evolution of horizontally-averaged variables in 21 LES at 2-, 4-, and 8-meter resolution.
    \item Random Gaussian \textit{uncertainty} $\eta \sim \mathcal{N}(0, \M)$ with covariance $\M$ associated with both $\G_m(\c)$ and $\y_m$. $\eta$ conflates uncertainty in $\y$ with ``structural'' uncertainty associated with imperfect forward maps $\G$.
\end{enumerate}

The elements of $\y$ are the discrete values of the horizontally-averaged temperature and velocity fields output from 21 LES coarse-grained to three grids with uniform 2-, 4-, and 8-meter spacing.
Each physical field is shifted, normalized, and weighted before being assembled into~$\y$.
Each forward map $G \left (\c \right )$ involves $3 \times 21 = 63$ simulations to find $U$, $V$, and $\Theta$ profiles for each LES case at the two model vertical resolutions.

\subsection{Ensemble Kalman dynamics}

Ensemble Kalman Inversion uses a dynamical system that governs the evolution of an ensemble of $N$ parameter sets, or ``particles'', $\CC \defn [\c^1, \c^2, \cdots, \c^N]$.
Here the superscript $\omega$ denotes the ``particle index'', which varies across the ensemble:
$\c^\omega_p$ is the $p^\mathrm{th}$ parameter value of the $\omega^\mathrm{th}$ particle.

Each parameter set $\c^\omega$ obeys the ordinary differential equation
\beq \label{eki-dynamic}
\frac{\rm d}{{\rm d} \t} \c^\omega= - \cov (\CC, \GC) \, \Gamma^{-1} \left ( \G^\omega - \y \right ) \com
\eeq
where $\G^\omega \defn \G(\c^\omega)$ is the forward map computed with the parameter set $\c^\omega$,
and $\t$ is the ``pseudotime''. %, and $\cov(\CC, \GC)$ is the parameter-map covariance matrix that couples the ensemble.
The matrix $\cov(\CC, \GC)$ in~\eqref{eki-dynamic} is the covariance matrix estimated from ensemble statistics at pseudotime $\t$, thus coupling the evolution of the ensemble.
For two ``ensemble matrices'' $\mathbf{A}$ and $\mathbf{B}$, where $\mathbf{A}$ for example is constructed from an ensemble of vectors $[A^1_i, A^2_i, \cdots, A^N_i]$,
the elements $\cov_{ij}(\mathbf{A}, \mathbf{B})$ are defined
\beq \label{ensemble-covariance}
\cov_{i j}\left (\mathbf{A}, \mathbf{B} \right ) \defn \frac{1}{N} \sum_{\omega=1}^N \Big ( A^\omega_i - \langle A \rangle_i \Big ) \Big ( B^\omega_j - \langle B \rangle_j \Big ) \com 
\quad \text{with} \quad
\langle C \rangle_i \defn \frac{1}{N} \sum_{\omega = 1}^N C^\omega_i \per
\eeq

For nearly-linear maps $\G_m(\c) \approx H_{m p} \c_p$,~\eqref{eki-dynamic} reduces to
\beq \label{linear-eki-dynamic}
\frac{\rm d}{{\rm d} \t} \c^\omega \approx - \cov(\CC, \CC) \, \nabla_{\c} \Phi^\omega \com
\eeq
where $\cov_{pq}(\mathbf{C}, \mathbf{C})$ is the $P \times P$ parameter-parameter covariance matrix \cite{kovachki2019ensemble}.
The ``EKI objective'' $\Phi^\omega$ associated with parameter set $\omega$ appears in~\eqref{linear-eki-dynamic}, where
\beq \label{objective}
\Phi(\G, \y; \c) \defn \big \| \M^{-1/2} \left [ \G(\c) - \y \right ] \big \|^2 \com
\eeq
and $\Phi^\omega \defn \Phi(\G, \y; \c^\omega)$.
$\Phi$ in~\eqref{objective} is a functional of $\G$ that measures the uncertain discrepancy between $\G(\c) - \y$.
The system~\eqref{linear-eki-dynamic} minimizes $\Phi$ using gradient descent preconditioned with $\cov(\CC, \CC)$, where the gradients $\nabla_{\c} \Phi$ are estimated from the parameter ensemble.

We integrate the EKI dynamical system~\eqref{eki-dynamic} in using a forward Euler discretization,
\beq \label{euler-eki-dynamic}
\c^\omega \, \big |_{n+1} = \c^\omega \, \big |_n - \Delta \t \, \Big [ \cov(\CC, \GC) \M^{-1} \left ( \G^\omega - \y \right ) \Big ]_n \com
\eeq
where $n$ is the pseudotime iteration, $\Delta \t$ is a pseudotime step size, and $\omega \in [1, N_e]$ is the ``ensemble index'' out of an ensemble with $N_e$ members.
The adaptive step size $\Delta \t$ is chosen at each iteration according to \citeA{kovachki2019ensemble}.
The initial parameter sets $\c^\omega$ at $\t=0$ are generated by randomly sampling the priors listed in table~\ref{favorite-parameters}.
%\ref{calibration-details} provides additional details about how we impose the parameter bounds in table~\ref{favorite-parameters}, and how we use Oceananigans \cite{OceananigansJOSS} in concert with two other Julia packages, EnsembleKalmanProcesses \cite{EnsembleKalmanProcesses.jl} and ParameterEstimocean \cite{ParameterEstimocean.jl} to integrate~\eqref{euler-eki-dynamic}.

EKI is practical for two reasons: \textit{(i)} it does not require explicit gradients of $\G$ with respect to parameters $\c$, and \textit{(ii)} the forward map evaluations $\G^\omega$ --- the most expensive part of integrating~\eqref{eki-dynamic} --- are independent and thus easily parallelized.
Reason \textit{(i)} means EKI is applicable to any simulation framework with changeable parameters $\c$.
Reason \textit{(ii)} means that considerable yet distributed resources can be leveraged efficiently: given sufficient distributed resources, the cost of a single EKI iteration depends only on the cost of a single forward map evaluation, independent of ensemble size.
This parallelizability benefits small problems such as calibration in a single column context and is decisive for large problems like global ocean calibration.

\subsection{Uncertainty covariance}
We associate the uncertainty $\M$ with the numerical fidelity of the large eddy simulations by defining
\beq
\M= \text{cov}\left ( [\y^\text{1m} \, \y^\text{2m} \, \y^\text{4m}] \right ) \com
\eeq
where $\y^\text{1m}, \y^\text{2m}, \y^\text{4m}$ denote observations obtained from LES with 1-, 2-, and 4-meter vertical resolution.

\subsection{Constrained and unconstrained parameters}
\label{appendix:bounds}

The dynamics~\eqref{euler-eki-dynamic} require normally-distributed parameters $\c_p$, which precludes the imposition of strict bounds such as non-negativity.
We therefore introduce the forward and inverse transforms,
\beq \label{logit-transformation}
\c_p = \log \frac{b - \tilde \c_p}{\tilde \c_p - a} \qquad \text{and} \qquad  \tilde \c_p = a + \frac{b - a}{1 + \exp(\c_p)} \com
\eeq
between ``constrained'' physical parameters $\tilde \c$ that are bounded between $(a, b)$, and unconstrained parameters $\c$.
The transformation~\eqref{logit-transformation} implies that if $\c_p$ is normally-distributed then $\tilde \c$ is bounded by $(a, b)$ with a scaled, shifted logit-normal distribution.

We denote the scaled, shifted logit-normal distribution bounded by $(a, b)$ as $\B(a, b)$ and use it to model the distribution of all of CATKE's free parameters.
The distributions $\B(a, b)$ formulated so their corresponding normal distributions have zero mean and unit variance.
When integrating~\eqref{euler-eki-dynamic}, the normally-distributed parameter sets $\c^\omega$ are transformed into their physical space counterparts $\tilde \c^\omega$ via~\eqref{logit-transformation} when evaluating $\G^\omega = \G(\c^\omega)$ and thus solving the single column equations~\eqref{u1d}--\eqref{c1d} and~\eqref{e1d}.

\subsection{Failure criterion handling}

Poor parameter choices $\c^\omega$ often lead to failed simulations of the single column system~\eqref{u1d}--\eqref{c1d} and~\eqref{e1d}.
In that case the forward map $\G^\omega$ is not informative and must be ignored when performing the Euler step~\eqref{euler-eki-dynamic}.

We first define the median and the ``median absolute deviation'' of the EKI objective samples, $\Phi^\omega \defn \Phi(\G, \y; \c^\omega)$,
\beq
\tilde \Phi \defn \text{median} \left ( \Phi^\omega \right )
\qquad \text{and} \qquad
s \defn \text{median}\left ( \big | \Phi^\omega - \tilde \Phi \big | \right )\com
\eeq
We mark a particle $\omega$ as ``failed'' if
\beq
\Phi^\omega > \tilde \Phi + 3 s \per
\eeq
This excludes both non-finite and just ``particularly anomalous'' $\Phi^\omega$.

\section*{Open Research Section}
% AGu requires an Availability Statement for the underlying data needed to understand, evaluate, and build upon the reported research at the time of peer review and publication.

% Authors should include an Availability Statement for the software that has a significant impact on the research. Details and templates are in the Availability Statement section of the Data and Software for Authors Guidance: \url{https://www.agu.org/Publish-with-AGu/Publish/Author-Resources/Data-and-Software-for-Authors#availability}

% It is important to cite individual datasets in this section and, and they must be included in your bibliography. Please use the type field in your bibtex file to specify the type of data cited. Options include [Dataset], [Software], [ComputationalNotebook], [Collection].

This work relied on the open-source software LESbrary.jl \cite{LESbrary.jl} and Oceananigans.jl \cite{OceananigansJOSS} to run the LES, Oceananigans.jl to run calibration simulations, and ParameterEstimocean.jl \cite{ParameterEstimocean.jl} and EnsembleKalmanProcesses.jl \cite{EnsembleKalmanProcesses.jl} for the Ensemble Kalman Inversion.
Visualizations were made using Makie.jl \cite{Makie.jl}.
Scripts for performing the calibration are available at  the GitHub repository \href{https://github.com/glwagner/SingleColumnModelCalibration.jl}{\texttt{github.com/glwagner/SingleColumnModelCalibration.jl}}.

\acknowledgments
This project is supported by Schmidt Sciences, LLC and by the National Science Foundation grant AGS-1835576.
N.C.C.~is additionally supported by the Australian Research Council under DECRA Fellowship DE210100749 and the Center of Excellence for the Weather of the 21st Century CE230100012.
Without implying their endorsement, we would also like to acknowledge stimulating and useful conversations with Bruno Deremble, Alex Legay, Qing Li, Brodie Pearson, Brandon Reichl, Roger Samelson, and Bill Young.

%% ------------------------------------------------------------------------ %%
%% References and Citations

%%%%%%%%%%%%%%%%%%%%%%%%%%%%%%%%%%%%%%%%%%%%%%%
%
% \bibliography{<name of your .bib file>} don't specify the file extension
%
% don't specify bibliographystyle

% In the References section, cite the data/software described in the Availability Statement (this includes primary and processed data used for your research). For details on data/software citation as well as examples, see the Data & Software Citation section of the Data & Software for Authors guidance
% https://www.agu.org/Publish-with-AGu/Publish/Author-Resources/Data-and-Software-for-Authors#citation

%%%%%%%%%%%%%%%%%%%%%%%%%%%%%%%%%%%%%%%%%%%%%%%

\end{document}